\def\spacingNumerator{5}
\def\spacingDenominator{4}

\def\ifundefined#1{\expandafter\ifx\csname#1\endcsname\relax}
\ifundefined{ftmagnification}  \def\ftmagnification{1200} \fi
\ifundefined{spacingNumerator}  \def\spacingNumerator{5} \fi
\ifundefined{spacingDenominator}  \def\spacingDenominator{4} \fi


\magnification\ftmagnification
\tolerance=10000
\hsize=17truecm\vsize=23truecm

\parindent=40pt
\mathsurround=0pt
     \multiply\baselineskip by \spacingNumerator
     \divide \baselineskip by \spacingDenominator 

%
%
\def\today{\ifcase\month\or January\or February\or March\or April\or
     May\or June\or July\or August\or September\or October\or November\or
     December\fi\space\number\day, \number\year}
%
%
\def\dst{\displaystyle}
\def\sst{\scriptstyle}
\def\tst{\textstyle}
\def\ssst{\scriptscriptstyle}
%
%
\def\frac#1#2{\dst {#1\over#2}}     
\def\sfrac#1#2{{\tst{#1\over#2}}}   

\def\deqalign#1{\vcenter{\openup1\jot \mathsurround=0pt \ialign{
                \strut\hfil$\displaystyle{##}$&&$\displaystyle{{}##}$\hfil
                \crcr
                #1\crcr}}}         

\def\meqalign#1{\vcenter{\openup1\jot \mathsurround=0pt \ialign{
                &\strut\hfil$\displaystyle{##}$&$\displaystyle{{}##}$\hfil&
                \quad$##$\crcr
                #1\crcr}}}         

%
%
\def\al{\alpha}
\def\be{\beta}
\def\ga{\gamma}
\def\de{\delta}

\def\ze{\zeta}
\def\et{\eta}
\def\th{\theta}
\def\ka{\kappa}
\def\la{\lambda}
\def\rh{\rho}
\def\si{\sigma}

\def\Ga{\Gamma}

\def\Si{\Sigma}

\def\Om{\Omega}   
%
%
\def\pmb#1{\setbox0=\hbox{#1}       
     \kern-.025em\copy0\kern-\wd0
     \kern.05em\copy0\kern-\wd0
     \kern-.025em\box0}             
\def\0{{\bf 0}}

\def\k{{\bf k}}

\def\t{{\bf t}}

\def\x{{\bf x}}
\def\y{{\bf y}}

\def\p{{\bf p}}

\def\cB{{\cal B}}
\def\cE{{\cal E}}
\def\cF{{\cal F}}
\def\cG{{\cal G}}

\def\cS{{\cal S}}

%
%
\font\tenfrak                 = eufm10
\font\sevenfrak               = eufm7
\font\fivefrak                = eufb5
\newfam\frakfam
     \textfont\frakfam=\tenfrak
     \scriptfont\frakfam=\sevenfrak   
     \scriptscriptfont\frakfam=\fivefrak
\def\frak{\fam\frakfam\tenfrak}
\font \tensans                = cmss10
\font \fivesans               = cmss10 at 5pt
\font \sevensans              = cmss10 at 7pt
\newfam\sansfam
     \textfont\sansfam=\tensans
     \scriptfont\sansfam=\sevensans
     \scriptscriptfont\sansfam=\fivesans
\def\sans{\fam\sansfam\tensans}
%
%
\def\bbbr{{\rm I\!R}}  
\def\bbbn{{\rm I\!N}}

\def\bbbone{{\mathchoice {\rm 1\mskip-4mu l} {\rm 1\mskip-4mu l}    
{\rm 1\mskip-4.5mu l} {\rm 1\mskip-5mu l}}}
\def\bbbc{{\mathchoice {\setbox0=\hbox{$\displaystyle\rm C$}\hbox{\hbox 
to0pt{\kern0.4\wd0\vrule height0.9\ht0\hss}\box0}}
{\setbox0=\hbox{$\textstyle\rm C$}\hbox{\hbox
to0pt{\kern0.4\wd0\vrule height0.9\ht0\hss}\box0}}
{\setbox0=\hbox{$\scriptstyle\rm C$}\hbox{\hbox
to0pt{\kern0.4\wd0\vrule height0.9\ht0\hss}\box0}}
{\setbox0=\hbox{$\scriptscriptstyle\rm C$}\hbox{\hbox
to0pt{\kern0.4\wd0\vrule height0.9\ht0\hss}\box0}}}}
\def\bbbq{{\mathchoice {\setbox0=\hbox{$\displaystyle\rm               
Q$}\hbox{\raise
0.15\ht0\hbox to0pt{\kern0.4\wd0\vrule height0.8\ht0\hss}\box0}}
{\setbox0=\hbox{$\textstyle\rm Q$}\hbox{\raise
0.15\ht0\hbox to0pt{\kern0.4\wd0\vrule height0.8\ht0\hss}\box0}}
{\setbox0=\hbox{$\scriptstyle\rm Q$}\hbox{\raise
0.15\ht0\hbox to0pt{\kern0.4\wd0\vrule height0.7\ht0\hss}\box0}}
{\setbox0=\hbox{$\scriptscriptstyle\rm Q$}\hbox{\raise
0.15\ht0\hbox to0pt{\kern0.4\wd0\vrule height0.7\ht0\hss}\box0}}}}
\def\bbbz{{\mathchoice {\hbox{$\sans\textstyle Z\kern-0.4em Z$}}       
{\hbox{$\sans\textstyle Z\kern-0.4em Z$}}
{\hbox{$\sans\scriptstyle Z\kern-0.3em Z$}}
{\hbox{$\sans\scriptscriptstyle Z\kern-0.2em Z$}}}}
%
%
\def\const{{\rm const}\,}

\def\half{\sfrac{1}{2}}

\def\optbar#1{\vbox{\ialign{##\crcr\hfil${\scriptscriptstyle(}\mkern -1mu
         \vrule height 1.2pt width 3pt depth -.8pt
         {\scriptscriptstyle)}$\hfil\crcr
          \noalign{\kern-1pt\nointerlineskip}$\hfil\displaystyle{#1}\hfil$\crcr}}}
\def\<{\left<}
\def\>{\right>}

\def\smprod{\mathop{\textstyle\prod}}
\def\smsum{\mathop{\textstyle\sum}}
\def\set#1#2{\big\{ \ #1\ \big|\ #2\ \big\}}
\def\eval#1{\big|\lower4pt\hbox{$\displaystyle\sst #1$}}
%
%
\font \tafontt                = cmbx10 scaled\magstep2
\font \tbfontt                = cmbx10 scaled\magstep1
\def\titlea#1{\centerline{\tafontt #1 }\vskip.5truein}
\def\titleb#1{\removelastskip\vskip.3truein%
\noindent{\tbfontt #1 }\vskip.25truein}

%
%
\def\newenvironment#1#2#3#4{\long\def#1##1##2{%
\removelastskip\penalty-100\vskip\baselineskip%
\noindent{#3#2\if!##1!.\else\unskip\ \ignorespaces
##1\unskip\fi\ }{#4\ignorespaces##2\vskip\baselineskip}}}
\newenvironment\lemma{Lemma}{\bf}{\it}
\newenvironment\proposition{Proposition}{\bf}{\it}
\newenvironment\theorem{Theorem}{\bf}{\it}
\newenvironment\corollary{Corollary}{\bf}{\it}
\newenvironment\example{Example}{\bf}{\rm}
\newenvironment\problem{Problem}{\bf}{\rm}
\newenvironment\definition{Definition}{\bf}{\rm}
\newenvironment\remark{Remark}{\bf}{\rm}
\newenvironment\hypothesis{Hypothesis}{\bf}{\it}
\newenvironment\convention{Convention}{\bf}{\it}

\def\Item{\vskip.1in\noindent}

%
%
\long\def\proof#1{\removelastskip\penalty-100\vskip\baselineskip\noindent{\bf
            Proof\if!#1!\else\ \ignorespaces#1\fi:\ }\ \ \ignorespaces}
\long\def\prf{\removelastskip\penalty-100\vskip\baselineskip\noindent{\bf
            Proof:\ }\ \ \ignorespaces}
\def\endproof{\hfill\vrule height .6em width .6em depth 0pt\goodbreak\vskip.25in }

\ifundefined{warnForwardRef}  \def\warnForwardRef{n} \fi
\newcount\chapno
\newcount\sectno
\newcount\equano
\newcount\theono
\newcount\probno

\def\IgNoRe#1{}

\chapno=0
\sectno=0
\equano=0
\theono=0
\probno=0
\def\eqhead{}
\def\frefwarning{\if\warnForwardRef y\immediate\write16{   Forward reference on line \the\inputlineno}\fi}
\def\qqqrefwarning{\immediate\write16{   ??? reference on line \the\inputlineno}}

\def\chap#1{\equano=0\sectno=0\theono=0\probno=0\global\advance\chapno by 1%
\def\eqhead{\ifcase\chapno\or I\or II\or III\or IV\or V\or VI\or VII\or
VIII\or IX\or X\or XI\or XII\or XIII\or XIV\or XV\or XVI\or XVII\or XVIII\or
XIX\or XX\or XXI\or XXII\or XXIII\or XXIV\or XXV\or XXVI\or XXVII\or XXVIII\or XXIX\or XXX\or XXXI\or XXXII\or XXXIII\or XXXIV\or XXXV\or XXXVI\or XXXVII\or XXXVIII\or XXXIX\fi.}%
\titlea{\eqhead \hglue 5pt #1}%
}

\def\sect#1{\global\advance\sectno by 1%
\titleb{\eqhead\number\sectno  \hglue 5pt #1}%
}%

\def\appendix#1#2{\equano=0\sectno=0\theono=0\probno=0\def\eqhead{#1.}
\titlea{Appendix #1: #2}%
}

\def\:#1{\def\temp{\expandafter\IgNoRe\string#1}%
\expandafter\ifx\csname\temp\endcsname\relax%
\expandafter\gdef#1{\qqqrefwarning ???}\fi#1}

\def\Eqn{{\hbox{\global\advance\equano by 1}}%
\eqno ({\rm \eqhead\number\equano})}%

\def\Eqno{{\hbox{\global\advance\equano by 1}}%
 ({\rm \eqhead\number\equano})}%

\def\EQN#1{\Eqn\edef\Zwi{\eqhead\number\equano}%
\global\let #1=\Zwi
}

\def\EQNO#1{\Eqno\edef\Zwi{\eqhead\number\equano}%
\global\let #1=\Zwi
}

\def\STM#1{{\global\advance \theono by 1}%
\eqhead\number\theono
\edef\Zwi{\eqhead\number\theono }
\global\let#1=\Zwi
}

\def\PRB#1{{\global\advance \probno by 1}%
\eqhead\number\probno
\edef\Zwi{\eqhead\number\probno }
\global\let#1=\Zwi
}

\def\PG#1{\def\Zwi{\number\pageno }
\global\let#1=\Zwi
}

\def\Stm{{\global\advance \theono by 1}%
\eqhead\number\theono
}

\def\Prb{{\global\advance \probno by 1}%
\eqhead\number\probno
}

\def\EDEF#1#2{
\def\tEmP{#1}\expandafter\gdef\tEmP{#2}
}



\def\suffix{ps}
\newcount\system
\global\system=3   

\def\ifundefined#1{\expandafter\ifx\csname#1\endcsname\relax}
\ifundefined{figdir}\def\figdir{}\fi
%
%
\newcount\firstline
\newdimen\pswidth  \newdimen\xleft
\newdimen\psheight \newdimen\ytop \newdimen\ybot
\newcount\justx \newcount\justy
\global\justx=0 \global\justy=0
\newdimen\vpos \newtoks\labeL 
\newread\labeLfile \newdimen\xcoord \newdimen\ycoord
\newif\ifdoit 
\newbox\labox
\newdimen\xdvikwid 
\newdimen\xdvikht
\newdimen\pspoints
\newdimen\rwi
\pspoints=1bp
\newcount\temp
\def\readdim#1{\global\read\labeLfile to \temp
\global #1=\temp pt}
%
%
%
%
\def\figcrop#1{\par
\openin\labeLfile=\figdir#1.lbl                                              
\global\read\labeLfile to\firstline\message{#1}               
\global\read\labeLfile to\temp
\readdim{\ybot}
\readdim{\xleft}
\readdim{\ytop}
\global\read\labeLfile to\justx
\global\read\labeLfile to\justy
\global\read\labeLfile to\labeL
\readdim{\pswidth}
\global\advance\pswidth by -\xleft
\readdim{\psheight}
\global\advance\ybot by -\psheight
\global\advance\psheight by -\ytop
\global\read\labeLfile to\justx
\global\read\labeLfile to\justy
\global\read\labeLfile to\labeL
\vbox to\psheight{\vfill
\ifnum\system=1
\fi 
\ifnum\system=2
\fi 
\ifnum\system=3
                                                 \fi         
\ifnum\system=4
\fi         
\ifnum\system=1
\hbox to \pswidth{\kern-\xleft\special{postscriptfile \figdir#1.\suffix }\hfil}\fi
\ifnum\system=2
\hbox to \pswidth{\kern-\xleft\special{ps: plotfile \figdir#1.\suffix }\hfil}\fi
\ifnum\system=3
\hbox to \pswidth{\kern-\xleft\includegraphics{\figdir#1.\suffix}\hfil}\fi
\ifnum\system=4
\hbox to \pswidth{\kern-\xleft\includegraphics{\figdir#1.\suffix}\hfil}\fi
\ifnum\system=5
\hbox to \pswidth{\kern-\xleft\includegraphics{\figdir#1.\suffix}\hfil}\fi 
\ifnum\system=6
   \xdvikwid=\pswidth
   \xdvikht=\psheight
   {\global\divide\xdvikwid by \pspoints}
   {\global\divide\xdvikht by \pspoints}
   \rwi=\xdvikwid
    {\global\multiply\rwi by 10}
\hbox to \pswidth{\kern-\xleft\includegraphics{\figdir#1.\suffix\space}\hfil}\fi                   
\vskip -\baselineskip
\vskip -\ybot 
\vskip-\psheight %
\hbox to\pswidth  {\hss}%
\parindent=0pt\offinterlineskip                                       
\vpos=0 pt%
\loop\readdim{\xcoord}                                 
\ifdim \xcoord < -999pt \doitfalse\else\doittrue\fi                        
\ifdoit \advance \xcoord by -\xleft
\readdim{\ycoord}
\advance \ycoord by -\ytop                              
\global\read\labeLfile to\justx                                       
\global\read\labeLfile to\justy                                       
\global\read\labeLfile to\labeL
\global\setbox\labox=\hbox{\labeL\hskip-0.3em}%
\advance\vpos by-\ycoord                                              
\vskip-\vpos \vpos=\ycoord                                         
\hbox to\pswidth{\hskip\xcoord %
\hbox to 0pt{\ifnum\justx>0\hss\fi%
\vbox to0pt{%
\ifnum\justy<2\vss\fi%
\copy\labox\kern0pt%
\ifnum\justy>0\vss\fi}%
\ifnum\justx<2\hss\fi}%
\hss}%
\repeat%
\advance\vpos by-\psheight%
\vskip-\vpos %
}\closein\labeLfile}
%
%
%
\def\figplace#1#2#3{
\openin\labeLfile=\figdir#1.lbl
\ifeof \labeLfile
       \immediate\write16{***Can't find \figdir#1.lbl; Skipping it.***}
\else  \closein\labeLfile
       \null\hskip#2\raise #3 \hbox{\figcrop{#1}}
\fi
}
%
%
%
%


    \def\squiggle{\raise2pt\hbox{${\scriptstyle\sim}$}}
    \def\stoday{\number\day\space\ifcase\month\or Jan\or Feb\or 
                      Mar\or Apr\or May\or Jun\or Jul\or Aug\or Sep\or 
                      Oct\or Nov\or Dec\fi, \number\year}

    \def\veps{{\varepsilon}}
    \def\smchoose#1#2{{\tst {#1\choose #2}}}
    \def\abcst{{\sst const}}
    \def\cst#1#2{{\rm const}^{#1}_{#2}\,}
    \def\Cont#1#2#3{\mathop{{\rm\ \, {\cal C}on}_{#3}}\limits_{#1\rightarrow#2}}
    \def\cont#1#2#3{\mathop{{\rm\ \, Con}_{#3}}\limits_{#1\rightarrow#2}}

    \def\abcst{{\sst const}}
    \def\cb{{\frak c}}
    \def\ib{{\rm b}}
    
    \def\cl{;}

    \def\dunion{\cup\kern-0.7em\cdot\kern0.45em}

    \def\cD{{\cal D}}
    \def\rD{{\rm D}}
    
    \def\cU{{\cal U}}
    \def\cV{{\cal V}}    
    \def\cW{{\cal W}}

    \def\fe{{\frak e}}

    \def\fN{{\frak N}}

    \def\bde{{\mathchoice{\pmb{$\de$}}{\pmb{$\de$}}
                              {\pmb{$\sst\de$}}{\pmb{$\ssst\de$}}}}

    \def\rw{\mathclose{:}}
    \def\lw{\mathopen{:}}
    \def\lW{\mathopen{{\tst{\hbox{.}\atop\raise 2.5pt\hbox{.}}}}}
    \def\rW{\mathclose{{\tst{{.}\atop\raise 2.5pt\hbox{.}}}}}
    \def\lww{\mathopen{{\tst{\raise 1pt\hbox{.}\atop\raise 1pt\hbox{.}}}}}
    \def\rww{\mathclose{{\tst{\raise 1pt\hbox{.}\atop\raise 1pt\hbox{.}}}}}

    \def\v{\pmb{$\vert$}}
    \def\V{\pmb{$\big\vert$}}

    \def\tn{|\kern-1pt|\kern-1pt|}
    \def\TN{\big|\kern-1.5pt\big|\kern-1.5pt\big|}
    \def\TTN{\Big|\kern-2pt\Big|\kern-2pt\Big|}

    \def\cnorm{\kern8pt\check{\kern-8pt\|}}
    \def\Cnorm{\kern8pt\check{\kern-8pt\big\|}}
    \def\CNorm{\kern8pt\check{\kern-8pt\Big\|}}

    \def\tnorm{\kern8pt\tilde{\kern-8pt\|}}
    \def\Tnorm{\kern8pt\tilde{\kern-8pt\big\|}}
    \def\TNorm{\kern8pt\tilde{\kern-8pt\Big\|}}

    \def\tv{\kern8pt\tilde{\kern-8pt\pmb{$\vert$}}}
    \def\tV{\kern8pt\tilde{\kern-8pt\pmb{$\big\vert$}}}
    \def\tVV{\kern8pt\tilde{\kern-8pt\pmb{$\Big\vert$}}}

    \def\jbar{{\mathchoice
                   {{\smash{\lower1ex\hbox{$\mathchar'26$}}\mkern-9mu j}}
                   {{\smash{\lower1ex\hbox{$\mathchar'26$}}\mkern-9mu j}}
                   {{\smash{\lower1.2ex\hbox{$\mathchar'26$}}\mkern-10.2mu j}}
                   {{\smash{\lower1.2ex\hbox{$\mathchar'26$}}\mkern-10.2mu j}}}}

\def\Eqnb{{\hbox{\global\advance\equano by 1}}%
\eqno ({\rm \eqhead\number\equano}}%

\def\EQNB#1{\Eqnb\edef\Zwi{\eqhead\number\equano}%
\global\let #1=\Zwi
}

   \font\sixrm=cmr6   \font\eightrm=cmr8  
   \font\sixi=cmmi6   \font\eighti=cmmi8  
  \font\sixsy=cmsy6  \font\eightsy=cmsy8 
  \font\sixbf=cmbx6  \font\eightbf=cmbx8 
                     \font\eightit=cmti8 
                     \font\eightsl=cmsl8 
                     \font\eighttt=cmtt8 

\font\eightfrak=eufm7 at 8pt

\def\eightpoint{\def\rm{\fam0\eightrm}
 \textfont0=\eightrm \scriptfont0=\sixrm \scriptscriptfont0=\fiverm
 \textfont1=\eighti \scriptfont1=\sixi \scriptscriptfont1=\fivei
 \textfont2=\eightsy \scriptfont2=\sixsy \scriptscriptfont2=\fivesy
 \textfont3=\tenex \scriptfont3=\tenex \scriptscriptfont3=\tenex
 \textfont\itfam=\eightit \def\it{\fam\itfam\eightit}%
 \textfont\slfam=\eightsl \def\sl{\fam\slfam\eightsl}%
 \textfont\ttfam=\eighttt \def\tt{\fam\ttfam\eighttt}%
 \textfont\frakfam=\eightfrak \def\frak{\fam\frakfam\tenfrak}%
 \textfont\bffam=\eightbf \scriptfont\bffam=\sixbf
 \scriptscriptfont\bffam=\fivebf \def\bf{\fam\bffam\eightbf}%
 \normalbaselineskip=9pt
 \setbox\strutbox=\hbox{\vrule height7pt depth2pt width0pt}%
 \let\sc=\sixrm \let\big=\eightbig \normalbaselines\rm}
\catcode`@=11
\def\footnote#1{\edef\@sf{\spacefactor\the\spacefactor}#1\@sf
     \insert\footins\bgroup\eightpoint
     \interlinepenalty100 \let\par=\endgraf
     \leftskip=0pt \rightskip=0pt
     \splittopskip=10pt plus 1pt minus 1pt \floatingpenalty=20000
     \smallskip\item{#1}\bgroup\strut\aftergroup\@foot\let\next}
\skip\footins=12pt plus 2pt minus 4pt
\dimen\footins=30pc
\catcode`@=12


  \IgNoRe{PG}
  \IgNoRe{STM Assertion }
  \IgNoRe{PG}
  \IgNoRe{PG}
  \IgNoRe{STM Assertion }
  \IgNoRe{PG}
  \IgNoRe{STM Assertion }
  \IgNoRe{STM Assertion }
  \IgNoRe{EQN}
  \IgNoRe{STM Assertion }
  \IgNoRe{STM Assertion }
  \IgNoRe{PG}
  \IgNoRe{STM Assertion }
  \IgNoRe{STM Assertion }
  \IgNoRe{EQN}
 \def\defGrasscontract{\frefwarning II.9} \IgNoRe{STM Assertion }
  \IgNoRe{STM Assertion }
  \IgNoRe{STM Assertion }
 \def\remcontract{\frefwarning II.12} \IgNoRe{STM Assertion }
 \def\lemcontract{\frefwarning II.13} \IgNoRe{STM Assertion }
  \IgNoRe{STM Assertion }
  \IgNoRe{PG}
  \IgNoRe{STM Assertion }
  \IgNoRe{STM Assertion }
  \IgNoRe{STM Assertion }
  \IgNoRe{STM Assertion }
  \IgNoRe{STM Assertion }
  \IgNoRe{STM Assertion }
  \IgNoRe{STM Assertion }
 \def\lemGrasscompatnorm{\frefwarning II.22} \IgNoRe{STM Assertion }
 \def\deffunctnorm{\frefwarning II.23} \IgNoRe{STM Assertion }
  \IgNoRe{STM Assertion }
  \IgNoRe{STM Assertion }
  \IgNoRe{STM Assertion }
  \IgNoRe{PG}
  \IgNoRe{EQN}
  \IgNoRe{STM Assertion }
 \def\theorII{\frefwarning II.28} \IgNoRe{STM Assertion }
  \IgNoRe{STM Assertion }
  \IgNoRe{PG}
  \IgNoRe{STM Assertion }
  \IgNoRe{STM Assertion }
 \def\corwicknorm{\frefwarning II.32} \IgNoRe{STM Assertion }
  \IgNoRe{STM Assertion }
  \IgNoRe{EQN}
  \IgNoRe{EQN}
  \IgNoRe{STM Assertion }
  \IgNoRe{PG}
  \IgNoRe{PG}
  \IgNoRe{STM Assertion }
  \IgNoRe{EQN}
  \IgNoRe{STM Assertion }
  \IgNoRe{STM Assertion }
  \IgNoRe{STM Assertion }
  \IgNoRe{EQN}
  \IgNoRe{STM Assertion }
  \IgNoRe{STM Assertion }
  \IgNoRe{STM Assertion }
  \IgNoRe{PG}
  \IgNoRe{STM Assertion }
  \IgNoRe{STM Assertion }
  \IgNoRe{PG}
 \def\remjointanalyticity{\frefwarning III.11} \IgNoRe{STM Assertion }
  \IgNoRe{STM Assertion }
  \IgNoRe{STM Assertion }
  \IgNoRe{PG}
  \IgNoRe{STM Assertion }
  \IgNoRe{STM Assertion }
 \def\lemprftwoA{\frefwarning IV.5} \IgNoRe{STM Assertion }
  \IgNoRe{STM Assertion }
 \def\lemprftwoB{\frefwarning IV.7} \IgNoRe{STM Assertion }
  \IgNoRe{STM Assertion }
  \IgNoRe{STM Assertion }
  \IgNoRe{STM Assertion }
  \IgNoRe{PG}
  \IgNoRe{STM Assertion }
  \IgNoRe{STM Assertion }
  \IgNoRe{STM Assertion }
  \IgNoRe{STM Assertion }
  \IgNoRe{STM Assertion }
  \IgNoRe{STM Assertion }
  \IgNoRe{STM Assertion }
  \IgNoRe{STM Assertion }
  \IgNoRe{PG}
  \IgNoRe{STM Assertion }
  \IgNoRe{STM Assertion }
  \IgNoRe{PG}
  \IgNoRe{PG}
  \IgNoRe{STM Assertion }
  \IgNoRe{STM Assertion }
  \IgNoRe{PG}
  \IgNoRe{PG}
  \IgNoRe{STM Assertion }
  \IgNoRe{STM Assertion }
  \IgNoRe{EQN}
  \IgNoRe{STM Assertion }
  \IgNoRe{PG}
  \IgNoRe{STM Assertion }
  \IgNoRe{STM Assertion }
  \IgNoRe{STM Assertion }
  \IgNoRe{PG}
  \IgNoRe{STM Assertion }
  \IgNoRe{STM Assertion }
  \IgNoRe{STM Assertion }
  \IgNoRe{EQN}
  \IgNoRe{STM Assertion }
  \IgNoRe{PG}
  \IgNoRe{EQN}
  \IgNoRe{STM Assertion }
  \IgNoRe{STM Assertion }
  \IgNoRe{STM Assertion }
  \IgNoRe{STM Assertion }
  \IgNoRe{EQN}
  \IgNoRe{EQN}
  \IgNoRe{PG}
  \IgNoRe{PG}
  \IgNoRe{STM Assertion }
  \IgNoRe{STM Assertion }
  \IgNoRe{STM Assertion }
  \IgNoRe{STM Assertion }
  \IgNoRe{STM Assertion }
  \IgNoRe{STM Assertion }
  \IgNoRe{STM Assertion }
  \IgNoRe{STM Assertion }
  \IgNoRe{PG}
  \IgNoRe{STM Assertion }
  \IgNoRe{STM Assertion }
  \IgNoRe{STM Assertion }
  \IgNoRe{STM Assertion }
  \IgNoRe{STM Assertion }
  \IgNoRe{STM Assertion }
  \IgNoRe{STM Assertion }
  \IgNoRe{STM Assertion }
  \IgNoRe{STM Assertion }
  \IgNoRe{STM Assertion }
  \IgNoRe{PG}
  \IgNoRe{STM Assertion }
  \IgNoRe{STM Assertion }
  \IgNoRe{STM Assertion }
  \IgNoRe{PG}
  \IgNoRe{STM Assertion }
  \IgNoRe{STM Assertion }
  \IgNoRe{STM Assertion }
  \IgNoRe{STM Assertion }
  \IgNoRe{PG}
  \IgNoRe{STM Assertion }
  \IgNoRe{STM Assertion }
  \IgNoRe{PG}
  \IgNoRe{STM Assertion }
  \IgNoRe{PG}
  \IgNoRe{STM Assertion }
  \IgNoRe{PG}
  \IgNoRe{STM Assertion }
  \IgNoRe{STM Assertion }
  \IgNoRe{STM Assertion }
  \IgNoRe{STM Assertion }
  \IgNoRe{PG}
  \IgNoRe{PG}
  \IgNoRe{STM Assertion }
  \IgNoRe{STM Assertion }
  \IgNoRe{EQN}
  \IgNoRe{EQN}
  \IgNoRe{STM Assertion }
  \IgNoRe{STM Assertion }
  \IgNoRe{STM Assertion }
  \IgNoRe{STM Assertion }
  \IgNoRe{PG}
  \IgNoRe{STM Assertion }
  \IgNoRe{STM Assertion }
  \IgNoRe{STM Assertion }
  \IgNoRe{STM Assertion }
  \IgNoRe{STM Assertion }
  \IgNoRe{STM Assertion }
  \IgNoRe{STM Assertion }
  \IgNoRe{PG}
  \IgNoRe{STM Assertion }
  \IgNoRe{EQN}
  \IgNoRe{EQN}
  \IgNoRe{PG}
  \IgNoRe{STM Assertion }
  \IgNoRe{EQN}
  \IgNoRe{STM Assertion }
  \IgNoRe{STM Assertion }
  \IgNoRe{STM Assertion }
  \IgNoRe{PG}
  \IgNoRe{STM Assertion }
  \IgNoRe{EQN}
  \IgNoRe{STM Assertion }
  \IgNoRe{PG}
  \IgNoRe{PG}


  \IgNoRe{STM Assertion }
  \IgNoRe{PG}
  \IgNoRe{EQN}
  \IgNoRe{EQN}
  \IgNoRe{EQN}
  \IgNoRe{EQN}
  \IgNoRe{EQN}
  \IgNoRe{STM Assertion }
  \IgNoRe{STM Assertion }
  \IgNoRe{STM Assertion }
  \IgNoRe{STM Assertion }
  \IgNoRe{STM Assertion }
  \IgNoRe{STM Assertion }
  \IgNoRe{STM Assertion }
  \IgNoRe{STM Assertion }
  \IgNoRe{STM Assertion }
  \IgNoRe{STM Assertion }
  \IgNoRe{STM Assertion }
  \IgNoRe{PG}
  \IgNoRe{PG}
  \IgNoRe{PG}
  \IgNoRe{PG}
  \IgNoRe{EQN}
  \IgNoRe{EQN}
  \IgNoRe{EQN}
  \IgNoRe{EQN}
  \IgNoRe{PG}
  \IgNoRe{PG}
  \IgNoRe{PG}
  \IgNoRe{EQN}
  \IgNoRe{PG}
  \IgNoRe{PG}
  \IgNoRe{EQN}
  \IgNoRe{EQN}
  \IgNoRe{EQN}
  \IgNoRe{EQN}
  \IgNoRe{EQN}
  \IgNoRe{EQN}
  \IgNoRe{EQN}
  \IgNoRe{EQN}
  \IgNoRe{EQN}
  \IgNoRe{EQN}
  \IgNoRe{PG}
  \IgNoRe{PG}
  \IgNoRe{EQN}
  \IgNoRe{EQN}
  \IgNoRe{EQN}
  \IgNoRe{STM Assertion }
  \IgNoRe{PG}
  \IgNoRe{EQN}
  \IgNoRe{EQN}
  \IgNoRe{EQN}
  \IgNoRe{EQN}
  \IgNoRe{STM Assertion }
  \IgNoRe{STM Assertion }
  \IgNoRe{STM Assertion }
  \IgNoRe{STM Assertion }
  \IgNoRe{STM Assertion }
  \IgNoRe{STM Assertion }
  \IgNoRe{STM Assertion }
  \IgNoRe{STM Assertion }
  \IgNoRe{STM Assertion }
  \IgNoRe{EQN}
  \IgNoRe{EQN}
  \IgNoRe{EQN}
  \IgNoRe{EQN}
  \IgNoRe{STM Assertion }
  \IgNoRe{EQN}
  \IgNoRe{STM Assertion }
  \IgNoRe{EQN}
  \IgNoRe{STM Assertion }
  \IgNoRe{PG}
  \IgNoRe{EQN}
  \IgNoRe{STM Assertion }
  \IgNoRe{STM Assertion }
  \IgNoRe{EQN}
  \IgNoRe{PG}
  \IgNoRe{PG}
  \IgNoRe{STM Assertion }
  \IgNoRe{STM Assertion }
  \IgNoRe{PG}
  \IgNoRe{STM Assertion }
  \IgNoRe{STM Assertion }
  \IgNoRe{STM Assertion }
  \IgNoRe{STM Assertion }
  \IgNoRe{STM Assertion }
  \IgNoRe{STM Assertion }
  \IgNoRe{STM Assertion }
  \IgNoRe{STM Assertion }
  \IgNoRe{PG}
  \IgNoRe{STM Assertion }
  \IgNoRe{STM Assertion }
  \IgNoRe{STM Assertion }
  \IgNoRe{STM Assertion }
  \IgNoRe{EQN}
  \IgNoRe{STM Assertion }
  \IgNoRe{STM Assertion }
  \IgNoRe{STM Assertion }
  \IgNoRe{EQN}
  \IgNoRe{STM Assertion }
  \IgNoRe{STM Assertion }
  \IgNoRe{STM Assertion }
  \IgNoRe{STM Assertion }
  \IgNoRe{PG}
  \IgNoRe{STM Assertion }
  \IgNoRe{STM Assertion }
  \IgNoRe{STM Assertion }
  \IgNoRe{STM Assertion }
  \IgNoRe{STM Assertion }
  \IgNoRe{STM Assertion }
  \IgNoRe{STM Assertion }
  \IgNoRe{STM Assertion }
  \IgNoRe{STM Assertion }
  \IgNoRe{EQN}
  \IgNoRe{EQN}
  \IgNoRe{EQN}
  \IgNoRe{STM Assertion }
  \IgNoRe{PG}
  \IgNoRe{STM Assertion }
  \IgNoRe{STM Assertion }
  \IgNoRe{STM Assertion }
  \IgNoRe{EQN}
  \IgNoRe{EQN}
  \IgNoRe{EQN}
  \IgNoRe{STM Assertion }
  \IgNoRe{STM Assertion }
  \IgNoRe{EQN}
  \IgNoRe{EQN}
  \IgNoRe{EQN}
  \IgNoRe{STM Assertion }
  \IgNoRe{PG}
  \IgNoRe{PG}
  \IgNoRe{STM Assertion }
  \IgNoRe{STM Assertion }
  \IgNoRe{PG}
  \IgNoRe{STM Assertion }
  \IgNoRe{STM Assertion }
  \IgNoRe{EQN}
  \IgNoRe{EQN}
  \IgNoRe{EQN}
  \IgNoRe{EQN}
  \IgNoRe{EQN}
  \IgNoRe{EQN}
  \IgNoRe{EQN}
  \IgNoRe{EQN}
  \IgNoRe{EQN}
  \IgNoRe{EQN}
  \IgNoRe{EQN}
  \IgNoRe{EQN}
  \IgNoRe{EQN}
  \IgNoRe{EQN}
  \IgNoRe{STM Assertion }
  \IgNoRe{EQN}
  \IgNoRe{PG}
  \IgNoRe{EQN}
  \IgNoRe{STM Assertion }
  \IgNoRe{EQN}
  \IgNoRe{STM Assertion }
  \IgNoRe{STM Assertion }
  \IgNoRe{EQN}
  \IgNoRe{EQN}
  \IgNoRe{EQN}
  \IgNoRe{STM Assertion }
  \IgNoRe{EQN}
  \IgNoRe{EQN}
  \IgNoRe{EQN}
  \IgNoRe{EQN}
  \IgNoRe{EQN}
  \IgNoRe{EQN}
  \IgNoRe{EQN}
  \IgNoRe{EQN}
  \IgNoRe{EQN}
  \IgNoRe{EQN}
  \IgNoRe{EQN}
  \IgNoRe{EQN}
  \IgNoRe{EQN}
  \IgNoRe{EQN}
  \IgNoRe{EQN}
  \IgNoRe{EQN}
  \IgNoRe{EQN}
  \IgNoRe{EQN}
  \IgNoRe{EQN}
  \IgNoRe{EQN}
  \IgNoRe{EQN}
  \IgNoRe{STM Assertion }
  \IgNoRe{PG}
  \IgNoRe{PG}
  \IgNoRe{STM Assertion }
  \IgNoRe{PG}
  \IgNoRe{EQN}
  \IgNoRe{EQN}
  \IgNoRe{EQN}
  \IgNoRe{EQN}
  \IgNoRe{STM Assertion }
  \IgNoRe{EQN}
  \IgNoRe{PG}
  \IgNoRe{EQN}
  \IgNoRe{EQN}
  \IgNoRe{EQN}
  \IgNoRe{EQN}
  \IgNoRe{EQN}
  \IgNoRe{EQN}
  \IgNoRe{EQN}
  \IgNoRe{EQN}
  \IgNoRe{EQN}
  \IgNoRe{EQN}
  \IgNoRe{EQN}
  \IgNoRe{EQN}
  \IgNoRe{STM Assertion }
  \IgNoRe{STM Assertion }
  \IgNoRe{EQN}
  \IgNoRe{EQN}
  \IgNoRe{PG}
  \IgNoRe{PG}
  \IgNoRe{STM Assertion }
  \IgNoRe{EQN}
  \IgNoRe{STM Assertion }
  \IgNoRe{PG}
  \IgNoRe{EQN}
  \IgNoRe{EQN}
  \IgNoRe{EQN}
  \IgNoRe{STM Assertion }
  \IgNoRe{STM Assertion }
  \IgNoRe{EQN}
  \IgNoRe{EQN}
  \IgNoRe{EQN}
  \IgNoRe{EQN}
  \IgNoRe{EQN}
  \IgNoRe{STM Assertion }
  \IgNoRe{EQN}
  \IgNoRe{EQN}
  \IgNoRe{EQN}
  \IgNoRe{EQN}
  \IgNoRe{STM Assertion }
  \IgNoRe{STM Assertion }
  \IgNoRe{EQN}
  \IgNoRe{STM Assertion }
  \IgNoRe{STM Assertion }
  \IgNoRe{STM Assertion }
  \IgNoRe{STM Assertion }
  \IgNoRe{PG}
  \IgNoRe{STM Assertion }
  \IgNoRe{STM Assertion }
  \IgNoRe{STM Assertion }
  \IgNoRe{STM Assertion }
 \def\defNPampGreen{\frefwarning XIII.9} \IgNoRe{STM Assertion }
  \IgNoRe{STM Assertion }
  \IgNoRe{STM Assertion }
  \IgNoRe{STM Assertion }
  \IgNoRe{STM Assertion }
  \IgNoRe{STM Assertion }
  \IgNoRe{STM Assertion }
  \IgNoRe{STM Assertion }
  \IgNoRe{STM Assertion }
  \IgNoRe{STM Assertion }
  \IgNoRe{STM Assertion }
  \IgNoRe{PG}
  \IgNoRe{STM Assertion }
  \IgNoRe{STM Assertion }
  \IgNoRe{STM Assertion }
  \IgNoRe{STM Assertion }
  \IgNoRe{STM Assertion }
  \IgNoRe{STM Assertion }
  \IgNoRe{STM Assertion }
  \IgNoRe{STM Assertion }
  \IgNoRe{STM Assertion }
  \IgNoRe{STM Assertion }
  \IgNoRe{STM Assertion }
  \IgNoRe{STM Assertion }
  \IgNoRe{EQN}
  \IgNoRe{STM Assertion }
  \IgNoRe{STM Assertion }
  \IgNoRe{STM Assertion }
  \IgNoRe{STM Assertion }
  \IgNoRe{STM Assertion }
  \IgNoRe{STM Assertion }
  \IgNoRe{EQN}
  \IgNoRe{STM Assertion }
  \IgNoRe{PG}
  \IgNoRe{PG}
  \IgNoRe{STM Assertion }
  \IgNoRe{STM Assertion }
  \IgNoRe{STM Assertion }
  \IgNoRe{EQN}
  \IgNoRe{STM Assertion }
  \IgNoRe{PG}
  \IgNoRe{EQN}
  \IgNoRe{STM Assertion }
  \IgNoRe{STM Assertion }
  \IgNoRe{EQN}
  \IgNoRe{EQN}
  \IgNoRe{EQN}
  \IgNoRe{EQN}
  \IgNoRe{EQN}
  \IgNoRe{EQN}
  \IgNoRe{EQN}
  \IgNoRe{EQN}
  \IgNoRe{EQN}
  \IgNoRe{EQN}
  \IgNoRe{EQN}
  \IgNoRe{EQN}
  \IgNoRe{EQN}
  \IgNoRe{EQN}
  \IgNoRe{EQN}
  \IgNoRe{EQN}
  \IgNoRe{EQN}
  \IgNoRe{EQN}
  \IgNoRe{EQN}
  \IgNoRe{EQN}
  \IgNoRe{STM Assertion }
  \IgNoRe{STM Assertion }
  \IgNoRe{PG}
  \IgNoRe{EQN}
  \IgNoRe{EQN}
  \IgNoRe{EQN}
  \IgNoRe{EQN}
  \IgNoRe{EQN}
  \IgNoRe{EQN}
  \IgNoRe{EQN}
  \IgNoRe{EQN}
  \IgNoRe{EQN}
  \IgNoRe{EQN}
  \IgNoRe{EQN}
  \IgNoRe{EQN}
  \IgNoRe{EQN}
  \IgNoRe{EQN}
  \IgNoRe{EQN}
  \IgNoRe{EQN}
  \IgNoRe{EQN}
  \IgNoRe{EQN}
  \IgNoRe{EQN}
  \IgNoRe{EQN}
  \IgNoRe{EQN}
  \IgNoRe{STM Assertion }
  \IgNoRe{PG}
  \IgNoRe{STM Assertion }
  \IgNoRe{STM Assertion }
  \IgNoRe{EQN}
  \IgNoRe{EQN}
  \IgNoRe{PG}
  \IgNoRe{EQN}
  \IgNoRe{EQN}
  \IgNoRe{EQN}
  \IgNoRe{EQN}
  \IgNoRe{EQN}
  \IgNoRe{EQN}
  \IgNoRe{EQN}
  \IgNoRe{EQN}
  \IgNoRe{STM Assertion }
  \IgNoRe{STM Assertion }
  \IgNoRe{STM Assertion }
  \IgNoRe{STM Assertion }
  \IgNoRe{STM Assertion }
  \IgNoRe{PG}
  \IgNoRe{PG}


\newcount\CHAPNO
\newcount\APPNO
\CHAPNO=0
\APPNO=1
\def\advCHAPNO{\advance\CHAPNO by 1}
\def\advAPPNO{\advance\APPNO by 1}

\def\caproman#1{\ifcase#1\or I\or II\or III\or IV\or V\or VI\or VII\or
VIII\or IX\or X\or XI\or XII\or XIII\or XIV\or XV\or XVI\or XVII\or XVIII\or
XIX\or XX\or XXI\or XXII\or XXIII\or XXIV\or XXV\or XXVI\or XXVII\or XXVIII\or XXIX\or XXX\or XXXI\or XXXII\or XXXIII\or XXXIV\or XXXV\or XXXVI\or XXXVII\or XXXVIII\or XXXIX\fi}%

\def\capletter#1{\ifcase#1\or A\or B\or C\or D\or E\or F\or G\or
H\or I\or J\or K\or L\or M\or N\or O\or P\or Q\or R\or
S\or T\or U\or V\or W\or X\or Y\or Z\fi}%

\newcount\cHintroI \cHintroI=\CHAPNO \advCHAPNO 
\newcount\cHintroOverview  \cHintroOverview=\CHAPNO \advCHAPNO 
                              \edef\CHintroOverview{\caproman\CHAPNO}  
\newcount\cHrenmap \cHrenmap=\CHAPNO \advCHAPNO 

 \advAPPNO

\newcount\cHintroII \cHintroII=\CHAPNO \advCHAPNO 
                              \edef\CHintroII{\caproman\CHAPNO}
\newcount\cHfirstscale \cHfirstscale=\CHAPNO \advCHAPNO
                              
\newcount\cHnewsectors \cHnewsectors=\CHAPNO \advCHAPNO
                              
\newcount\cHphladders \cHphladders=\CHAPNO \advCHAPNO
                              
\newcount\cHfinitescale \cHfinitescale=\CHAPNO \advCHAPNO
                              
\newcount\cHstep \cHstep=\CHAPNO \advCHAPNO
                              
\newcount\cHrecurs \cHrecurs=\CHAPNO \advCHAPNO
                              
 \advAPPNO

\newcount\cHintroIII \cHintroIII=\CHAPNO \advCHAPNO
                              
\newcount\cHtildefinitescale \cHtildefinitescale=\CHAPNO \advCHAPNO
                              
\newcount\cHtildenewsectors \cHtildenewsectors=\CHAPNO \advCHAPNO
                              
\newcount\cHtildephladders \cHtildephladders=\CHAPNO \advCHAPNO
                              
\newcount\cHtildestep  \cHtildestep=\CHAPNO \advCHAPNO

 \advAPPNO
 \advAPPNO


  \IgNoRe{PG}
 \def\eqnOSintroI{\frefwarning I.1} \IgNoRe{EQN}
 \def\defOSmultideriv{\frefwarning II.1} \IgNoRe{STM Assertion }
  \IgNoRe{PG}
 \def\lemOSleibniz{\frefwarning II.2} \IgNoRe{STM Assertion }
  \IgNoRe{STM Assertion }
  \IgNoRe{STM Assertion }
  \IgNoRe{STM Assertion }
 \def\exOSSymmNorm{\frefwarning II.6} \IgNoRe{STM Assertion }
 \def\lemOSelloneinfty{\frefwarning II.7} \IgNoRe{STM Assertion }
  \IgNoRe{EQN}
  \IgNoRe{STM Assertion }
 \def\defOSFmn{\frefwarning II.9} \IgNoRe{STM Assertion }
  \IgNoRe{STM Assertion }
 \def\defOScontnorm{\frefwarning III.1} \IgNoRe{STM Assertion }
  \IgNoRe{STM Assertion }
  \IgNoRe{STM Assertion }
 \def\exOSelloneinftycontr{\frefwarning III.4} \IgNoRe{STM Assertion }
  \IgNoRe{STM Assertion }
  \IgNoRe{PG}
  \IgNoRe{STM Assertion }
  \IgNoRe{STM Assertion }
  \IgNoRe{STM Assertion }
 \def\defOSgrnorm{\frefwarning III.9} \IgNoRe{STM Assertion }
 \def\thmOSroptheorII{\frefwarning III.10} \IgNoRe{STM Assertion }
  \IgNoRe{STM Assertion }
 \def\defIntBndsS{\frefwarning IV.1} \IgNoRe{STM Assertion }
  \IgNoRe{STM Assertion }
  \IgNoRe{EQN}
  \IgNoRe{PG}
  \IgNoRe{PG}
 \def\propIntBndsII{\frefwarning IV.3} \IgNoRe{STM Assertion }
  \IgNoRe{STM Assertion }
 \def\propIntBndsIV{\frefwarning IV.5} \IgNoRe{STM Assertion }
 \def\defOSderivmom{\frefwarning IV.6} \IgNoRe{STM Assertion }
  \IgNoRe{STM Assertion }
 \def\propOSpropbnd{\frefwarning IV.8} \IgNoRe{STM Assertion }
  \IgNoRe{PG}
  \IgNoRe{EQN}
  \IgNoRe{EQN}
  \IgNoRe{EQN}
  \IgNoRe{EQN}
  \IgNoRe{EQN}
  \IgNoRe{EQN}
  \IgNoRe{STM Assertion }
 \def\defOScbzero{\frefwarning IV.10} \IgNoRe{STM Assertion }
 \def\propOSrealfirstpropbound{\frefwarning IV.11} \IgNoRe{STM Assertion }
 \def\lemOSscalednorm{\frefwarning V.1} \IgNoRe{STM Assertion }
  \IgNoRe{PG}
 \def\thmOSinsulators{\frefwarning V.2} \IgNoRe{STM Assertion }
  \IgNoRe{EQN}
  \IgNoRe{STM Assertion }
  \IgNoRe{STM Assertion }
  \IgNoRe{STM Assertion }
  \IgNoRe{STM Assertion }
  \IgNoRe{PG}
  \IgNoRe{STM Assertion }
 \def\corOSappMonoidIV{\frefwarning A.5} \IgNoRe{STM Assertion }
  \IgNoRe{STM Assertion }
  \IgNoRe{STM Assertion }
  \IgNoRe{PG}
 \def\eqnOSjdef{\frefwarning VI.1} \IgNoRe{EQN}
 \def\eqnOSphijpsi{\frefwarning VI.2} \IgNoRe{EQN}
 \def\pgOSVI{\frefwarning 1} \IgNoRe{PG}
 \def\defOSrengrpmap{\frefwarning VII.1} \IgNoRe{STM Assertion }
 \def\remOSgenfnrengrp{\frefwarning VII.2} \IgNoRe{STM Assertion }
 \def\lemOStworengrpmaps{\frefwarning VII.3} \IgNoRe{STM Assertion }
 \def\eqnOSshiftwick{\frefwarning VII.1} \IgNoRe{EQN}
 \def\pgOSVII{\frefwarning 3} \IgNoRe{PG}
 \def\defOSextimpr{\frefwarning VII.4} \IgNoRe{STM Assertion }
 \def\lemOSextimpr{\frefwarning VII.5} \IgNoRe{STM Assertion }
 \def\eqnOSextimpr{\frefwarning VII.2} \IgNoRe{EQN}
 \def\propOSextimpr{\frefwarning VII.6} \IgNoRe{STM Assertion }
 \def\corOSextimpr{\frefwarning VII.7} \IgNoRe{STM Assertion }
 \def\lemOSscalednormextimpr{\frefwarning VII.8} \IgNoRe{STM Assertion }
 \def\defOSscales{\frefwarning VIII.1} \IgNoRe{STM Assertion }
 \def\pgOSVIII{\frefwarning 8} \IgNoRe{PG}
 \def\remOSlargej{\frefwarning VIII.2} \IgNoRe{STM Assertion }
 \def\convOSI{\frefwarning VIII.3} \IgNoRe{STM Assertion }
 \def\defOSextendedshell{\frefwarning VIII.4} \IgNoRe{STM Assertion }
 \def\defOSCTmSpace{\frefwarning VIII.5} \IgNoRe{STM Assertion }
 \def\thmOSfirststep{\frefwarning VIII.6} \IgNoRe{STM Assertion }
 \def\remOSthmV{\frefwarning VIII.7} \IgNoRe{STM Assertion }
 \def\remOSneedsectors{\frefwarning VIII.8} \IgNoRe{STM Assertion }
 \def\defOSfourtrans{\frefwarning IX.1} \IgNoRe{STM Assertion }
 \def\pgOSIX{\frefwarning 16} \IgNoRe{PG}
 \def\remOStransinv{\frefwarning IX.2} \IgNoRe{STM Assertion }
 \def\defOSftcov{\frefwarning IX.3} \IgNoRe{STM Assertion }
 \def\defOSfourtransII{\frefwarning IX.4} \IgNoRe{STM Assertion }
 \def\lemOSjhat{\frefwarning IX.5} \IgNoRe{STM Assertion }
 \def\lemOSprepintup{\frefwarning IX.6} \IgNoRe{STM Assertion }
 \def\eqnOSlemsixtzero{\frefwarning IX.1} \IgNoRe{EQN}
 \def\defOSamptransinv{\frefwarning X.1} \IgNoRe{STM Assertion }
 \def\defOSdiffdecay{\frefwarning X.2} \IgNoRe{STM Assertion }
 \def\pgOSX{\frefwarning 22} \IgNoRe{PG}
 \def\remOSdiffdecay{\frefwarning X.3} \IgNoRe{STM Assertion }
 \def\defOSdiffdecaynorm{\frefwarning X.4} \IgNoRe{STM Assertion }
 \def\remOSdiffdecaynorm{\frefwarning X.5} \IgNoRe{STM Assertion }
 \def\lemOSelloneinftyamp{\frefwarning X.6} \IgNoRe{STM Assertion }
 \def\remOSelloneinftyamp{\frefwarning X.7} \IgNoRe{STM Assertion }
 \def\defOScheckcF{\frefwarning X.8} \IgNoRe{STM Assertion }
 \def\corOSmomcontrintbound{\frefwarning X.9} \IgNoRe{STM Assertion }
 \def\lemOSTscalednorm{\frefwarning X.10} \IgNoRe{STM Assertion }
 \def\lemOSTZsourceterm{\frefwarning X.11} \IgNoRe{STM Assertion }
 \def\eqnOSTextimpr{\frefwarning X.1} \IgNoRe{EQN}
 \def\thmOSTfirststep{\frefwarning X.12} \IgNoRe{STM Assertion }
 \def\defOSsymmetries{\frefwarning B.1} \IgNoRe{STM Assertion }
 \def\pgOSB{\frefwarning 35} \IgNoRe{PG}
 \def\remOSgrassmannsymmetries{\frefwarning B.2} \IgNoRe{STM Assertion }
 \def\remOSsymmetryConsequences{\frefwarning B.3} \IgNoRe{STM Assertion }
 \def\remOSphiJpsi{\frefwarning B.4} \IgNoRe{STM Assertion }
 \def\remOSrengrppreserves{\frefwarning B.5} \IgNoRe{STM Assertion }
 \def\lemOSphipsistruct{\frefwarning B.6} \IgNoRe{STM Assertion }
 \def\lemOSappGrassI{\frefwarning C.1} \IgNoRe{STM Assertion }
 \def\lemOSappGrassII{\frefwarning C.2} \IgNoRe{STM Assertion }
 \def\pgOSC{\frefwarning 39} \IgNoRe{PG}
 \def\remOSappGrassII{\frefwarning C.3} \IgNoRe{STM Assertion }
 \def\pgOSIIref{\frefwarning 41} \IgNoRe{PG}
  \IgNoRe{PG}
 \def\defOSsectors{\frefwarning XII.1} \IgNoRe{STM Assertion }
  \IgNoRe{STM Assertion }
  \IgNoRe{PG}
  \IgNoRe{STM Assertion }
  \IgNoRe{STM Assertion }
  \IgNoRe{STM Assertion }
  \IgNoRe{STM Assertion }
  \IgNoRe{STM Assertion }
  \IgNoRe{STM Assertion }
  \IgNoRe{STM Assertion }
  \IgNoRe{STM Assertion }
  \IgNoRe{STM Assertion }
  \IgNoRe{STM Assertion }
  \IgNoRe{STM Assertion }
  \IgNoRe{STM Assertion }
  \IgNoRe{STM Assertion }
  \IgNoRe{STM Assertion }
  \IgNoRe{STM Assertion }
  \IgNoRe{STM Assertion }
  \IgNoRe{EQN}
  \IgNoRe{STM Assertion }
  \IgNoRe{STM Assertion }
  \IgNoRe{PG}
  \IgNoRe{STM Assertion }
  \IgNoRe{EQN}
  \IgNoRe{EQN}
  \IgNoRe{STM Assertion }
  \IgNoRe{EQN}
  \IgNoRe{EQN}
  \IgNoRe{STM Assertion }
  \IgNoRe{EQN}
  \IgNoRe{STM Assertion }
  \IgNoRe{EQN}
  \IgNoRe{STM Assertion }
  \IgNoRe{STM Assertion }
  \IgNoRe{STM Assertion }
  \IgNoRe{STM Assertion }
  \IgNoRe{PG}
  \IgNoRe{STM Assertion }
  \IgNoRe{STM Assertion }
  \IgNoRe{STM Assertion }
  \IgNoRe{STM Assertion }
  \IgNoRe{EQN}
  \IgNoRe{EQN}
  \IgNoRe{STM Assertion }
  \IgNoRe{STM Assertion }
  \IgNoRe{PG}
 \def\thOSrengroupestimate{\frefwarning XV.3} \IgNoRe{STM Assertion }
  \IgNoRe{STM Assertion }
  \IgNoRe{STM Assertion }
  \IgNoRe{EQN}
  \IgNoRe{STM Assertion }
  \IgNoRe{EQN}
  \IgNoRe{STM Assertion }
  \IgNoRe{STM Assertion }
  \IgNoRe{EQN}
  \IgNoRe{STM Assertion }
  \IgNoRe{EQN}
  \IgNoRe{EQN}
  \IgNoRe{EQN}
  \IgNoRe{EQN}
  \IgNoRe{STM Assertion }
  \IgNoRe{STM Assertion }
  \IgNoRe{STM Assertion }
  \IgNoRe{STM Assertion }
  \IgNoRe{PG}
  \IgNoRe{STM Assertion }
  \IgNoRe{STM Assertion }
  \IgNoRe{STM Assertion }
  \IgNoRe{STM Assertion }
  \IgNoRe{STM Assertion }
  \IgNoRe{STM Assertion }
  \IgNoRe{STM Assertion }
  \IgNoRe{STM Assertion }
  \IgNoRe{STM Assertion }
  \IgNoRe{STM Assertion }
  \IgNoRe{EQN}
  \IgNoRe{EQN}
  \IgNoRe{STM Assertion }
  \IgNoRe{PG}
  \IgNoRe{STM Assertion }
  \IgNoRe{STM Assertion }
  \IgNoRe{EQN}
  \IgNoRe{STM Assertion }
 \def\lemOStildesourceterm{\frefwarning XVII.5} \IgNoRe{STM Assertion }
  \IgNoRe{EQN}
  \IgNoRe{EQN}
  \IgNoRe{EQN}
  \IgNoRe{EQN}
  \IgNoRe{EQN}
  \IgNoRe{STM Assertion }
  \IgNoRe{STM Assertion }
  \IgNoRe{STM Assertion }
  \IgNoRe{EQN}
  \IgNoRe{EQN}
  \IgNoRe{EQN}
  \IgNoRe{EQN}
  \IgNoRe{EQN}
  \IgNoRe{EQN}
  \IgNoRe{STM Assertion }
  \IgNoRe{STM Assertion }
  \IgNoRe{STM Assertion }
  \IgNoRe{PG}
  \IgNoRe{STM Assertion }
  \IgNoRe{STM Assertion }
  \IgNoRe{EQN}
  \IgNoRe{EQN}
  \IgNoRe{STM Assertion }
  \IgNoRe{EQN}
  \IgNoRe{EQN}
  \IgNoRe{STM Assertion }
  \IgNoRe{EQN}
  \IgNoRe{EQN}
  \IgNoRe{STM Assertion }
  \IgNoRe{STM Assertion }
  \IgNoRe{PG}
  \IgNoRe{STM Assertion }
  \IgNoRe{STM Assertion }
  \IgNoRe{STM Assertion }
  \IgNoRe{PG}
  \IgNoRe{STM Assertion }
  \IgNoRe{STM Assertion }
  \IgNoRe{STM Assertion }
  \IgNoRe{STM Assertion }
  \IgNoRe{PG}
  \IgNoRe{STM Assertion }
  \IgNoRe{STM Assertion }
  \IgNoRe{STM Assertion }
  \IgNoRe{STM Assertion }
  \IgNoRe{STM Assertion }
  \IgNoRe{STM Assertion }
  \IgNoRe{STM Assertion }
  \IgNoRe{STM Assertion }
  \IgNoRe{STM Assertion }
  \IgNoRe{STM Assertion }
  \IgNoRe{STM Assertion }
  \IgNoRe{STM Assertion }
  \IgNoRe{STM Assertion }
  \IgNoRe{STM Assertion }
  \IgNoRe{PG}
  \IgNoRe{STM Assertion }
  \IgNoRe{PG}
  \IgNoRe{STM Assertion }
  \IgNoRe{STM Assertion }
  \IgNoRe{PG}
  \IgNoRe{STM Assertion }
  \IgNoRe{STM Assertion }
  \IgNoRe{EQN}
  \IgNoRe{PG}
  \IgNoRe{EQN}
  \IgNoRe{STM Assertion }
  \IgNoRe{STM Assertion }
  \IgNoRe{PG}
  \IgNoRe{EQN}
  \IgNoRe{EQN}
  \IgNoRe{EQN}
  \IgNoRe{EQN}
  \IgNoRe{EQN}
  \IgNoRe{STM Assertion }
  \IgNoRe{STM Assertion }
  \IgNoRe{EQN}
  \IgNoRe{STM Assertion }
  \IgNoRe{PG}
  \IgNoRe{PG}
  \IgNoRe{PG}
  \IgNoRe{STM Assertion }
  \IgNoRe{EQN}
  \IgNoRe{STM Assertion }
  \IgNoRe{PG}
  \IgNoRe{STM Assertion }
  \IgNoRe{EQN}
  \IgNoRe{STM Assertion }
  \IgNoRe{STM Assertion }
  \IgNoRe{PG}
  \IgNoRe{EQN}
  \IgNoRe{EQN}
  \IgNoRe{EQN}
  \IgNoRe{STM Assertion }
  \IgNoRe{STM Assertion }
  \IgNoRe{STM Assertion }
  \IgNoRe{EQN}
  \IgNoRe{STM Assertion }
  \IgNoRe{STM Assertion }
  \IgNoRe{STM Assertion }
  \IgNoRe{STM Assertion }
  \IgNoRe{STM Assertion }
  \IgNoRe{STM Assertion }
  \IgNoRe{PG}
  \IgNoRe{STM Assertion }
  \IgNoRe{STM Assertion }
  \IgNoRe{STM Assertion }
  \IgNoRe{STM Assertion }
  \IgNoRe{STM Assertion }
  \IgNoRe{STM Assertion }
  \IgNoRe{STM Assertion }
  \IgNoRe{PG}
  \IgNoRe{PG}
 \def\pgOSIInot{\frefwarning 42} \IgNoRe{PG}
  \IgNoRe{PG}
  \IgNoRe{PG}


\newcount\CHAPNO
\newcount\APPNO
\CHAPNO=0
\APPNO=1
\def\advCHAPNO{\advance\CHAPNO by 1}
\def\advAPPNO{\advance\APPNO by 1}

\def\caproman#1{\ifcase#1\or I\or II\or III\or IV\or V\or VI\or VII\or
VIII\or IX\or X\or XI\or XII\or XIII\or XIV\or XV\or XVI\or XVII\or XVIII\or
XIX\or XX\or XXI\or XXII\or XXIII\or XXIV\or XXV\or XXVI\or XXVII\or XXVIII\or XXIX\or XXX\or XXXI\or XXXII\or XXXIII\or XXXIV\or XXXV\or XXXVI\or XXXVII\or XXXVIII\or XXXIX\fi}%

\def\capletter#1{\ifcase#1\or A\or B\or C\or D\or E\or F\or G\or
H\or I\or J\or K\or L\or M\or N\or O\or P\or Q\or R\or
S\or T\or U\or V\or W\or X\or Y\or Z\fi}%

\newcount\cHintroI \cHintroI=\CHAPNO \advCHAPNO 
                              
\newcount\cHnorms  \cHnorms=\CHAPNO \advCHAPNO 
                              \edef\CHnorms{\caproman\CHAPNO}
\newcount\cHproprengrp \cHproprengrp=\CHAPNO \advCHAPNO 
                              
\newcount\cHcovbounds  \cHcovbounds=\CHAPNO \advCHAPNO 
                              
\newcount\cHinsulator \cHinsulator=\CHAPNO \advCHAPNO

 \advAPPNO

\newcount\cHintroII \cHintroII=\CHAPNO \advCHAPNO 
                              \edef\CHintroII{\caproman\CHAPNO}
\newcount\cHamputate \cHamputate=\CHAPNO \advCHAPNO
                              \edef\CHamputate{\caproman\CHAPNO}
\newcount\cHscales \cHscales=\CHAPNO \advCHAPNO
                              \edef\CHscales{\caproman\CHAPNO}
\newcount\cHfourier \cHfourier=\CHAPNO \advCHAPNO
                              \edef\CHfourier{\caproman\CHAPNO}
\newcount\cHmomentum \cHmomentum=\CHAPNO \advCHAPNO
                              \edef\CHmomentum{\caproman\CHAPNO}

\edef\APappSymmetries{\capletter\APPNO} \advAPPNO
\edef\APappGrass{\capletter\APPNO} \advAPPNO

\newcount\cHintroIII \cHintroIII=\CHAPNO \advCHAPNO
                              
\newcount\cHsectors \cHsectors=\CHAPNO \advCHAPNO
                              \edef\CHsectors{\caproman\CHAPNO}
\newcount\cHsecpropbounds \cHsecpropbounds=\CHAPNO \advCHAPNO
                              
\newcount\cHladdersNotn  \cHladdersNotn=\CHAPNO \advCHAPNO
                              
\newcount\cHestren  \cHestren=\CHAPNO \advCHAPNO
                              
\newcount\cHsecmomnorm \cHsecmomnorm=\CHAPNO \advCHAPNO
                              
\newcount\cHmomestren \cHmomestren=\CHAPNO \advCHAPNO

 \advAPPNO

\newcount\cHintroIV  \cHintroIV=\CHAPNO \advCHAPNO
                              
\newcount\cHcomparison   \cHcomparison=\CHAPNO \advCHAPNO
                              
\newcount\cHsumsmom  \cHsumsmom=\CHAPNO \advCHAPNO
                              
\newcount\cHsectorsmom   \cHsectorsmom=\CHAPNO \advCHAPNO
                              
\newcount\cHppladsect    \cHppladsect=\CHAPNO \advCHAPNO

 \advAPPNO

\chapno=\cHintroII


{\nopagenumbers
\multiply\baselineskip by \spacingDenominator\divide \baselineskip by\spacingNumerator

\null\vskip3truecm

%
%
\centerline{\tafontt Single Scale Analysis of Many Fermion Systems}

\vskip0.1in
\centerline{\tbfontt Part 2: The First Scale}

\vskip0.75in
\centerline{Joel Feldman{\parindent=.15in\footnote{$^{*}$}{Research supported 
in part by the
 Natural Sciences and Engineering Research Council of Canada and the Forschungsinstitut f\"ur Mathematik, ETH Z\"urich}}}
\centerline{Department of Mathematics}
\centerline{University of British Columbia}
\centerline{Vancouver, B.C. }
\centerline{CANADA\ \   V6T 1Z2}
\centerline{feldman@math.ubc.ca}
\centerline{http:/\hskip-3pt/www.math.ubc.ca/\squiggle
feldman/}
\vskip0.3in
\centerline{Horst Kn\"orrer, Eugene Trubowitz}
\centerline{Mathematik}
\centerline{ETH-Zentrum}
\centerline{CH-8092 Z\"urich}
\centerline{SWITZERLAND}
\centerline{knoerrer@math.ethz.ch, trub@math.ethz.ch}
\centerline{http:/\hskip-3pt/www.math.ethz.ch/\squiggle
knoerrer/}

\vskip0.75in
\noindent
%
{\bf Abstract.\ \ \ } 
The first renormalization group map arising from the momentum space 
decomposition of a weakly coupled system of fermions at temperature zero 
differs from all subsequent maps. Namely, the component of momentum dual to 
temperature may be arbitrarily large -- there is no ultraviolet cutoff. 
The methods of Part 1 are supplemented to control this special case.

\vfill
\eject


\titleb{Table of Contents}
\halign{\hfill#\ &\hfill#\ &#\hfill&\ p\ \hfil#&\ p\ \hfil#\cr
\noalign{\vskip0.05in}
\S VI&\omit Introduction to Part 2                         \span&\:\pgOSVI\cr
\noalign{\vskip0.05in}
\S VII&\omit Amputated and Nonamputated Green's Functions  \span&\:\pgOSVII\cr
\noalign{\vskip0.05in}
\S VIII&\omit Scales                                   \span&\:\pgOSVIII\cr
\noalign{\vskip0.05in}
\S IX&\omit The Fourier Transform                    \span&\:\pgOSIX\cr
\noalign{\vskip0.05in}
\S X&\omit Momentum Space Norms                         \span&\:\pgOSX\cr
\noalign{\vskip0.05in}
{\bf Appendices}\span\cr
\noalign{\vskip0.05in}
\S B&\omit Symmetries                                 \span&\:\pgOSB\cr
\noalign{\vskip0.05in}
\S C&\omit Some Standard Grassmann Integral Formulae  \span&\:\pgOSC\cr
\noalign{\vskip0.05in}
 &\omit References                                    \span&\:\pgOSIIref \cr
\noalign{\vskip0.05in}
 &\omit Notation                                      \span&\:\pgOSIInot \cr
}
\vfill\eject
\multiply\baselineskip by \spacingNumerator\divide \baselineskip by\spacingDenominator}
\pageno=1


\chap{Introduction to Part 2}\PG\pgOSVI

We continue our analysis of
models for weakly interacting fermions in  $d$-dimensions given in terms of
\item{$\bullet$} a single particle dispersion relation $e(\k)$ on $\bbbr^d$,
\item{$\bullet$} an ultraviolet cutoff $U(\k)$ on $\bbbr^d$,
\item{$\bullet$} an interaction.

\noindent
From now on, we fix $r\ge 2$ and assume that the dispersion relation is 
at least $r+d+1$ times differentiable. 
As discussed in part 1, formally, the generating functional for the  
connected amputated Green's functions is
$$
\cG_{\rm amp}(\phi) =  \log\sfrac{1}{Z} 
\int  e^{\cV(\psi+\phi)}\,d\mu_{C}(\psi)
$$
where $Z=\int e^{\cV(\psi)} d\mu_C(\psi)$.
In this Grassmann integral, there are anticommuting fields
$\psi(\xi)$, where $\xi=(x_0,\x,\si,a) \in\cB= \bbbr \times \bbbr^d \times \{\uparrow,\downarrow\} \times \{0,1\}$. See the beginning of \S\CHnorms.
 The covariance of the Grassmann Gaussian measure $d\mu_C$ is the Fourier transform $C(\xi,\xi')$ of
$$
C(k_0,\k) = \frac{U(\k)}{\imath k_0 - e(\k)}
$$
as in Proposition \propOSpropbnd.
The interaction  is 
$$
\cV(\psi) = \int_{(\bbbr\times\bbbr^2\times\{\uparrow,\downarrow\})^4}  \hskip-.7in
V_0(x_1,x_2,x_3,x_4)\, \psi({\sst(x_1,1)})\psi({\sst(x_2,0)})\psi({\sst(x_3,1)})\psi({\sst(x_4,0)})\
dx_1dx_2dx_3dx_4
$$
We shall, at various places, assume that $V_0$ has a number of symmetries, 
that we abbreviate by single letters --- translation invariance (T), spin independence (S), conservation of particle number (N), ``$k_0$--reversal reality'' (R) and ``bar/unbar exchange invariance'' (B). Precise definitions and a discussion of the properties of these symmetries are given in Appendix 
\APappSymmetries.

Formally, the Green's functions of the many fermion system are
$$
S_{2n}(x_1,y_1,\cdots,x_n,y_n) 
= \sfrac{1}{Z} \int \smprod_{i=1}^n \psi({\sst(x_i,0)})\psi({\sst (y_i,1)})\,
e^{\cV(\psi)} d\mu_C(\psi)
$$
where $Z=\int e^{\cV(\psi)} d\mu_C(\psi)$.
The generating functional for these Green's functions is
$$
\cS(\phi) =  \sfrac{1}{Z} 
\int e^{\phi J\psi} e^{\cV(\psi)}\,d\mu_{C}(\psi)
$$
where the operator $J$ has kernel 
$$
J\big((x_0,\x,\si,a),(x'_0,\x',\si',a')\big)
= \de(x_0-x'_0)\de(\x-\x')\de_{\si,\si'}
    \cases{1& if $a=1,\ a'=0$\cr
          -1& if $a=0,\ a'=1$\cr
           0& otherwise}
\EQN\eqnOSjdef$$
so that the source term has the form
$$
\phi J\psi 
=  \int d\xi\,d\xi' \ \phi(\xi)J(\xi,\xi')\psi(\xi')
=  \int dx \ \bar\phi(x)\psi(x)+\bar\psi(x)\phi(x)
=\psi J\phi
\EQN\eqnOSphijpsi$$
The generating functional for the  connected Green's functions is
$$
\cG(\phi) =  \log\sfrac{1}{Z} 
\int e^{\phi J\psi} e^{\cV(\psi)}\,d\mu_{C}(\psi)
$$
and the connected Green's functions themselves are determined by
$$
\cG(\phi) = 
\sum_{n=1}^\infty \sfrac{1}{(n!)^2} \int\smprod_{i=1}^n dx_idy_i\ 
G_{2n}(x_1,y_1,\cdots,x_n,y_n) 
\smprod_{i=1}^n \bar\phi(x_i)\phi(y_i) 
$$
The relation between the connected Green's functions and the amputated
connected Green's functions is
$$
G_{2n}(x_1,y_1,\cdots,x_n,y_n)  
= \int \smprod_{i=1}^n dx'_i \, dy'_i \,
\big( \smprod_{i=1}^n C(x_i,x'_i)\, C(y'_i,y_i) \big)\,
G_{2n}^{\rm amp}(x'_1,y'_1,\cdots,x'_n,y'_n) 
$$
for $n\ge 2$, and
$$
G_2(x,y) -C(x,y) 
= \int dx' dy'\,C(x,x')\, C(y',y)\,G_2^{\rm amp}(x',y') 
$$

In a multiscale analysis we shall estimate the position space supremum norm of 
connected Green's functions  and the momentum space supremum norm of 
connected amputated Green's functions. 
We fix $r_0\ge 2$ and control the Green's functions, including up to $r_0$ derivatives in the $k_0$ direction.
In \S\CHamputate, we introduce a
variant, $\tilde \Om$, of the renormalization group map $\Om$ for use with the connected Green's functions. In \S\CHscales, we introduce the scale decomposition  that will be used for the multiscale analysis. Using the results of Part 1, we discuss the map $\tilde \Om$, for the first few scales. 
The discussion will be sufficiently general to allow the absorption of a (renormalization) counterterm in the dispersion relation.
In \S\CHmomentum, we introduce norms for use with the amputated Green's functions and discuss the map $\Om$, for the first few scales. 
Notation tables are provided at the end of the paper.

\vfill\eject

\chap{ Amputated and Nonamputated Green's Functions}\PG\pgOSVII

\definition{\STM\defOSrengrpmap}{
The (unamputated) renormalization group map $\tilde\Om_C$ with respect to 
the covariance $C$ associates the Grassmann function
$$
\tilde \Om_C(\cW)(\phi,\psi) 
= \log \sfrac{1}{Z}\int e^{\phi J\ze}\,e^{\cW(\phi,\psi +\ze)} d\mu_C(\ze) 
\qquad{\rm where}\quad Z= \int e^{W(0,\ze)} d\mu_C(\ze)\ne 0
$$
to the Grassmann function $\cW(\phi,\psi)$. As was the case with $\Om_C(\cW)$,
Theorem \theorII\ [FKTr1] implies that, under hypotheses that we will make
explicit later, the formal Taylor expansion of $\tilde \Om_C(\cW)$
converges to an analytic function of $\cW$.
}

\remark{\STM\remOSgenfnrengrp}{
\Item (i)
In the situation described in the introduction, the generating functional for the connected Green's functions is
$$
\cG(\phi) = \tilde \Om_C(\cV)(\phi,0)
$$
\Item (ii) $\tilde\Om$ obeys the semigroup property
$$
\tilde\Om_{C_1+C_2}=\tilde\Om_{C_1}\circ\tilde\Om_{C_2}
$$

}

\noindent In order to use the results 
of part 1 and [FKTr1], we note the following relationship between 
$\tilde\Om_C$ and the renormalization group map
$$
\Om_C(\cW)(\phi,\psi) = \log \sfrac{1}{Z}\int e^{\cW(\phi,\psi +\ze)} d\mu_C(\ze) 
\qquad{\rm where}\quad Z= \int e^{W(0,\ze)} d\mu_C(\ze)
$$
of part 1.

\lemma{\STM\lemOStworengrpmaps}{
$$\eqalign{
\tilde \Om_C(\cW)(\phi,\psi) 
&=\sfrac{1}{ 2}\phi JCJ \phi+\Om_C(\cW)(\phi,\psi+CJ\phi) \cr
& \hskip -30pt = \sfrac{1}{ 2}\phi JCJ \phi+ \Om_C(\cW)(\phi,\psi) +
\int \lW \Om_C(\cW)(\phi,\psi+\ze) -  \Om_C(\cW)(\phi,\psi) \rW_\ze\ 
\lw e^{\phi J\ze} \rw_\ze\,d\mu_C(\ze) \cr
}$$
where for any kernel $B(\et,\et')$,
$B\phi = \int d\eta\, B(\eta,\eta')\phi(\eta')\,$ and
$\phi B\phi = \int d\eta d\eta'\, \phi(\eta)B(\eta,\eta')\phi(\eta')\,$. 
}

\prf
By Lemma \:\lemOSappGrassI, with $\phi$ replaced by $J\phi$, and 
(\eqnOSphijpsi),
$$\eqalign{
\tilde \Om_C(\cW)(\phi,\psi)
&=\log e^{-{1\over 2}(J\phi) C(J\phi)}\sfrac{1}{Z}\int
e^{\cW(\phi,\ze+\psi+CJ\phi)}\  d\mu_{C}(\ze) \cr
&=\log e^{{1\over 2}\phi JCJ \phi}\,e^{\Om_C(\cW)(\phi,\psi+CJ\phi)} \cr
&=\sfrac{1}{ 2}\phi JCJ \phi+\Om_C(\cW)(\phi,\psi+CJ\phi) \cr}
$$
Also by Lemma \:\lemOSappGrassI
$$\eqalign{
\Om_C(\cW)(\phi,\psi+CJ\phi) 
&= \int \lw \Om_C(\cW)(\phi,\psi+\ze+CJ\phi)\rw_\ze\,d\mu_C(\ze) \cr
&\hskip-20pt= e^{{1\over 2}(J\phi) C(J\phi)} 
\int \lw \Om_C(\cW)(\phi,\psi+\ze)\rw_\ze\ e^{\ze J\phi}\,d\mu_C(\ze) \cr
&\hskip-20pt= \int \lw \Om_C(\cW)(\phi,\psi+\ze)\rw_\ze\ \lw e^{\ze J\phi} \rw_\ze\,d\mu_C(\ze) \cr
&\hskip-20pt= \Om_C(\cW)(\phi,\psi) +
\int \lW \Om_C(\cW)(\phi,\psi+\ze) -  \Om_C(\cW)(\phi,\psi) \rW_\ze\ 
\lw e^{\phi J\ze} \rw_\ze\,d\mu_C(\ze) \cr
}\EQN\eqnOSshiftwick$$
\endproof

The aim of the next section is to estimate the (unamputated) 
renormalization group map, $\tilde \Om$, with respect to the norms 
of Definition \defOSgrnorm. 
The difference between the maps $\Omega_C$ and $\tilde \Omega_C$ lies in the source terms as is described in Lemma \lemOStworengrpmaps. The estimates for 
this difference are similar to, but easier than, the estimates
for the map $\Omega_C$ itself.

\definition{\STM\defOSextimpr (External Improving)}{
Let $\|\cdot\|$ be a family of symmetric seminorms on the spaces 
$\cF_m(n)$. We say that  the covariance $C$ is $\Ga$--external improving 
with respect to this family of seminorms if, for each $m\ge 0,\ n\ge 1$, there is an $i$ with
$1\le i \le n$ such that
$$
\Big\| {\rm Ant_{ext}}\int\!\! d\ze\,d\ze'\ J({\sst\et_{m+1},\ze})C({\sst\ze,\ze'})\,
f({\sst\et_1,\cdots,\et_m\,;\,\xi_1\cdots,\xi_{i-1},\ze',\xi_{i},\cdots,\xi_{n-1}}) \Big\|
\le \Ga\,\|f\|
$$
for all $f \in \cF_m(n)$. Recall that ${\rm Ant_{ext}}$ was introduced in
Definition \defOSFmn\ of [FKTo1].
Observe that the function on the left hand side is in $\cF_{m+1}(n-1)$.
 }

\lemma{\STM\lemOSextimpr }{
Let $\|\cdot\|$ be a family of symmetric seminorms and
let the covariance $C$ be $\Ga$--external improving with 
respect to this family of seminorms. Let $f(\phi,\psi,\ze)$ be of degree
$p'$ in $\ze$. The integral $\int
\lW f(\phi,\psi,\ze)\rW_{\ze,C}\ 
\lW\big[\phi J\ze\big]^p\rW_{\ze,C}
d\mu_C(\ze)$ vanishes unless $p=p'$ and then
$$
\Big\|\int
\lW f(\phi,\psi,\ze)\rW_{\ze,C}\ 
\lW\big[\phi J\ze\big]^p\rW_{\ze,C}
d\mu_C(\ze)\Big\|
\le p!\Ga^p\,\|f\|
$$
}
\prf
Observe that $\phi J\psi={\tst \int}d\xi d\xi'\,\phi(\xi)J(\xi,\xi')\psi(\xi')\in A_1\otimes V$. 
By Definition \defOSextimpr\ and Definition \defOScontnorm\ of [FKTo1],
$$
\big\| \Cont{i}{n+1}{C} \big({\rm Ant_{ext}} (h\otimes \phi J\psi)\big) \big\|
\le \Ga\,\|h\|
\EQN\eqnOSextimpr$$
for all $h \in A_m\otimes V^{\otimes n}$,  $m\ge 0,\ n\ge 1$ and some
$1\le i \le n$. Observe that 
$h\otimes \phi J\psi\in(A_m\otimes V^{\otimes n})\otimes(A_1\otimes V)
\cong A_m\otimes A_1\otimes V^{\otimes n+1}$, so that 
${\rm Ant_{ext}} (h\otimes \phi J\psi) \in A_{m+1}\otimes V^{\otimes n+1}$ and
$ \Cont{i}{n+1}{C} \big({\rm Ant_{ext}} (h\otimes \phi J\psi)\big)\in
A_{m+1}\otimes V^{\otimes n-1}$.

Set
$$
g(\phi,\psi,\ze,\ze')=f(\phi,\psi,\ze)\ 
\big[\phi J\ze'\big]^p
$$
By Lemma \lemcontract\ and Remark \remcontract\ of [FKTr1], $p$ times,
starting with $f(\xi,\xi',\xi'')=\lw g(\phi,\psi,\xi',\xi'')\rw_{\xi''}$ 
$$\eqalign{
\int
\lW f(\phi,\psi,\ze)\rW_{\ze,C}\ 
\lW\big[\phi J\ze\big]^p\rW_{\ze,C}
d\mu_C(\ze)
&=\int\big[\lw g(\phi,\psi,\ze,\ze')\rw_{\ze,\ze'}\big]_{\ze'=\ze}
d\mu_C(\ze)\cr
&\hskip-1in=\int\big[\lw \cont{\ze}{\ze'}{C} 
g(\phi,\psi,\ze,\ze')\rw_{\ze,\ze'}\big]_{\ze'=\ze}
d\mu_C(\ze)\cr
&\hskip-1in=\int\big[\lw {\cont{\ze}{\ze'}{C}}^{p'} 
g(\phi,\psi,\ze,\ze')\rw_{\ze,\ze'}\big]_{\ze'=\ze}
d\mu_C(\ze)\cr
&\hskip-1in= {\cont{\ze}{\ze'}{C}}^{p'} g(\phi,\psi,\ze,\ze')\cr
}$$
if $p'\ge p$, since then ${\cont{\ze}{\ze'}{C}}^{p'} g(\phi,\psi,\ze,\ze')$
is independent of $\ze$ and $\ze'$. If $p'>p$, 
${\cont{\ze}{\ze'}{C}}^{p'} g(\phi,\psi,\ze,\ze')=0$. If $p'<p$,
${\cont{\ze}{\ze'}{C}}^{p'} g(\phi,\psi,\ze,\ze')$ is of degree 0 in 
$\ze$ and of degree $p-p'>0$ in $\ze'$ and the integral
$=\int\big[\lw {\cont{\ze}{\ze'}{C}}^{p'} 
g(\phi,\psi,\ze,\ze')\rw_{\ze,\ze'}\big]_{\ze'=\ze}
d\mu_C(\ze)$ again vanishes. 
It now suffices to apply Definition \defGrasscontract\ of [FKTr1]
and (\eqnOSextimpr), $p$ times.
\endproof
\proposition{\STM\propOSextimpr}{
Let $\ga>0$ and $\al\ge 1$ obey $\sfrac{\ga}{\al}\le\sfrac{1}{3}$.
Let 
$$
\cW'(\phi,\psi) = \cW(\phi,\psi+CJ\phi)
$$
If $C$ is $\ga \ib$--external improving
$$
N\big(\cW'-\cW;\cb,\ib,\al\big)
\le  \sfrac{\ga}{\al}\ N\big(\cW;\cb,\ib,2\al\big)
$$

}
\prf
Write $\cW(\phi,\psi) = \sum_{m,n}\cW_{m,n}(\phi,\psi)$ 
with $\cW_{m,n}\in A_m[n]$ and 
$$
\cW(\phi,\psi+\ze) =  \sum_{m,n}\cW_{m,n}(\phi,\psi+\ze)
=\sum_{m,n}\smsum_{p=0}^n\cW_{m,n-p,p}(\phi,\psi,\ze)
$$ 
with $\cW_{m,n-p,p}\in A_m[n-p,p]$. By Lemma \lemGrasscompatnorm.iii of [FKTr1]
$$
\|\cW_{m,n-p,p}\|\le {\tst{n\choose p}}\|\cW_{m,n}\|
$$
 Then, by (\eqnOSshiftwick),
$$\meqalign{
\cW'(\phi,\psi)-\cW(\phi,\psi)
 &= \cW(\phi,\psi+CJ\phi)-\cW(\phi,\psi)\cr
&=\int \lW \cW(\phi,\psi+\ze)-\cW(\phi,\psi)\rW_{\ze,C}
\ \lw e^{\phi J\ze}\rW_{\ze,C}\ d\mu_C(\ze)\cr
&=\sum_{m,p\ge 0\atop n\ge1}\smsum_{p'=1}^n\sfrac{1}{p!}
\int \lw \cW_{m,n-p',p'}(\phi,\psi,\ze)\rW_{\ze,C}
\ \lw (\phi J\ze)^p\rW_{\ze,C}\ d\mu_C(\ze)\cr
&=\sum_{m\ge 0\atop n\ge1}\smsum_{p=1}^n\sfrac{1}{p!}
\int \lW \cW_{m,n-p,p}(\phi,\psi,\ze)\rW_{\ze,C}
\ \lW (\phi J\ze)^p\rW_{\ze,C}\ d\mu_C(\ze)\cr
}$$
By Lemma \lemOSextimpr
$$\eqalign{
\Big\|\int
 \lW \cW_{m,n-p,p}(\phi,\psi,\ze)\rW_{\ze,C}
\ \lW (\phi J\ze)^p\rW_{\ze,C}\ d\mu_C(\ze)\Big\|
&\le p!(\ga\ib)^p\,\|\cW_{m,n-p,p}\|\cr
&\le p!{\tst{n\choose p}}(\ga\ib)^p\,\|\cW_{m,n}\|
}$$
so that
$$\eqalign{
N\big(\cW'-\cW;\cb,\ib,\al\big)
&\le \sfrac{\cb}{\ib^2}\sum_{m\ge 0\atop n\ge1}\smsum_{p=1}^n 
\sfrac{1}{p!}\al^{n-p}\ib^{n-p}
\Big\|\int
 \lW \cW_{m,n-p,p}\rW_{\ze,C}
\ \lW (\phi J\ze)^p\rW_{\ze,C}\ d\mu_C(\ze)\Big\|
\cr
&\le\sfrac{\cb}{\ib^2}\sum_{m\ge 0\atop n\ge 1}\smsum_{p=1}^n
{\tst{n\choose p}}\big(\sfrac{\ga}{\al}\big)^{p}\al^n\ib^{n}
 \big\| W_{m,n} \big\|\cr
&=\sfrac{\cb}{\ib^2}\sum_{m\ge 0\atop n\ge 1}
\Big[\big(1+\sfrac{\ga}{\al}\big)^{n}-1\Big]\al^n\ib^{n}
 \big\| W_{m,n} \big\|\cr
}$$
Applying
$$
\big(1+\sfrac{\ga}{\al}\big)^{n}-1
\le \sfrac{\ga}{\al}n\big(1+\sfrac{\ga}{\al}\big)^{n-1}
\le \sfrac{\ga}{\al}\big(\sfrac{3}{2}\big)^n\big(1+\sfrac{\ga}{\al}\big)^{n-1}
\le\sfrac{\ga}{\al}2^{n}
$$
we have
$$\eqalign{
N\big(\cW'-\cW;\cb,\ib,\al\big)
&\le \sfrac{\cb}{\ib^2}\sum_{m\ge 0\atop n\ge 1}
\sfrac{\ga}{\al}2^{n} \al^n\ib^{n} \big\| W_{m,n} \big\|\cr
&\le \sfrac{\ga}{\al} N\big(\cW;\cb,\ib,2\al\big)
}$$
\endproof

\corollary{\STM\corOSextimpr}{
Let $\ga,\ga'>0$ and $\al>1$ with $\sfrac{\ga}{\al}\le\sfrac{1}{6}$.
Let 
$$
\cW'_\ka(\phi,\psi) = \cW(\phi,\psi+C_\ka J\phi)
$$
If $C_0$ is $\ga \ib$--external improving and
$\sfrac{d\hfill}{d\ka}C_\ka\big|_{\ka=0}$ is $\ga' \ib$--external improving
$$
N\big(\sfrac{d\hfill}{d\ka}\cW'_\ka\big|_{\ka=0};\cb,\ib,\al\big)
\le  2\sfrac{\ga'}{\al}\ N\big(\cW;\cb,\ib,2\al\big)
$$

}
\prf
Define $D_z=C_0+z\sfrac{d\hfill}{d\ka}C_\ka\big|_{\ka=0}$ and
$\cW''_z(\phi,\psi) = \cW(\phi,\psi+D_zJ\phi)-\cW(\phi,\psi)$. Then
$\sfrac{d\hfill}{d\ka}\cW'_\ka\big|_{\ka=0}
=\sfrac{d\hfill}{dz}\cW''_z\big|_{z=0}$. Furthermore, applying the triangle
inequality directly to the Definition \defOSextimpr\ of ``external improving'',
we see that $D_z$ is $(\ga+|z|\ga')\ib$--external improving. As
$\sfrac{\ga+|z|\ga'}{\al}\le\sfrac{1}{3}$ for all $|z|\le \sfrac{\al}{6\ga'}$,
Proposition \propOSextimpr\ implies that
$$
N\big(\cW''_z;\cb,\ib,\al\big)
\le  \sfrac{1}{3}\ N\big(\cW;\cb,\ib,2\al\big)
$$
for all $|z|\le \sfrac{\al}{6\ga'}$. The Corollary now follows by the Cauchy
integral theorem, applied with contour the circle of radius 
$\sfrac{\al}{6\ga'}$ centered on the origin.
\endproof

Similarly to Lemma \lemOSscalednorm, we have
\lemma{\STM\lemOSscalednormextimpr}{
Let  $ \rho_{m;n}$ be a  sequence of nonnegative real numbers such that 
$\rho_{m;n'} \le \rho_{m;n}$ for $n'\le n$. Define
for $f\in\cF_m(n)$
$$
\|f\| =\rho_{m;n} \,\|f\|_{1,\infty}
$$
where $\|f\|_{1,\infty}$ is the $L_1$--$L_\infty$--norm 
introduced in Example \exOSSymmNorm. 
Let $C$ be a covariance and $\Ga$ obey 
$$\meqalign{
\Ga &\ge \sfrac{\rho_{1\,;\,n-1}}{\rho_{0;n}}\,\tn C\tn_{1,\infty}
\qquad && &{\rm for\ all}\ n\ge 1\cr 
\Ga & \ge 
\sfrac{\rho_{m+1\,;\,n-1}}{\rho_{m;n}}\,\tn C\tn_\infty  
\qquad && &{\rm for\ all}\ m,n\ge 1
}$$
Then $C$ is $\Ga$--external improving with respect to the family of seminorms 
$\|\cdot\|$, in the sense of Definition \defOSextimpr.

}

\prf
Let $f\in\cF_m(n)$ and set
$$
g({\sst\et_1,\cdots,\et_{m+1}\,;\,\xi_1,\cdots,\xi_{n-1}})
={\rm Ant_{ext}}\int\!\! d\ze d\ze'\ J({\sst\et_{m+1},\ze})C({\sst\ze,\ze'})\,
f({\sst\et_1,\cdots,\et_m\,;\,\xi_1\cdots,\xi_{i-1},\ze',\xi_{i},\cdots,\xi_{n-1}})
$$
If $m=0$
$$
\tn g \tn_{1,\infty}
\le \tn f\tn_{1,\infty}\tn JC\tn_{1,\infty} 
= \tn f\tn_{1,\infty}\tn C\tn_{1,\infty} 
$$
by Lemma \lemOSelloneinfty. If $m\ne 0$,
$$
\tn g \tn_{1,\infty}
\le \tn f\tn_{1,\infty} \tn JC\tn_\infty
= \tn f\tn_{1,\infty} \tn C\tn_\infty
$$
Since $g\in\cF_{m+1}$, $\| g \|_{1,\infty}=\tn g \tn_{1,\infty}$.
The Lemma now follows from the hypothesis on $\Ga$.
\endproof

\vfill\eject

\chap{ Scales}\PG\pgOSVIII

From now on we discuss the situation that the dispersion relation $e(\k)$ has zeroes on the support of the ultraviolet cutoff $U(\k)$; in other words, that the Fermi surface $F$ is not empty. Then a single scale analysis as for 
insulators is not possible because
there is an infrared problem due to the singularity of the 
propagator $\sfrac{U(\k)}{\imath k_0 -e(\k)}$ on the set
$$
\{(k_0,\k)\in \bbbr\times\bbbr^d\ \big| k_0 =0,\ e(\k)=0\}
$$
which can be canonically identified with the Fermi surface. This singularity causes the $L_1$--$L_\infty$ norm (in position space) of the propagator to be infinite.
 
To analyze the singularity at the Fermi surface, we introduce scales by slicing momentum space into shells around the Fermi surface. 
We choose a  ``scale parameter'' $M>1$ and a function 
$\nu\in C^\infty_0([\sfrac{1}{M},\,2M])$ that 
takes values in $[0,1]$, is identically 1 on $[\sfrac{2}{M},M]$ and obeys
$$
\sum_{j=0}^\infty \nu\big(M^{2j}x\big) = 1
$$
for $0<x<1$. 

The scale parameter $M$ is chosen sufficiently big (depending on the 
dispersion relation $e(\k)$ and the ultraviolet cutoff $U(\k)$). The function $\nu$
may be constructed by choosing a function
$\varphi\in C_0^\infty\big((-2,2)\big)$ that is identically one on $[-1,1]$
and setting $\nu(x)=\varphi(x/M)-\varphi(Mx)$ for $x>0$ and zero otherwise. Then $\nu(x)$ vanishes for
$x\ge 2M$ and $x\le\sfrac{1}{M}$ and is identically one for $\sfrac{2}{M}\le
x\le M$ and $\sum_{j=0}^\infty \nu\big(M^{2j}x\big)=\varphi(x/M)$ for $x>0$.

\definition{\STM\defOSscales}{
\Item i) 
For $j\ge 1$, the $j^{\rm th}$ scale function on $\bbbr\times\bbbr^d$ is defined as
$$ 
\nu^{(j)}(k)=\nu\left(M^{2j}(k_0^2+e(\k)^2)\right) 
$$
By construction, $\nu^{(j)}$ is identically one on
$$
\set{k=(k_0,\k) \in \bbbr\times\bbbr^d}
{\sqrt{\sfrac{2}{M}}\, \sfrac{1}{M^j}\le |ik_0-e(\k)|\le \sqrt{M} \sfrac{1}{M^j} }
$$
The support of $\nu^{(j)}$ is called the $j^{\rm th}$ shell. By construction, it is contained in
$$
\set{k\in \bbbr\times\bbbr^d}
{\sfrac{1}{\sqrt{M}}\, \sfrac{1}{M^j}\le |ik_0-e(\k)|\le \sqrt{2M} \sfrac{1}{M^j} }
$$
The momentum $k$ is said to be of scale $j$ if $k$ lies in the $j^{\rm th}$ shell.
\Item ii) 
For real $j\ge 1$, set
$$
\nu^{(\ge j)}(k)=\varphi\big(M^{2j-1}(k_0^2+e(\k)^2)\big)
$$
with the function $\varphi$ introduced just before this definition.
By construction, $\nu^{(\ge j)}$ is identically 1 on 
$$
\set{k \in \bbbr\times\bbbr^d} {|ik_0-e(\k)|\le \sqrt{M} \sfrac{1}{M^j} }
$$ 
Observe that if $j$ is an integer, then for $|ik_0-e(\k)|>0$
$$
\nu^{(\ge j)}(k)=\smsum_{i\ge j}\nu^{(i)}(k)
$$
The support of $\nu^{(\ge j)}$ is called the $j^{\rm th}$ neighbourhood of 
the Fermi surface. By construction, it  is contained in
$$
\set{k\in \bbbr\times\bbbr^d}{|ik_0-e(\k)|\le \sqrt{2M} \sfrac{1}{M^j} }
$$
}

\remark{\STM\remOSlargej}{
Since the scale parameter  $M>1$, the shells near the Fermi curve have $j$ near 
$+\infty$, and the neighbourhoods shrink as $j \rightarrow \infty$. 
}

\noindent
{\bf Conventions \STM\convOSI}
\Item{i)}
We choose $M$ so big that $\nu^{(\ge 1)}(k) \le U(\k)$ for all 
$k =(k_0,\k)\in \bbbr\times\bbbr^d$.
\Item{ii)} We also use the notations
$$\eqalign{
\nu^{(\le j)}(k) &=\smsum_{i=0}^j \nu^{(i)}(k)  \cr
\nu^{( < j)}(k) &= \nu^{(\le j-1)}(k) \qquad,\qquad 
  \nu^{( > j)}(k) = \nu^{(\ge j+1)}(k)   \cr
}$$
\Item{iii)} Generic constants that depend only on the dispersion 
relation $e(\k)$ and the ultraviolet cutoff $U(\k)$ will be denoted by 
``$\abcst$". Generic constant that may also depend on the scale parameter $M$,
but still not on the scale $j$, will be denoted ``$\const$".

\vskip 1cm

For technical discussions we also need a set of functions that ``envelope'' the
various shells. We set $\tilde \nu=\varphi(x/M^2)-\varphi(M^2x)$ 
for $x>0$ and zero otherwise. It is in 
$C^\infty_0\big((\sfrac{1}{M^2},\,2M^2)\big)$, takes values 
in $[0,1]$ and is identically 1 on $[\sfrac{2}{M^2},\,M^2]$ and hence 
on the support of $\nu$, assuming that $M\ge 2$.

\definition{\STM\defOSextendedshell}{
\Item i) 
For $j\ge 1$, the $j^{\rm th}$ extended scale function on 
$\bbbr\times\bbbr^d$ is defined as
$$ 
\tilde\nu^{(j)}(k)=\tilde\nu\left(M^{2j}(k_0^2+e(\k)^2)\right) 
$$
The support of $\tilde\nu^{(j)}$ is called the $j^{\rm th}$ extended shell. 
It is (for $j\ge 2$) contained in the union of the
$(j-1)^{\rm st}$, $j^{\rm th}$ and $(j+1)^{\rm st}$ shells. In fact,
if $M\ge 2$,  $\tilde\nu^{(j)}$ is identically 1 on the $j^{\rm th}$ shell
and, if $j,M\ge 2$,  $\nu^{(j-1)}+\nu^{(j)}+\nu^{(j+1)}$ is identically 1 on 
the $j^{\rm th}$ extended shell.
\Item ii) 
By definition, the $j^{\rm th}$ extended neighbourhood is the union of the 
$i^{\rm th}$ extended shells with $i\ge j$. It is (for $j\ge 2$) contained 
in the $(j-1)^{\rm st}$ neighbourhood of the Fermi surface. The function
$$
\tilde\nu^{(\ge j)}(k) = \varphi(M^{2j-2}(k_0^2+e(\k)^2))
$$
is supported on the $j^{\rm th}$ extended neighbourhood and identically one
on the $j^{\rm th}$ neighbourhood.
\Item{iii)}
Set $\bar\nu^{(\ge j)}(k) = \varphi(M^{2j-3}(k_0^2+e(\k)^2))=\nu^{(\ge
j-1)}$. 
Then $\bar\nu^{(\ge j)}(k)$ is identically one on the $j^{\rm th}$ extended neighbourhood. The support of $\bar\nu^{(\ge j)}$ is called the $j^{\rm th}$ doubly extended neighbourhood and is contained in
$\ 
\set{k\in \bbbr\times\bbbr^d}{|ik_0-e(\k)|\le \sqrt{2}M^{3/2} \sfrac{1}{M^j} }
\ $.
}

Observe that the ultraviolet cutoff $U(\k)$ does not depend on $k_0$, 
so that the propagator $\sfrac{U(\k)}{\imath k_0 -e(\k)}$ is not compactly
supported. However, we can use the results of part 1 to integrate
out the ultraviolet part of the model in the $k_0$--direction and  to pass to a
model  whose propagator is supported in the second neighbourhood. We choose to pass to a model supported in the second, rather than first, neighbourhood
because the second doubly extended neighbourhood can be made arbitrarily 
small by choosing $M$ sufficiently large.

In the renormalization group analysis we shall add a counterterm $\de e(\k)$
to the dispersion relation $e(\k)$.

\definition{\STM\defOSCTmSpace}{ 
Let $\mu>0$. The space of counterterms, $\cE_\mu$, consists of all  functions $\de e(\k)$
on $\bbbr^d$ that are supported in $\set{\k\in\bbbr^d}{U(\k)\ne 0}$
and obey
$$ 
\| \de \hat e\|_{1,\infty} < \mu
+\smsum_{\de\ne \0}\infty\, t^\de
$$
where $\de \hat e$ was defined just before Definition \defOScbzero\ 
and the norm $\|\ \cdot\ \|_{1,\infty}$ was defined in Example 
\exOSSymmNorm.
}

Recall, from Definition \defOScbzero\ that
$\
\cb_0=\sum\limits_{|\bde|\le r\atop |\de_0|\le r_0}t^\de
+\sum\limits_{|\bde|> r\atop {\rm or\ }|\de_0|> r_0}\infty\, t^\de
\in\fN_{d+1}
\ $ 
and 
$\ 
\fe_0(X) = \sfrac{\cb_0}{1- X}
\ $,
for $X \in \fN_{d+1}$ with $X_\0<1$.

\theorem{\STM\thmOSfirststep}{ 
Fix $j_0\ge 1$ and set, for $\de e\in\cE_\mu$,
$$
C_0(k;\de e) = \sfrac{U(\k) - \nu^{(> j_0)}(k)}{\imath k_0 - e(\k)+\de e(\k)}
$$
Define the covariance $C_0(\de e)$ by
$$
C_0(\xi,\xi';\de e)
=\cases{ \de_{\si,\si'} \int\sfrac{d^{d+1}k}{(2\pi)^{d+1}}\, 
e^{\imath<k,x-x'>_-}C_0(k;\de e)
& if $a=0, \,a'=1$ \cr
\noalign{\vskip.05in}
0 & if $a=a'$\cr
\noalign{\vskip.05in}
-C_0(\xi',\xi;\de e)  & if $a=1, \,a'=0$ \cr
}$$
for $\xi = (x,a)=(x_0,\x,\si,a),\ \xi' = (x',a')=(x'_0,\x',\si',a')$. 
Then there are ($M$ and $j_0$--dependent) constants
$\ib,\ \be_0,\ \veps_0,\ \const\,$ and $\mu>0$ such that, for all 
$\be\ge\be_0$ and $\veps\le \veps_0$, the following holds:

{\parindent=.25in\item{}
Choose a system $\vec\rho=\big(\rho_{m;n}\big)_{m,n\in\bbbn_0}$, of positive real numbers obeying
$\rho_{m;n-1} \le \rho_{m;n}$,  $\rho_{m+1;n-1} \le \rho_{m;n}$ and 
$\rho_{m+m';n+n'-2} \le \rho_{m;n}\,\rho_{m';n'}$.
For an even Grassmann function 
$$
\cW(\phi,\psi) = \smsum_{m,n\ge 0\atop m+n\ {\rm even}} 
\int_{\cB^{m+n}}\hskip-15pt {\sst d\et_1\cdots d\et_m\ d\xi_1\cdots d\xi_n}\ 
W_{m,n}({\sst\et_1\cdots \et_m,\xi_1,\cdots,\xi_n})\,
\phi{\sst(\et_1)}\cdots\phi{\sst(\et_m)}\,\psi{\sst(\xi_1)}\cdots{\sst\psi(\xi_n)}
$$
with kernels $W_{m,n}$ that are separately antisymmetric under 
permutations of their $\et$ and $\xi$ arguments and 
$X \in \fN_{d+1}$ with $X_\0<1$, set
$$
N_0(\cW \cl \be;X,\vec\rho) = \fe_0(X)
\smsum_{m+n \ge 2 \atop m+n\ {\rm even}} \be^n \rho_{m;n}
\| W_{m,n}\|_{1,\infty}
$$
Let  $X \in \fN_{d+1}$ with $X_\0<\sfrac{1}{4}$.
The formal Taylor series $\tilde\Om_{C_0(\de e)}(\cV)$
converges to an analytic map  on 
$\set{\big(\cV(\psi),\de e\big)}{\cV\hbox{ even},\  
N_0(\cV \cl 32\be;X,\vec\rho)_\0\le \veps\fe_0(X)_\0,\ \de e\in\cE_\mu,\ 
\tn\de\hat e\tn_{1,\infty}\le X_\0}$.
Furthermore, for all $\de e\in\cE_\mu$ with
$ \|\de\hat e\|_{1,\infty}\le X$ and all
even Grassmann functions $\cV(\psi)$ with $\,N_0(\cV \cl 32\be;X,\vec\rho) 
\le \veps\fe_0(X)\,$ , one
has
$$
N_0\big(\tilde\Om_{C_0(\de e)}(\cV)(\phi,\psi) - \cV(\psi)
-\half\phi JC_0(\de e)J\phi\,\cl \be;X,\vec\rho\big) 
\le \sfrac{\veps\ib}{\be}\fe_0(X) 
$$
and
$$
N_0\big(\sfrac{d\hfill}{ds}\big[\tilde\Om_{C_0(\de e+s\de e')}(\cV)(\phi,\psi)
-\half\phi JC_0(\de e+s\de e')J\phi\big]_{s=0}\cl \be;X,\vec\rho\big) 
\le \sfrac{\veps\ib}{\be}\fe_0(X)
 \|\de\hat e'\|_{1,\infty}
$$

}
}

\prf
By Proposition \propIntBndsIV\ (with $\chi = \nu^{(> j_0)}$ and $e$ 
replaced by $e-\de e$) and Proposition \propOSrealfirstpropbound, 
there is a constant $\cst{}{1}$ such that
$
S\big(C_0(\de e)\big)\le\cst{}{1},\ 
\tn C_0(\de e)\tn_\infty\le \cst{}{1}$ and 
$$
\| C_0(\de e)\|_{1,\infty}
\ \le \sfrac{1}{4}\cst{}{1}\fe_0({\sst \|\de \hat e\|_{1,\infty}})
$$
Set $\|\ \cdot\ \| = \rho_{m;n}\|\ \cdot\ \|_{1,\infty}$ for functions 
on $\cB^m\times\cB^n$. Set $\ib=4\cst{}{1}$ and 
$\cb = \cst{}{1}\fe_0(X)$. By
Lemma \lemOSscalednorm, with respect to this family of seminorms,
$\ib$ is an integral bound for $C_0(\de e)$ and $\cb$
is a contraction bound for $C_0(\de e)$. Furthermore, by Lemma
\lemOSscalednormextimpr, $C_0(\de e)$ is ${\rm const}_1$--external improving.

For any Grassmann function $\cW(\phi,\psi)$ let $N(\cW ;\cb,\ib, \al)$ be the 
norm of Definition \defOSgrnorm\ with respect to the family of seminorms 
$\|\cdot\|$. Set $\al=\sfrac{\be}{\ib}$. Then, as $\cb=\sfrac{\ib}{4}\fe_0(X)$,
$\ 
N(\cW ;\cb,\ib, \al)\ 
=\ \sfrac{1}{4\ib}N_0(\cW \cl \be;X,\vec\rho)
\ $
and, if $\cV = \lw \cV' \rw_{C_0(\de e)}$,
$$\eqalign{
N(\cV' ;\cb,\ib, \al) &\le \sfrac{1}{4\ib}\,N_0(\cV \cl 2\be;X,\vec\rho) \cr
N(\cV'-\cV ;\cb,\ib, \al) &\le \sfrac{\ib}{2\be^2}\,N_0(\cV \cl 2\be;X,\vec\rho) \cr
}$$
by Corollary \corwicknorm\ of [FKTr1].

To prove the first part of the Theorem, set 
$\,\cV = \lw \cV' \rw_{C_0(\de e)}$. 
Then the hypotheses of Theorem \thmOSroptheorII\ with $C=C_0(\de e)$
 and $\cW=\cV'$ are fulfilled.
Therefore,
$$\eqalign{
\sfrac{1}{4\ib}
N_0\big(\Om_{C_0(\de e)}(\cV)-\cV\cl\be;X,\vec\rho\big) 
&\le  N\big(\Om_{C_0(\de e)}(\lw \cV'\rw_{C_0(\de e)})-\cV';\cb,\ib,\al\big)  
+ N\big(\cV'-\cV;\cb,\ib,\al\big) \cr
&\le \sfrac{2}{\al^2}\, \sfrac{N(\cV';\cb,\ib, 8\al)^2}
{1-{4\over\al^2}N(\cV';\cb,\ib, 8\al ) }
+ N\big(\cV'-\cV;\cb,\ib,\al\big) \cr
&\le \sfrac{2}{\al^2}\, \sfrac{{1\over 16\ib^2}N_0(\cV;16\be;X,\vec\rho)^2}
{1-{4\over\al^2}{1\over 4\ib}N_0(\cV;16\be;X,\vec\rho ) }
+ \sfrac{\ib}{2\be^2}\,N_0\big(\cV\cl 2\be;X,\vec\rho\big) \cr
&\le \sfrac{\veps^2}{8\be^2}\, \sfrac{\fe_0(X)^2}
{1-{\veps\ib\over\be^2}\fe_0(X) }
+ \sfrac{\veps\ib}{2\be^2}\fe_0(X)\cr
&\le \sfrac{\veps\ib}{\be^2}\fe_0(X)\cr
}$$
since $\veps_0$ is chosen so that $\sfrac{\ib}{\be^2}\veps<\sfrac{1}{4}$ and,
by Corollary \corOSappMonoidIV.ii,
$\sfrac{\fe_0^2(X)}{1-{1\over4}\fe_0(X) }\le\abcst\ \fe_0(X)$.
By Lemma \lemOStworengrpmaps\ and Proposition \propOSextimpr,
$$\eqalignno{
&N_0\big(\tilde\Om_{C_0(\de e)}(\cV)
                -\half\phi JC_0(\de e)J\phi-\cV\cl\be;X,\vec\rho\big)\cr
&\hskip.5in = N_0\big(\Om_{C_0(\de e)}(\cV)(\phi,\psi
               +C_0(\de e)J\phi)-\cV(\psi)\cl\be;X,\vec\rho\big)\cr
&\hskip.5in\le N_0\big(\Om_{C_0(\de e)}(\cV)(\phi,\psi
     +C_0(\de e)J\phi)-\Om_{C_0(\de e)}(\cV)(\phi,\psi)\cl\be;X,\vec\rho\big)\cr
&\hskip3in
+N_0\big(\Om_{C_0(\de e)}(\cV)(\phi,\psi)-\cV(\psi)\cl\be;X,\vec\rho\big)\cr
&\hskip.5in
\le\sfrac{1}{4\al} N_0\big(\Om_{C_0(\de e)}(\cV)\cl 2\be;X,\vec\rho\big)
+\sfrac{4\veps\ib^2}{\be^2}\fe_0(X)\cr
&\hskip.5in
\le\sfrac{1}{4\al} 
\Big[N_0\big(\Om_{C_0(\de e)}(\cV)-\cV\cl 2\be;X,\vec\rho\big)
            +N_0\big(\cV\cl 2\be;X,\vec\rho\big)\Big]
+\sfrac{4\veps\ib^2}{\be^2}\fe_0(X)\cr
&\hskip.5in
\le\sfrac{\ib}{4\be} 
\Big[\sfrac{\veps\ib^2}{\be^2}\fe_0(X)+\veps\fe_0(X)\Big]
+\sfrac{4\veps\ib^2}{\be^2}\fe_0(X)\cr
&\hskip.5in\le \sfrac{\veps\ib}{\be}\fe_0(X)\cr
}$$
The joint analyticity in $\cV$ and $\de e$ follows from
Proposition \propOSrealfirstpropbound\ and Remark \remjointanalyticity\ 
of [FKTr1].

Finally, we prove the bound on 
$\sfrac{d\hfill}{ds}\big[\tilde\Om_{C_0(\de e+s\de e')}(\cV)(\phi,\psi)
-\half\phi JC_0(\de e+s\de e')J\phi\big]_{s=0}$.
As
$$
\sfrac{d\hfill}{ds}C_0(k;\de e+s\de e')\big|_{s=0}
=-\sfrac{U(\k)-\nu^{(> j_0)}(k)}{[\imath k_0-e(\k)+\de e(\k)]^2}\de e'(\k)
$$
Proposition \propIntBndsII.i, Proposition \propOSpropbnd.i and
Proposition  \propOSrealfirstpropbound.ii give that
$$\eqalign{
S\big(\sfrac{d\hfill}{ds}C_0(\de e+s\de e')\big|_{s=0}\big)
                   &\le\cst{}{1}\sqrt{\tn \de\hat e'\tn_{1,\infty}}\cr
\TN \sfrac{d\hfill}{ds}C_0(\de e+s\de e')\big|_{s=0}\TN_\infty
                      &\le \cst{}{1}\tn \de\hat e'\tn_{1,\infty}\cr
\big\| \sfrac{d\hfill}{ds}C_0(\de e+s\de e')\big|_{s=0}\big\|_{1,\infty}
&\le \half\cst{}{1}\fe_0({\sst \|\de \hat e\|_{1,\infty}})\,
              \| \de\hat e'\|_{1,\infty}\cr
}$$
Set $\ib'=4\cst{}{1}\sqrt{\tn \de\hat e'\tn_{1,\infty}}$ and 
$\cb' = \cst{}{1}\fe_0(X)\,\| \de\hat e'\|_{1,\infty}$. By
Lemma \lemOSscalednorm, $\half \ib'$ is an integral bound for 
$\sfrac{d\hfill}{ds}C_0(\de e+s\de e')\big|_{s=0}$ and $\cb'$
is a contraction bound for $\sfrac{d\hfill}{ds}C_0(\de e+s\de e')\big|_{s=0}$. 
Furthermore, by Lemma \lemOSscalednormextimpr, 
$\sfrac{d\hfill}{ds}C_0(\de e+s\de e')\big|_{s=0}$ is 
$\cst{}{1}\tn \de\hat e'\tn_{1,\infty}$--external improving.

Define $C_\ka=C_0(\de e+\ka\de e')$ and $\cW_\ka$ by 
$\lw \cW_\ka\rw_{C_\ka}=\cV$. Even when $\al$ is replaced by $2\al$,
the hypotheses of Lemmas  \lemprftwoA.i and \lemprftwoB.i of [FKTr1], 
with $\mu=1$, are satisfied. By these two Lemmas, followed by Corollary  
\corwicknorm.iii of [FKTr1],
$$\eqalign{
N\big(\sfrac{d\hfill}{ds}\Om_{C_s}(\cV)\big|_{s=0}\cl\cb,\ib, \al\big) 
&=N\big(\sfrac{d\hfill}{ds}\Om_{C_s}\big(\lw \cW_s\rw_{C_s}\big)\big|_{s=0}
              \cl\cb,\ib, \al\big) \cr
&\le N\big(\sfrac{d\hfill}{ds}\Om_{C_s}
\big(\lw\cV'\rw_{C_s}\big)\big|_{s=0}\cl\cb,\ib, \al\big)  
+N\big(\sfrac{d\hfill}{ds}\Om_{C_0}\big(\lw \cW_s\rw_{C_0}\big)\big|_{s=0}
              \cl\cb,\ib, \al\big) \cr
&\le \sfrac{1}{2\al^2}\, \sfrac{N(\cV';\cb,\ib, 8\al)}
{1-{4\over\al^2}N(\cV';\cb,\ib, 8\al ) }
\cst{}{1}\fe_0(X)\,\| \de\hat e'\|_{1,\infty}\cr
&\hskip.25in+\Big\{1+\sfrac{2}{\al^2}\, \sfrac{N(\cV';\cb,\ib, 8\al)}
                  {1-{4\over\al^2}N(\cV';\cb,\ib, 8\al ) } \Big\} 
N\big(\sfrac{d\hfill}{ds}\cW_s\big|_{s=0} \cl\cb,\ib, 2\al\big)\cr
&\le \sfrac{1}{2\al^2}\, \sfrac{N(\cV';\cb,\ib, 8\al)}
{1-{4\over\al^2}N(\cV';\cb,\ib, 8\al ) }
\cst{}{1}\fe_0(X)\,\| \de\hat e'\|_{1,\infty}\cr
&\hskip.25in+\Big\{1+\sfrac{2}{\al^2}\, \sfrac{N(\cV';\cb,\ib, 8\al)}
                  {1-{4\over\al^2}N(\cV';\cb,\ib, 8\al ) } \Big\} 
\sfrac{\| \de\hat e'\|_{1,\infty}}{(2\al-1)^2}N\big(\cV \cl\cb,\ib, 4\al\big)\cr
&\le \const\sfrac{\veps}{\al^2}\fe_0(X)\,\| \de\hat e'\|_{1,\infty}\cr
}$$
as above. 
By Lemma \lemOStworengrpmaps, Proposition \propOSextimpr\ and Corollary
\corOSextimpr
$$\eqalign{
&N_0\big(\sfrac{d\hfill}{ds}\big[\tilde\Om_{C_0(\de e+s\de e')}(\cV)(\phi,\psi)
-\half\phi JC_0(\de e+s\de e')J\phi\big]_{s=0}\cl\be;X,\vec\rho\big)\cr
&\hskip.5in= N_0\Big(
\sfrac{d\hfill}{ds}\Om_{C_0(\de e+s\de e')}(\cV)\big(\phi,\psi
               +C_0(\de e+s\de e')J\phi\big)\big|_{s=0}\cl\be;X,\vec\rho\Big)\cr
&\hskip.5in\le N_0\Big(\sfrac{d\hfill}{ds}\Om_{C_0(\de e+s\de e')}(\cV)
\big(\phi,\psi+C_0(\de e)J\phi\big)\big|_{s=0}\cl\be;X,\vec\rho\Big)\cr
&\hskip2in
+N_0\Big(\sfrac{d\hfill}{ds}\Om_{C_0(\de e)}(\cV)\big(\phi,\psi
               +C_0(\de e+s\de e')J\phi\big)\big|_{s=0}\cl\be;X,\vec\rho\Big)\cr
&\hskip.5in
\le\sfrac{1}{4\al} N_0\big(\sfrac{d\hfill}{ds}\Om_{C_0(\de e+s\de e')}(\cV)
                        \cl 2\be;X,\vec\rho\big)
+\sfrac{1}{2\al} N_0\big(\Om_{C_0(\de e)}(\cV)\cl 2\be;X,\vec\rho\big)
                          \tn \de\hat e'\tn_{1,\infty}
\cr
&\hskip.5in\le \const\sfrac{\veps}{\al^3}\fe_0(X)\| \de\hat e'\|_{1,\infty}
+\sfrac{\veps}{2\al}\fe_0(X)\| \de\hat e'\|_{1,\infty}\cr
&\hskip.5in\le \sfrac{\veps\ib}{\be}\fe_0(X)\| \de\hat e'\|_{1,\infty}\cr
}$$
as above.
\endproof

\remark{\STM\remOSthmV}{
\Item{i)} 
Let 
$$
\cV(\psi) = \smsum_{n\ {\rm even}} \ 
\int_{\cB^{n}}\hskip-5pt {\sst d\xi_1\cdots d\xi_n}\ 
V_{n}({\sst\xi_1,\cdots,\xi_n})\,\psi{\sst(\xi_1)}\cdots{\sst\psi(\xi_n)}
$$
 as in Theorem \thmOSfirststep. Since $\cb_0^2\le\abcst\,\cb_0$, if
$$
\smsum (32\be)^n \rho_{0;n}\| V_{n}\|_{1,\infty}
\le \sfrac{\veps}{\abcst}\cb_0
$$
then the hypothesis $\,N_0(\cV \cl 32\be;X,\vec\rho) \le \veps\fe_0(X)\,$ is satisfied.
\Item{ii)} 
Observe that $\Om_{C_0}(\cV)$ does not depend on $\phi$, while
$\tilde\Om_{C_0}(\cV)$ does.
\Item{iii)}
In the applications we have in mind, there is a small constant $\la>0$ (the
coupling constant) and a small number $\upsilon>0$ such that
$$
\rho_{m;n} = \cases{ \sfrac{1}{\la^{(1-\upsilon)(m+n-2)/2}} & if $m+n\ge 4$ \cr
                 \sfrac{1}{\la^{(1-\upsilon)}} & if $m+n=2$ \cr
}$$
Then the hypotheses on $\rho_{m;n}$ in the Theorem are fulfilled.
}

\remark{\STM\remOSneedsectors}{
The norms of Theorem \thmOSfirststep\ are too coarse for a multi scale 
analysis of many fermion systems. See [FKTf1, \S\CHintroOverview, subsection 7]. In the notation of this 
paper, this may be seen as follows. For simplicity, set $r=r_0=0$. Let
$$
C^{(j)}(k) = \sfrac{\nu^{(j)}(k)}{\imath k_0 -e(\k)}
$$
The condition
$$
\rho_{0;n+n'-2} \le \rho_{0;n}\,\rho_{0;n'}
$$
with $n=n'=2$ implies that $\rho_{0;2} \ge 1$ and hence $\rho_{0;n} \ge 1$ 
for all $n\ge 2$. From Proposition \propIntBndsII.i one deduces that
$\sfrac{\const}{M^{j/2}}$ is an integral bound for $C^{(j)}$. A direct application
of Proposition \propOSpropbnd.i\ and Lemma \lemOSscalednorm\ gives the poor 
estimate $M^{dj}\,t^\0+\smsum\limits_{|\de|>0}\infty\,t^\de$ for a contraction bound for $C^{(j)}$. A more careful
argument in which one 
\item{$\bullet$}
decomposes $C^{(j)}=\smsum_{s\in\Si}C^{(j)}_s$ into
$M^{(d-1)j/2}$ terms each having the projection of $\k$ onto the Fermi
surface restricted to a roughly rectangular region of side $M^{-j/2}$ 
(see Definition \defOSsectors\ of [FKTo3] for the precise construction)
\item{$\bullet$} applies Proposition \propOSpropbnd.i\ to obtain
$\tn C^{(j)}_s\tn_{1,\infty}\le\const M^j$ for all $s\in\Si$

\noindent
yields $\const\,M^{{d+1 \over 2}j}\,t^\0+\smsum\limits_{|\de|>0}\infty\,t^\de$ as a realistic contraction bound. 
Thus, for an even Grassmann function
$$
\cW(\psi) = \smsum_{n=0 }^\infty 
\int {\sst d\xi_1\cdots d\xi_{2n}}\, W_{2n}({\sst \xi_1,\cdots,\xi_{2n}})\,
\psi({\sst \xi_1}) \cdots \psi({\sst \xi_{2n}})
$$
the norm $N(\cW;\cb,\ib,\al)$ of Definition \defOSgrnorm\ has 
$$\eqalign{
N(\cW;\cb,\ib,\al)_\0
& = \const M^{{d+3 \over 2}j}\Big\{ \sfrac{\al^2}{M^j}\,\rho_{0;2}\, 
\tn W_2\tn_{1,\infty}\ 
+\ \sfrac{\al^4}{M^{2j}}\,\rho_{0;4}\,\tn W_4\tn_{1,\infty} \cr 
& \hskip 5.5cm+ \, \smsum_{n\ge 3} \big(\const\sfrac{\al^2\,}{M^j}\big)^n
\rho_{0;2n}\,\tn W_{2n}\tn_{1,\infty} \Big\}
}$$
In particular, for $N(\cW;\cb,\ib,\al)_\0$ to be order one, it is necessary that
$\tn W_4\tn_{1,\infty} $ is of order $\sfrac{1}{M^{{d-1 \over 2}j}}$. For $d\ge2$,
this is not even the case for the original interaction $\cV$, with all
momenta restricted to the $j^{\rm th}$ shell.

}

\vfill\eject

\chap{ The Fourier transform}\PG\pgOSIX

Theorem \thmOSfirststep\ and its higher scale analog, Theorem 
\:\thOSrengroupestimate, shall be used in a renormalization group
flow to obtain estimates on the suprema of connected Green's functions in position space. We wish to also obtain estimates on the suprema of 
certain connected amputated Green's functions in momentum space. In section 
\CHmomentum, we introduce norms tailored to that purpose and prove an 
analog of Theorem \thmOSfirststep. In this Section, we set up notation and 
concepts for the passage between position space and momentum space that will 
be needed to do that.

In Remark \:\remOSneedsectors, we pointed out that estimates based on
integral and contraction bounds of
the propagator $\sfrac{\nu^{(\ge j)}}{\imath k_0 - e(\k)}$ are worse than
those one expects by naive power counting in momentum space. The reason is that
conservation of momentum is not exploited effectively. In the case $d=2$, 
this difficulty can be completely
overcome by introducing sectors, see [FMRT], [FKTr2 \S X], [FKTf1 \S \CHintroOverview, subsection 8] and Section \CHsectors. 
They localize momenta into small pieces of the shells introduced 
in Definition \:\defOSscales. The discussion of this section is also useful for 
that purpose.

To systematically deal with Fourier transforms, we call
$$
\check \cB= \bbbr \times \bbbr^d \times \{\uparrow, \downarrow\}\times\{0,1\}
$$
``momentum space''. For 
$ \check\xi = (k,\si',a') = (k_0,\k,\si',a') \in \check\cB$ and
$\xi = (x,a) = (x_0,\x,\si,a) \in \cB$ we define the inner product
$$
\<\check\xi,\xi\>\ =\ \de_{\si',\si} \de_{a',a}\,(-1)^a\,\<k,x\>_-\  
=\ \de_{\si',\si} \de_{a',a}\,(-1)^a\, \big(- k_0 x_0+\k_1\x_1+\cdots+\k_d\x_d \big)
$$
 ``characters''
$$\deqalign{
E_+(\check\xi,\xi)\ &=\ \de_{\si',\si} \de_{a',a}e^{\imath\<\check\xi,\xi\>}
&=\ \de_{\si',\si}\de_{a',a}e^{\imath(-1)^a
                          \big(-k_0 x_0+\k_1\x_1+\cdots+\k_d\x_d \big)}\cr
E_-(\check\xi,\xi)\ &=\ \de_{\si',\si} \de_{a',a}e^{-\imath\<\check\xi,\xi\>}
&=\ \de_{\si',\si}\de_{a',a}e^{-\imath(-1)^a
                            \big(-k_0 x_0+\k_1\x_1+\cdots+\k_d\x_d \big)}\cr
}$$
and integrals
$$
\int d\xi\ \bullet\ = \sum_{a\in\{0,1\}\atop\si\in\{\uparrow,\downarrow\}}
\int_{\bbbr\times\bbbr^d} dx_0\, d^d\x\ \bullet\ \qquad\qquad
\int d\check\xi\ \bullet\ = \sum_{a\in\{0,1\}\atop\si\in\{\uparrow,\downarrow\}}
\int_{\bbbr\times\bbbr^d} dk_0\, d^d\k\ \bullet\ 
$$
For $\check\xi = (k,\si,a), \,\check\xi' = (k',\si',a') \in \check \cB$ we set
$$
\check\xi+\check\xi' = (-1)^a\,k + (-1)^{a'}\,k'\ \in\bbbr \times \bbbr^d
$$

\definition{\STM\defOSfourtrans (Fourier transforms)}{
Let $f({\sst \eta_1,\cdots,\eta_m;\,\xi_1,\cdots,\xi_n})$
be a translation invariant function on $\cB^m \times \cB^n$.
\Item{i)}
The total Fourier transform $\check f$ of $f$ is defined by 
$$\eqalign{
&\check f({\sst
\check\eta_1,\cdots,\check\eta_m;\,\check\xi_1,\cdots,\check\xi_n }) 
\ (2\pi)^{d+1}\de({\sst \check\eta_1+\cdots+\check\eta_m+\check\xi_1+\cdots+\check\xi_n}) \cr
&\hskip2in= \int \smprod_{i=1}^m E_+(\check\eta_i,\eta_i)\,d\et_i\,
\smprod_{j=1}^n E_+(\check\xi_j,\xi_j)\,d\xi_j\,
f({\sst \eta_1,\cdots,\eta_m;\,\xi_1,\cdots,\xi_n }) 
\cr 
}$$ 
or, equivalently, by 
$$\eqalign{
&f({\sst \eta_1,\cdots,\eta_m;\,\xi_1,\cdots,\xi_n }) \cr
&= \int \smprod_{i=1}^m
\sfrac{ E_-(\check\eta_i,\eta_i)\,d\check\et_i}{(2\pi)^{d+1}}\,
\smprod_{j=1}^n \sfrac{E_-(\check\xi_j,\xi_j)\,d\check\xi_j }{(2\pi)^{d+1}}\,
\check f({\sst
\check\eta_1,\cdots,\check\eta_m;\,\check\xi_1,\cdots,\check\xi_n }) 
\ (2\pi)^{d+1}\de({\sst \check\eta_1+\cdots+\check\eta_m+\check\xi_1+\cdots+\check\xi_n}) \cr 
}$$ 
$\check f$ is defined on the set
$ \set{ ({\sst
\check\eta_1,\cdots,\check\eta_m;\,\check\xi_1,\cdots,\check\xi_n }) \in
\check \cB^m \times \cB^n }
{ {\sst \check\eta_1+\cdots+\check\eta_m+\check\xi_1+\cdots+\check\xi_n=0}}$.

If $m=0$, $n=2$ and $f({\sst \xi_1,\xi_2 })$ conserves particle number and is spin independent and antisymmetric, we define $\check f(k)$ by
$$
\check f({\sst (k,\si,1),(k,\si',0)}) =\de_{\si,\si'}\check f(k)
$$ 
or equivalently by 
$$
\check f(k)=\int dy\ e^{\imath<k,y>_-}f((0,\si,1),(y,\si,0))
$$
\Item{ii)}
If $n\ge 1$, the partial Fourier transform $f^\sim$ is defined by
$$
f^\sim({\sst\check\eta_1,\cdots,\check\eta_m;\,\xi_1,\cdots,\xi_n }) 
= \int \Big(\smprod_{i=1}^m E_+(\check\eta_i,\eta_i)\,
d\eta_i\Big)\  f({\sst \eta_1,\cdots,\eta_m;\,\xi_1,\cdots,\xi_n })
$$
or, equivalently, by
$$
f({\sst \eta_1,\cdots,\eta_m;\,\xi_1,\cdots,\xi_n })
= \int \Big(\smprod_{i=1}^m E_-(\check\eta_i,\eta_i)\,
\sfrac{d\check\eta_i}{(2\pi)^{d+1}}\Big)\, f^\sim({\sst
\check\eta_1,\cdots,\check\eta_m;\,\xi_1,\cdots,\xi_n }) 
$$ 
If $n=0$, we set $f^\sim =\check f$. 
}

\remark{\STM\remOStransinv}{
Translation invariance of $f$ implies that for all 
$t\in\bbbr \times \bbbr^d$
$$
f^\sim({\sst \check\eta_1,\cdots,\check\eta_m;\,\xi_1+t,\cdots,\xi_n+t })
= e^{\imath \<\check\eta_1+\cdots+\check\eta_m,t\>_-}\, 
f^\sim({\sst \check\eta_1,\cdots,\check\eta_m;\,\xi_1,\cdots,\xi_n })
$$
This is what we mean when we say that ``$f^\sim({\sst \check\eta_1,\cdots,\check\eta_m;\,\xi_1,\cdots,\xi_n })$ is translation invariant''.

}

There will be two different situations in which we wish to associate
a two--point function $B(\xi,\xi')$ in position/spin/particle--hole
space, $\cB$, to a function $B(k)$ in momentum space that has no spin/particle--hole dependence. In the first case, treated in Definition
\:\defOSftcov\ below, $B(\xi,\xi')$ is a propagator and so is spin--independent
and particle number conserving, so that $B(\xi,\xi')$ vanishes unless 
one of $\xi,\xi'$ is particle and the other is hole. In the second case,
treated in Definition \:\defOSfourtransII\ below, convolution with 
$B(\xi,\xi')$ corresponds to pure multiplication by $B(k)$ in momentum space.
This is used to, for example, introduce partitions of unity in momentum
space. In this case $B(\xi,\xi')$ is diagonal in the particle/hole
indices.

\definition{\STM\defOSftcov (Fourier transforms of covariances)}{
If $C(k)$ is a function on $\bbbr \times \bbbr^d$, we say that the 
{\bf covariance} $C(\xi,\xi')$ on $\cB \times \cB$, defined by
$$
C\big((x_0,\x,\si,a),(x'_0,\x',\si',a') \big)
=\cases{ \de_{\si,\si'} \int\sfrac{d^{d+1}k}{(2\pi)^{d+1}}\, e^{\imath<k,x-x'>_-}C(k)
& if $a=0, \,a'=1$ \cr
\noalign{\vskip.05in}
0 & if $a=a'$\cr
\noalign{\vskip.05in}
-C\big((x',\si',a'),(x,\si,a) \big)  & if $a=1,\,a'=0$ \cr
}$$
is the Fourier transform of $C(k)$. 

}

\noindent
As in part (ii) of Proposition \:\propIntBndsII, we use the notation

\definition{\STM\defOSfourtransII}{
If $\chi(k)$ is a function on $\bbbr\times \bbbr^d$, we define the Fourier
transform $\hat\chi$ by 
$$
\hat\chi(\xi,\xi') 
= \de_{\si,\si'} \de_{a,a'}\int e^{(-1)^a\imath<k,x-x'>_-}\,\chi(k)\,
\sfrac{d^{d+1}k}{(2\pi)^{d+1}} 
$$
for $\xi=(x,a)=(x_0,\x,\si,a),\,\xi'=(x',a') =(x'_0,\x',\si',a')\in \cB$.

}

\lemma{\STM\lemOSjhat}{
\Item i) 
Let $C(k)$ be a function on $\bbbr\times\bbbr^d$ and $C(\xi,\xi')$ the 
associated covariance in the sense of Definition \defOSftcov. Then
$$
CJ =\hat C\qquad
(JCJ)^{\check{}}(k)=C(k)
$$
where $J$ was defined in (\eqnOSjdef), $\hat C$ was defined in Definition 
\defOSfourtransII\ and  $\check f(k)$ was defined in Definition \defOSfourtrans.i.
\Item ii) 
Let $\chi(k)$ and $\chi'(k)$ be functions on $\bbbr\times\bbbr^d$. Then
$$
\int d\xi''\ \hat\chi(\xi,\xi'')\hat\chi'(\xi'',\xi')
=\widehat{\chi\chi'}(\xi,\xi')
$$
\Item iii) 
Let $\chi(k)$ be a function on $\bbbr\times\bbbr^d$. Then
$$\eqalign{
(J\hat\chi)^{\check{}}(k)&=\chi(k)\cr
(J\hat\chi J)(\xi,\xi')&=-\hat \chi(\xi',\xi)\cr
}$$

}
\prf  The proof of this Lemma consists of a number of three or four line
 computations.
\endproof

In a renormalization group analysis we will adjust the counterterms in such a way that, at each scale, the Fourier transform of the two point function is small on $\{\,k=(k_0,\k) \in \bbbr\times\bbbr^d\ \big|\ k_0=0\,\}$. Then the absolute value of the Fourier transform of the two point function at a point $(k_0,\k)$ can be estimated in terms of $|k_0|$ and the $k_0$ derivative of the Fourier transform of the two point function. The following Lemma is used to make an analogous estimate in position space.

For a function $f(x)$ on $\bbbr\times\bbbr^d$, we define
$$
\|f\|_{L^1} = \smsum_{\de \in \bbbn_0\times\bbbn_0^d} \sfrac{1}{\de!}
\bigg[ \int |x^\de f(x)| d^{d+1}x \bigg]\,t^\de\ \in\ \fN_{d+1}
$$

\lemma{\STM\lemOSprepintup}{
Let $u(\xi,\xi')$ be a translation invariant function on $\cB^2$ that satisfies
$\check u ( {\sst ((0,\k),\si,a), ((0,\k'),\si',a')})=0$. Furthermore let
$\chi(k)$ be a function on $\bbbr\times\bbbr^d$. For $x \in \bbbr\times\bbbr^d$ set
$$
\chi'(x) =\int e^{ \imath<k,x>_- }\chi(k) \sfrac{d^{d+1}k}{(2\pi)^{d+1}} 
$$
so that for $\xi =(x,\si,a),\ \xi' =(x',\si',a')  \in \cB$
$$
\hat\chi(\xi,\xi') = \de_{\si,\si'}\de_{a,a'} \,\chi'({\sst (-1)^a}(x-x'))
$$
Then
\Item i)
$$
\Big\| \int d\eta\,\hat\chi(\xi,\eta)\,u(\eta,\xi') \Big\|_{1,\infty}
\le \big\| \sfrac{\partial \chi'}{\partial x_0} \big\|_{L^1}\,
  \big\| \cD_{1,2}^{(1,0,\cdots,0)} u \big\|_{1,\infty}
+\smsum_{\de\in\bbbn_0\times\bbbn_0^d\atop \de_0\ne 0}\infty t^\de
$$
\Item ii)
$$\eqalign{
\Big\|\cD_{1,2}^{(1,0,\cdots,0)} 
 \int d\eta\,\hat\chi(\xi,\eta)\,u(\eta,\xi') \Big\|_{1,\infty}
&\le  \abcst\Big( \| \hat \chi\|_{1,\infty} +
 \big\|x_0 \sfrac{\partial \chi'}{\partial x_0} \big\|_{L^1} \Big)
  \big\| \cD_{1,2}^{(1,0,\cdots,0)} u \big\|_{1,\infty}\cr
&\hskip2in+\smsum_{\de\in\bbbn_0\times\bbbn_0^d\atop \de_0>r_0}\infty t^\de
}$$
}

\prf
i) 
Fix  $\xi =(x,\si,a),\ \xi' =(x',\si',a')  \in \cB$. By translation invariance 
$$\eqalign{
\int d\eta\,\hat\chi\big(\xi,\eta)\,u(\eta,\xi'\big)
& = \int dy \,\chi'( {\sst (-1)^a}(x-y))\, u\big( (y,\si,a),(x',\si',a')\big)\cr
& = \int dy \,\chi'( {\sst (-1)^a}(x-y))\,
u\big( (y-x',\si,a),(0,\si',a')\big)\cr
& = \int dy \,\chi'( {\sst (-1)^a}y)\, v(x-x'-y)\cr
}$$
where $\,v(y) = u\big( (y,\si,a),(0,\si',a')\big)\,$. By hypothesis
$$
\int dy_0\,v(y_0,\y)= 0\qquad {\rm for\ all\ } \y \in \bbbr^d
$$
Therefore
$$\eqalign{
\int& d\eta\,\hat\chi\big(\xi,\eta)\,u(\eta,\xi'\big)
 = \int dy \,
\Big( \chi'( {\sst (-1)^a}y) -  \chi'( {\sst (-1)^a}(x_0-x_0',\y)) \Big) \  v(x-x'-y)\cr
& = \int dy \,
\sfrac{ \chi'( {\sst (-1)^a}y) -  \chi'( {\sst (-1)^a}(x_0-x_0',\y))}
{(x_0-x_0')-y_0} \ \big[ (x_0-x_0'-y_0)\,  v(x-x'-y)\big]\cr
& ={\sst  (-1)^{a+1}} \int dy \,\int_0^1 ds\,
\sfrac{ \partial \chi'}{\partial x_0}
\big( {\sst (-1)^a}(sy_0+(1-s)(x_0-x_0'),\y )\big) \cr
& \hskip 7cm
\cD_{1,2}^{(1,0,\cdots,0)} u\big( (x-x'-y,\si,a),(0,\si',a')\big)\cr
& ={\sst  (-1)^{a+1}} \int dy \,\int_0^1 ds\,
\sfrac{ \partial \chi'}{\partial x_0}
\big( {\sst (-1)^a}(sy_0+(1-s)(x_0-x_0'),\y )\big) \
\cD_{1,2}^{(1,0,\cdots,0)} u\big( (x-y,\si,a),\xi'\big)\cr
}$$
Consequently, for fixed $\xi' \in \cB$
$$\eqalign{
\int d\xi\ \Big| \int& d\eta\,\hat\chi\big(\xi,\eta)\,u(\eta,\xi'\big)\, \Big| \cr
&\le \smsum_{\si,a}\int_0^1 ds  \int \ d\x \, d\y  \int dx_0\,dy_0 \ 
\Big| \sfrac{ \partial \chi'}{\partial x_0}
\big( {\sst (-1)^a}(sy_0+(1-s)(x_0-x_0'),\y )\big)  \Big| \cr
&\hskip 6cm
\Big| \cD_{1,2}^{(1,0,\cdots,0)} u\big( (x_0-y_0,\x-\y,\si,a),\xi'\big) \Big|\cr
&=  \smsum_{\si,a}\int_0^1 ds  \int \ d\x \, d\y  \int d\al\,d\be \ 
\Big| \sfrac{ \partial \chi'}{\partial x_0}
\big( {\sst (-1)^a}(\be,\y )\big)  \Big| \cr
&\hskip 6cm
\Big| \cD_{1,2}^{(1,0,\cdots,0)} u\big( (\al,\x-\y,\si,a),\xi'\big) \Big|\cr
&\le  \big\| \sfrac{\partial \chi'}{\partial x_0} \big\|_{L^1}\,
  \big\| \cD_{1,2}^{(1,0,\cdots,0)} u \big\|_{1,\infty}
}\EQN\eqnOSlemsixtzero$$
Here we used, for each fixed $s$, the change of variables 
$\,\al=x_0-y_0,\ \be=sy_0+(1-s)x_0\,$. The integral
$\,\int d\xi' \Big| \int d\eta\,\hat\chi\big(\xi,\eta)\,u(\eta,\xi'\big)\, \Big|\,$, for fixed $\xi \in \cB$, is treated similarly.

By Leibniz's rule (Lemma \lemOSleibniz) and (\eqnOSlemsixtzero),
$$\eqalign{
&\smsum_{\de_0= 0}
\sfrac{t^\de}{\de!}
\int d\xi\ \Big|\cD_{1,2}^\de \int d\eta\,\hat\chi\big(\xi,\eta)\,u(\eta,\xi'\big)\, \Big| \cr
&\hskip1in\le \smsum_{\de_0= 0}
\smsum_{\al,\be\in\bbbn_0\times\bbbn_0^d\atop\al+\be=\de}
\smchoose{\de}{\al,\be}\sfrac{t^\de}{\de!}
\int d\xi\ \Big| \int d\eta\,\big(\cD_{1,2}^\al\hat\chi\big)\big(\xi,\eta)
\,\big(\cD_{1,2}^\be u\big)(\eta,\xi'\big)\, \Big| \cr
&\hskip1in\le \smsum_{\al_0= 0}\smsum_{\be_0= 0}
\sfrac{t^\al}{\al!}
\sfrac{t^\be}{\be!}\Big(
\big\| \sfrac{\partial \hfill}{\partial x_0} x^\al\chi'\big\|_{L^1}\,
  \big\| \cD_{1,2}^{(1,0,\cdots,0)}\cD_{1,2}^\be u \big\|_{1,\infty} 
\big|_{t=0}\Big)\cr
&\hskip1in= \smsum_{\al_0= 0}\smsum_{\be_0= 0}
\sfrac{t^\al}{\al!}
\sfrac{t^\be}{\be!}\Big(
\big\| x^\al\sfrac{\partial \hfill}{\partial x_0} \chi'\big\|_{L^1}\,
  \big\| \cD_{1,2}^\be\cD_{1,2}^{(1,0,\cdots,0)} u \big\|_{1,\infty} 
\big|_{t=0}\Big)\cr
&\hskip1in\le  \big\| \sfrac{\partial \chi'}{\partial x_0} \big\|_{L^1}\,
  \big\| \cD_{1,2}^{(1,0,\cdots,0)} u \big\|_{1,\infty}
}$$

\Item ii)
By Leibniz's rule (Lemma \lemOSleibniz), part (i) of this Lemma (applied to
$\sfrac{\partial \chi}{\partial k_0}$) and Lemma \lemOSelloneinfty
$$\eqalign{
&\sum_{\de\in\bbbn_0\times\bbbn_0^d}\sfrac{t^\de}{\de!}
\TTN\cD_{1,2}^\de\cD_{1,2}^{(1,0,\cdots,0)}
 \int d\eta\,\hat\chi(\xi,\eta)\,u(\eta,\xi') \TTN_{1,\infty}\cr
&\hskip0.2in\le 
\sum_{\de\in\bbbn_0\times\bbbn_0^d}
\sum_{\al,\be\in\bbbn_0\times\bbbn_0^d\atop\al+\be=\de+(1,0,\cdots,0)}
\smchoose{\de+(1,0,\cdots,0)}{\al,\be}\sfrac{t^\de}{\de!}
\TTN \int d\eta\ \cD_{1,2}^\al\hat\chi(\xi,\eta)\,
\cD_{1,2}^\de\cD_{1,2}^\be u(\eta,\xi') \TTN_{1,\infty}\cr
&\hskip0.2in\le 
\sum_{\de\in\bbbn_0\times\bbbn_0^d}
\sum_{\al',\be\in\bbbn_0\times\bbbn_0^d\atop{\al'+\be=\de,\ \be_0=0}}
\smchoose{\de+(1,0,\cdots,0)}{\al'+(1,0,\cdots,0),\be}\sfrac{t^\de}{\de!}
\TTN \int d\eta\ \cD_{1,2}^{\al'}\cD_{1,2}^{(1,0,\cdots,0)}\hat\chi(\xi,\eta)\,
\cD_{1,2}^\be u(\eta,\xi') \TTN_{1,\infty}\cr
&\hskip0.3in+
\sum_{\de\in\bbbn_0\times\bbbn_0^d}
\sum_{\al,\be'\in\bbbn_0\times\bbbn_0^d\atop\al+\be'=\de}
\smchoose{\de+(1,0,\cdots,0)}{\al,\be'+(1,0,\cdots,0)}\sfrac{t^\de}{\de!}
\TTN \int d\eta\ \cD_{1,2}^\al\hat\chi(\xi,\eta)\,
\cD_{1,2}^{\be'}\cD_{1,2}^{(1,0,\cdots,0)} u(\eta,\xi') \TTN_{1,\infty}\cr
&\hskip0.2in\le 
\sum_{\al',\be\in\bbbn_0\times\bbbn_0^d\atop{\be_0=0}}(\al'_0+1)
\sfrac{t^{\al'}}{\al'!}\sfrac{t^\be}{\be!}
\TTN \int d\eta\ \cD_{1,2}^{\al'}\cD_{1,2}^{(1,0,\cdots,0)}\hat\chi(\xi,\eta)\,
\cD_{1,2}^\be u(\eta,\xi') \TTN_{1,\infty}\cr
&\hskip0.3in+
\sum_{\al,\be'\in\bbbn_0\times\bbbn_0^d}(\al_0+\be'_0+1)
\sfrac{t^\al}{\al!}\sfrac{t^{\be'}}{\be'!}
\TTN \int d\eta\ \cD_{1,2}^\al\hat\chi(\xi,\eta)\,
\cD_{1,2}^{\be'}\cD_{1,2}^{(1,0,\cdots,0)} u(\eta,\xi') \TTN_{1,\infty}\cr
&\hskip0.2in\le 
\sum_{\al',\be\in\bbbn_0\times\bbbn_0^d\atop{\be_0=0}}(\al'_0+1)
\sfrac{t^{\al'}}{\al'!}\sfrac{t^\be}{\be!}
\big\| \sfrac{\partial \hfill}{\partial x_0} x^{\al'+(1,0,\cdots,0)}\chi'\big\|_{L^1}\big|_{t=0}\,
\TN \cD_{1,2}^{(1,0,\cdots,0)}\cD_{1,2}^\be u \TN_{1,\infty}\cr
&\hskip0.3in+
\sum_{\al,\be'\in\bbbn_0\times\bbbn_0^d}(\al_0+\be'_0+1)
\sfrac{t^\al}{\al!}\sfrac{t^{\be'}}{\be'!}
\TN \cD_{1,2}^\al\hat\chi\TN_{1,\infty}\,
\TN\cD_{1,2}^{\be'}\cD_{1,2}^{(1,0,\cdots,0)}u \TN_{1,\infty}\cr
&\hskip0.2in\le (r_0+1)\Big( (r_0+1)\| \hat \chi\|_{1,\infty} +
 \big\|x_0 \sfrac{\partial \chi'}{\partial x_0} \big\|_{L^1}
   +\| \hat \chi\|_{1,\infty} \Big)
  \big\| \cD_{1,2}^{(1,0,\cdots,0)} u \big\|_{1,\infty}
+\smsum_{\de\in\bbbn_0\times\bbbn_0^d\atop \de_0>r_0}\infty t^\de
}$$
since
$
\sfrac{ \partial\ }{\partial x_0}\big(x^{\al'+(1,0,\cdots,0)}\,\chi'(x)\big)
=(\al'_0+1)x^{\al'}\,\chi'(x)
+x^{\al'}x_0\sfrac{ \partial\ }{\partial x_0}\chi'(x)
$.
\endproof

\vfill\eject

\chap{ Momentum Space Norms}\PG\pgOSX

In this section, we introduce momentum space norms designed to 
control amputated Green's functions in
momentum space. The set of momentum conserving $m$--tuples of momenta is
$$
\check \cB_m = \set{(\check \eta_1,\cdots,\check \eta_m)\in \check \cB^m}
{\check \eta_1+\cdots+\check \eta_m=0}
$$
(we use the addition introduced before Definition \defOSfourtrans).
We are particularly interested in the two and four point functions. In the
renormalization group analysis we shall control the external fields in momentum
space, while the fields that are going to be integrated out are still treated in
position space. That is, we will estimate partial Fourier
transforms of functions on
$\cB^m\times \cB^n$ as in Definition \defOSfourtrans.ii. Motivated by 
Remark \remOStransinv\ we define

\definition{\STM\defOSamptransinv}{
A function $f$ on $\check \cB^m \times \cB^n$ is called translation
invariant, if for all $t\in\bbbr \times \bbbr^d$
$$
f({\sst \check\eta_1,\cdots,\check\eta_m;\,\xi_1+t,\cdots,\xi_n+t })
= e^{\imath \<\check\eta_1+\cdots+\check\eta_m,t\>_-}\, 
f({\sst \check\eta_1,\cdots,\check\eta_m;\,\xi_1,\cdots,\xi_n })
$$

}

\noindent
Generalizing Definition \defOSmultideriv\ we set

\definition{\STM\defOSdiffdecay (Differential--decay operators)}{
Let $m,n\ge 0$. If $n\ge 1$, 
let $f$ be a function on $\check \cB^m \times \cB^n$. If $n=0$, let
$f$ be a function on $\check \cB_m $.
\Item{i)} 
For $1\le j \le m$ and a multiindex $\de$ set
$$\eqalign{
&\rD^\de_j
f\,({\sst(p_1,\tau_1,b_1),\cdots,(p_m,\tau_m,b_m);\,\xi_1,\cdots,\xi_n}) \cr
& \hskip 1.5cm= 
\big[\imath(-1)^{b_j}\big]^{\de_0}
\smprod_{\ell=1}^d\big[-\imath(-1)^{b_j}\big]^{\de_\ell}\ 
\sfrac{\partial^{\de_0}\hfill}{\partial p_{j,0}^{\de_0}}\,
\sfrac{\partial^{\de_1}\hfill}{\partial \p_{j,1}^{\de_1}} \cdots
\sfrac{\partial^{\de_d}\hfill}{\partial \p_{j,d}^{\de_d}}\,
f({\sst(p_1,\tau_1,b_1),\cdots,(p_m,\tau_m,b_m);\,\xi_1,\cdots,\xi_n})
}$$

\Item{ii)}
Let $1\le i\ne j\le m+n$ and $\de$ a multiindex. Set
$$\deqalign{
\rD_{i;j}^\de f & = (\rD_i-\rD_j)^\de\,f  
           &{\rm if}\ 1\le i <j \le m \cr
\rD_{i;j}^\de f & = (\rD_i-\xi_{j-m})^\de\,f   
           &{\rm if}\ 1\le i  \le m,\ m+1\le j \le m+n \cr
\rD_{i;j}^\de f & = (\xi_i-\rD_{j-m})^\de\,f   
           &{\rm if}\ m+1\le i  \le m+n,\ 1\le j \le m \cr
\rD_{i;j}^\de f & = (\xi_{i-m}-\xi_{j-m})^\de\,f \ = \cD_{i-m,j-m}^\de f  \hskip 1cm 
           &{\rm if}\  m+1\le i<j \le m+n \cr
}$$

\Item{iii)}
A differential--decay operator (dd--operator) of type $(m,n)$, with 
$m+n\ge 2$, is an operator $\rD$
of the form
$$
\rD = \rD^{\de^{(1)}}_{i_1;j_1}\, \cdots  \rD^{\de^{(r)}}_{i_r;j_r}
$$
with $1\le i_\ell\ne j_\ell\le m+n$ for all $1\le\ell\le r$. 
A dd--operator of type $(1,0)$
is an operator of the form
$\ 
\rD = \rD^{\de^{(1)}}_{1}\, \cdots  \rD^{\de^{(r)}}_{1}
= \rD^{\de^{(1)}+\cdots+\de^{(r)}}_{1}
.$
The total order of $\rD$ is $\de(\rD) =\de^{(1)}+\cdots+\de^{(r)}$.

}

\remark{\STM\remOSdiffdecay}{
\Item{i)}
Let $\rD$ be a differential--decay operator. If $f$ is a translation invariant
function on $\check\cB_m \times \cB^n$, then $\rD\,f$ is again translation 
invariant.

\Item{ii)} For a translation invariant function $\varphi$ on $\cB^m \times \cB^n$
$$
\rD_{i;j}(\varphi^\sim) = (\cD_{i,j}\varphi)^\sim
$$
In particular, Leibniz's rule also applies for differential--decay operators.

\Item{iii)} Let $f$ be a translation invariant function on 
$\check\cB^m \times \cB$. Then, for $\xi =(x_0,\x,\si,a) \in \cB$,
$$
f(\check\eta_1,\cdots,\check\eta_m;\xi)  = 
 e^{\imath \<\check\eta_1+\cdots+\check\eta_m,(x_0,\x)\>_-}\,
    f(\check\eta_1,\cdots,\check\eta_m;(0,\si,a))  
$$
Consequently, for $1\le i\le m$ and a multiindex $\de$
$$
\rD_{i;m+1}^\de f({\sst \check\eta_1,\cdots,\check\eta_m;\xi})
= e^{\imath \<\check\eta_1+\cdots+\check\eta_m,(x_0,\x)\>_-}\,
    \rD_i^\de f(\check\eta_1,\cdots,\check\eta_m;(0,\si,a))  
$$

}

\definition{\STM\defOSdiffdecaynorm}{
For  a function $f$ on $\check\cB_m$, set
$$
\| f\tnorm 
= \smsum\limits_{\de \in \bbbn_0\times\bbbn_0^2} \sfrac{1}{\de !}\ 
\max\limits_{\rD\ {\rm dd-operator} 
\atop{\rm with\ } \de(\rD)=\de}\
\sup\limits_{\check\eta_1,\cdots,\check\eta_m \in \check \cB}
\big| \rD f\,
({\sst \check\eta_1,\cdots,\check\eta_m}) \big|\ t^\de
$$
Let $f$ be a function on $\check\cB^m\times \cB^n$ with $n\ge 1$. Set
$$
\| f\tnorm 
= \smsum_{\de\in \bbbn_0\times\bbbn_0^2} \sfrac{1}{\de!}
\max_{\rD\, {\rm dd-operator} \atop{\rm with\ } \de(\rD) =\de}\
\sup_{\check\eta_1,\cdots,\check\eta_m \in \check \cB}
\TN \rD\, f\,({\sst\check\eta_1,\cdots,\check\eta_m; \xi_1,\cdots,\xi_n})
\TN_{1,\infty}
\ t^\de
$$
when $m\le p \le m+n\,$. The norm
$\tn\,\cdot\,\tn_{1,\infty}$ of Example \exOSSymmNorm\ refers to the variables 
${\sst \xi_1,\cdots,\xi_n}$. That is,
$$
\TN \rD\, f\,({\sst\check\eta_1,\cdots,\check\eta_m; \xi_1,\cdots,\xi_n})
\TN_{1,\infty}
=\max\limits_{1\le j_0 \le n}\ 
\sup\limits_{\xi_{j_0} \in \cB}\  
\int \prod\limits_{j=1,\cdots, n \atop j\ne j_0} d\xi_j\, 
| \rD\, f\,({\sst\check\eta_1,\cdots,\check\eta_m; \xi_1,\cdots,\xi_n}) |
$$
}

\remark{\STM\remOSdiffdecaynorm}{
In the case $m=0$ the norm $\| \,\cdot\,\|_{1,\infty}$ of Example
\exOSSymmNorm\ and the norm $\| \,\cdot\,\tnorm$ of Definition 
\defOSdiffdecaynorm\ agree.
}

\noindent In analogy to Lemma \lemOSelloneinfty, we have

\lemma{\STM\lemOSelloneinftyamp}{
Let $f$ be a translation invariant function on $\check\cB^{m}\times
\cB^{n}$, 
$f'$ a translation invariant function on $\check\cB^{m'}\times
\cB^{n'}$ and 
$1\le \mu \le n,\  1\le \nu\le n'$.

\noindent 
If $n\ge 2$ or $n'\ge 2$ define the function 
$g$ on $\check \cB^{m+m'}\times \cB^{n+n'-2}$ by
$$\eqalign{
&g({\sst \check\eta_1,\cdots,\check\eta_{m+m'};
\xi_1,\cdots,\xi_{\mu-1},\,\xi_{\mu+1},\cdots,\xi_n,\,
\xi_{n+1},\cdots,\xi_{n+\nu-1},\,\xi_{n+\nu+1},\cdots,\xi_{n+n'}}) \cr
&\hskip .7cm  =\int_\cB {\sst d\zeta}\, f( {\sst \check\eta_1,\cdots,\check\eta_m;\, 
\xi_1,\cdots,\xi_{\mu-1},\,\zeta,\,\xi_{\mu+1},\cdots,\xi_n })\, 
f'( {\sst \check\eta_{m+1},\cdots,\check\eta_{m+m'};
\,\xi_{n+1},\cdots,\xi_{n+\nu-1},\,\zeta,\,\xi_{n+\nu+1},\cdots,\xi_{n+n'}})
\cr
}$$
If $n=n'=1$, define the function $g$ on $\check \cB_{m+m'}$ by
$$
g({\sst \check\eta_1,\cdots,\check\eta_{m+m'}})\ (2\pi)^{d+1}
 \de({\sst \check\eta_1+\cdots+\check\eta_{m+m'}}) 
= 
\int_\cB {\sst d\zeta}\,
f( {\sst \check\eta_1,\cdots,\check\eta_m;\,\zeta})\
 f'( {\sst\check\eta_{m+1},\cdots,\check\eta_{m+m'};\,\zeta})
$$
Then 
$$
\| g\,\tnorm \le \| f\tnorm \ \| f'\tnorm\ \ \cases{4& if $n=n'=1$\cr
                                                1& otherwise\cr}
$$
}

\prf
If $n \ge 2$ or $n' \ge 2$, the proof is analogous to that of Lemma
\lemOSelloneinfty. Therefore we only discuss the case $n=n'=1$. In this
case, by Remark \remOSdiffdecay.iii
$$\eqalign{
&\int_\cB {\sst d\xi}\,
f( {\sst \check\eta_1,\cdots,\check\eta_m;\,\xi})\,
 f'( {\sst\check\eta_{m+1},\cdots,\check\eta_{m+m'};\,\xi}) \cr
&\ \ =\int \hskip -.1cm{\sst dx_0} \hskip -.1cm
   \int\hskip -.1cm {\sst d\x} \hskip-.2cm 
\smsum_{\si \in \{\uparrow,\downarrow\} \atop b \in \{0,1\}}\,  
 f ({\sst \check\eta_1,\cdots,\check\eta_m;\,(0,\si,b)})\ 
 e^{\imath \<\check\eta_1+\cdots+\check\eta_{m+m'},(x_0,\x)\>_-}
 f' ({\sst \check\eta_{m+1},\cdots,\check\eta_{m+m'};\,(0,\si,b)}) \cr 
&\ \ = \smsum_{\si \in \{\uparrow,\downarrow\} \atop b \in \{0,1\}} \,
 f ({\sst \check\eta_1,\cdots,\check\eta_m;(0,\si,b)})\
f'({\sst \check\eta_{m+1},\cdots,\check\eta_{m+m'};(0,\si,b)}) \  
(2\pi)^{d+1}\de( {\sst\check\eta_1+\cdots+\check\eta_{m+m'}})
}$$
Consequently
$$
g({\sst \check\eta_1,\cdots,\check\eta_{m+m'}}) = 
\smsum_{\si \in \{\uparrow,\downarrow\} \atop b \in \{0,1\}}\,
f ({\sst \check\eta_1,\cdots,\check\eta_m;(0,\si,b)})\
f' ({\sst \check\eta_{m+1},\cdots,\check\eta_{m+m'};\,(0,\si,b)})
$$
The claim now follows by iterated application of the product rule 
for derivatives and Remark \remOSdiffdecay.iii. The factor of 4 comes from
the sum over $\si$ and $b$ and is required only when $n=n'=1$.

\endproof

\remark{\STM\remOSelloneinftyamp}{
Let
$$\eqalign{
F( \check\eta_1,\cdots,\check\eta_m)
= \smsum\limits_{\de \in \bbbn_0\times\bbbn_0^2} \sfrac{1}{\de !}
\max\limits_{\rD\ {\rm dd-operator} 
\atop{\rm with\ } \de(\rD)=\de}
\TN \rD f( \check\eta_1,\cdots,\check\eta_m;\,\cdot\,,\dots,\,\cdot\,) \TN_{1,\infty} t^\de
}$$
so that
$$
\| f\tnorm 
= \sup_{\check\eta_1,\cdots,\check\eta_m \in \check \cB} 
F(\check\eta_1,\cdots,\check\eta_m)
$$
(with the supremum of the formal power series $F$ taken componentwise)
and define $G(\check\eta_1,\cdots,\check\eta_{m+m'})$ and 
$F'( \check\eta_{m+1},\cdots,\check\eta_{m+m'})$ similarly.
The proof
of Lemma \lemOSelloneinftyamp\ actually shows that
$$
G(\check\eta_1,\cdots,\check\eta_{m+m'})
\le F( \check\eta_1,\cdots,\check\eta_m)
F'( \check\eta_{m+1},\cdots,\check\eta_{m+m'})
\ \ \cases{4& if $n=n'=1$\cr
                                                1& otherwise\cr}
$$
for all $(\check\eta_1,\cdots,\check\eta_{m+m'})\in\check\cB^{m+m'}$.
}

\definition{\STM\defOScheckcF }{

\Item{i)}  
For $n\ge 1$, denote by  $\check\cF_m(n)$ the space
of all translation invariant, complex valued functions 
$\ 
f({\sst\check\eta_1,\cdots,\check\eta_m;\,\xi_1,\cdots,\xi_n} )
\ $
on $\check\cB^m \times   \cB^n$ that are antisymmetric in their 
external ($=\check\eta$) variables. Let  $\check\cF_m(0)$ be the space
of all antisymmetric, complex valued functions 
$\ 
f({\sst\check\eta_1,\cdots,\check\eta_m} )
\ $
on $\check\cB_m $.

\Item{ii)}  
Let $C(\xi,\xi')$ be any skew symmetric function on $\cB^2$.
Let $f\in \check\cF_m(n)$ and $1\le i < j\le n$. We define ``contraction'',
for $n\ge 2$, by
$$\eqalign{
&\Cont{i}{j}{C} f\, ({\sst \check\eta_1,\cdots,\check\eta_m;
\xi_1,\cdots,\xi_{i-1},\xi_{i+1},\cdots,\xi_{j-1},\xi_{j+1},\cdots,\xi_n})
= (-1)^{j-i+1} \!\!\int\!\! {\sst d\xi_i d\xi_j}\  C({\sst \xi_i,\xi_j})\,
f({\sst \check\eta_1,\cdots,\check\eta_m;\xi_1,\cdots,\xi_n}) 
}$$  
and, for $n=2$, by
$$\eqalign{
&\Cont{1}{2}{C} f\, ({\sst \check\eta_1,\cdots,\check\eta_m})\ 
(2\pi)^{d+1} \de({\sst \check\eta_1+\cdots+\check\eta_m})
=  \int {\sst d\xi_1\,d\xi_2}\  C({\sst \xi_1,\xi_2})\,
f({\sst \check\eta_1,\cdots,\check\eta_m;\xi_1,\xi_2}) 
}$$ 
}

$\check\cF_m(n)$ consists of the partial Fourier transforms $\varphi^\sim$
(as in Definition \defOSfourtrans.ii) of translation invariant 
functions $\varphi\in\cF_m(n)$ as in Definition \defOSFmn. Also, 
$\ \Cont{i}{j}{C} \varphi^\sim = (\Cont{i}{j}{C} \varphi)^\sim\ $, where
$\Cont{i}{j}{C} \varphi$ is defined in Definition \defOScontnorm.

\corollary{\STM\corOSmomcontrintbound}{
Let  $C(\xi,\xi') \in \cF_0(2)$ be an antisymmetric function. 
Let $m,m'\ge 0$, $n,n'\ge 1$
and $f\in\check\cF_m(n),\ f'\in\check\cF_{m'}(n')$. Then
$$
\| \Cont{1}{n+1}{C}\, {\rm Ant}_{\rm ext}(f\otimes f')\tnorm\
\le 4\| C\|_{1,\infty}\| f\tnorm\ \| f'\tnorm
$$

}

\prf The claim follows by iterated application of Lemma \lemOSelloneinftyamp\ 
and the observation that $\| C\tnorm=\| C\|_{1,\infty}$ by Remark
\remOSdiffdecaynorm.
\endproof

\noindent
We shall prove an analog of Theorem \thmOSfirststep, for the momentum 
space norms of Definition \defOSdiffdecaynorm. By way of preparation,
we first formulate the following variant of Lemma \lemOSscalednorm.

\lemma{\STM\lemOSTscalednorm}{
Let  $ \rho_{m;n}$ be a  sequence of nonnegative real numbers such that 
$\rho_{m;n'} \le \rho_{m;n}$ for $n'\le n$. Define (locally) for 
$f\in\check\cF_m(n)$
$$
\|f\| =\rho_{m;n} \,\|f\tnorm
$$
where $\|f\tnorm$ is the norm of Definition \defOSdiffdecaynorm. 
\Item{i)} 
The seminorms $\|\cdot\|$ are symmetric.
\Item{ii)}
For a covariance $C$, let $S(C)$ be the quantity introduced in Definition
\defIntBndsS. Then
$2S(C)$ is an integral bound for the covariance $C$ with 
respect to the family of seminorms $\|\cdot\|$.
\Item{iii)} 
Let $C$ be a covariance. Assume that for all $m,m'\ge 0$ and $n,n'\ge1$
$$
\rho_{m+m';\,n+n'-2} \le \rho_{m;n}\,\rho_{m';n'} 
$$
and let $\cb$ obey 
$$
\cb \ge 4\| C\|_{1,\infty}
$$
Then $\cb$ is a contraction bound for the covariance $C$ with 
respect to the family of seminorms $\|\cdot\|$.

}

\prf
Parts (i) and (ii) are trivial. To prove part (iii), let 
$f\in \check\cF_m(n),\ f'\in \check\cF_{m'}(n')$ and $1\le i \le n, \ 1\le j\le n'$.
If $n\ge 2$ or $n'\ge 2$ define the function 
$g$ on $\check \cB^{m+m'}\times \cB^{n+n'-2}$ by
$$\eqalign{
&g({\sst \check\eta_1,\cdots,\check\eta_{m+m'};
\xi_1,\cdots,\xi_{i-1},\,\xi_{i+1},\cdots,\xi_n,\,
\xi_{n+1},\cdots,\xi_{n+j-1},\,\xi_{n+j+1},\cdots,\xi_{n+n'}}) \cr
&\hskip 3cm  =\int {\sst d\zeta\,d\zeta'}\, f( {\sst \check\eta_1,\cdots,\check\eta_m;\, 
\xi_1,\cdots,\xi_{i-1},\,\zeta,\,\xi_{i+1},\cdots,\xi_n })\, C({\sst \zeta,\zeta'})
\cr
& \hskip 6cm
f'( {\sst \check\eta_{m+1},\cdots,\check\eta_{m+m'};
\,\xi_{n+1},\cdots,\xi_{n+j-1},\,\zeta',\,\xi_{n+j+1},\cdots,\xi_{n+n'}})\cr
}$$
If $n=n'=1$, define the function $g$ on $\check \cB_{m+m'}$ by
$$
g({\sst \check\eta_1,\cdots,\check\eta_{m+m'}})\ (2\pi)^{d+1}
 \de({\sst \check\eta_1+\cdots+\check\eta_{m+m'}}) 
= 
\int {\sst d\zeta\,d\zeta'}\, 
f( {\sst \check\eta_1,\cdots,\check\eta_m;\,\zeta})\  C({\sst \zeta,\zeta'})\ 
 f'( {\sst\check\eta_{m+1},\cdots,\check\eta_{m+m'};\,\zeta'})
$$
or equivalently, by
$$
g({\sst \check\eta_1,\cdots,\check\eta_{m+m'}}) 
= 
\smsum_{\si \in \{\uparrow,\downarrow\} \atop b \in \{0,1\}}\,
\int {\sst d\zeta'}\, 
f( {\sst \check\eta_1,\cdots,\check\eta_m;\,(0,\si,b)})\  
C({\sst (0,\si,b),\zeta'})\ 
 f'( {\sst\check\eta_{m+1},\cdots,\check\eta_{m+m'};\,\zeta'})
$$
Then
$$
\Cont{i}{n+j}{C}{\rm Ant_{ext}}(f\otimes f') = {\rm Ant_{ext}}\, g
$$
and therefore
$$
\big\| \Cont{i}{n+j}{C}{\rm Ant_{ext}}(f\otimes f') \big\| 
\le \|g\|
$$
As
$$
\|g\tnorm\ 
\le\ 4\|f\tnorm\ \ \| C\|_{1,\infty}\  \|f'\tnorm
$$
and consequently
$$\eqalign{
\big\| \Cont{i}{n+j}{C}{\rm Ant_{ext}}(f\otimes f') \big\|
&\le 4\rho_{m+m';n+n'-2}\,\| C\|_{1,\infty}\,\|f\tnorm\, \|f'\tnorm \cr
&\le \cb\, \rho_{m;n}\, \|f\tnorm\ \rho_{m';n'}\|f'\tnorm \cr 
&=\, \cb\, \|f\|\, \|f'\|\cr 
}$$

\endproof

\noindent
The analog of ``external improving'' for the current setting is
the following Lemma. We shall later, in Lemma \lemOStildesourceterm\ of [FKTo3],
prove a general scale version.
Let $\big(\rho_{m;n}\big)_{m,n\in\bbbn_0}$ be  a system of positive real 
numbers and $X \in \fN_{d+1}$ with $X_\0<1$.
For an even Grassmann function 
$$
\cW(\phi,\psi) = \smsum_{m,n\ge 0\atop m+n\ {\rm even}} 
\int_{\cB^{m+n}}\hskip-15pt {\sst d\et_1\cdots d\et_m\ d\xi_1\cdots d\xi_n}\ 
W_{m,n}({\sst\et_1\cdots \et_m,\xi_1,\cdots,\xi_n})\,
\phi{\sst(\et_1)}\cdots\phi{\sst(\et_m)}\,\psi{\sst(\xi_1)}\cdots{\sst\psi(\xi_n)}
$$
with kernels $W_{m,n}$ that are separately antisymmetric under 
permutations of their $\et$ and $\xi$ arguments, define 
$$
N_0^\sim(\cW \cl \be;X,\vec\rho) = \fe_0(X)
\smsum_{m+n \ge 2 \atop m+n\ {\rm even}} \be^{m+n} \rho_{m;n}
\| W^\sim_{m,n}\tnorm
$$
where $\|\ \cdot\ \tnorm$ is the norm of Definition \defOSdiffdecaynorm

\lemma{\STM\lemOSTZsourceterm}{
Let $\big(\rho_{m;n}\big)_{m,n\in\bbbn_0}$ be  a system of positive real 
numbers and $X \in \fN_{d+1}$ with $X_\0<1$.
There are constants $\abcst$ and $\Ga_0$, independent
of $M$, $X$ and $\vec\rho$ such that the following holds for all $\Ga\le\Ga_0$.
Let $C(k)$ be a function that obeys  
$\ 
\|C(k)\tnorm \le  \Ga \sfrac{\rho_{m;n}}{\rho_{m+1\,;\,n-1}}\,\fe_0(X)
\ $ 
for  all $ m\ge0$, $n\ge 1$ and let $C(\xi,\xi')$ be the covariance 
associated to it by Definition \defOSftcov.
\Item i)
Let $\cW(\phi,\psi)$ be a Grassmann function and set
$$
\cW'(\phi,\psi)=\cW(\phi,\psi+CJ\phi)
$$
Then 
$$
N^\sim_0(\cW'-\cW;\be;\,X,\vec\rho)
\le \abcst\,\Ga \, N^\sim_0(\cW;2\be;\,X,\vec\rho)
$$

\Item ii)
Assume that there is a $Y\in\fN_{d+1}$ such that
$\ 
\|C'(k)\tnorm \le  \sfrac{\rho_{m;n}}{\rho_{m+1\,;\,n-1}}\,Y\,\fe_0(X)
\ $ 
for  all $ m\ge0$, $n\ge 1$. Set
$$
\cW'_s(\phi,\psi)=\cW(\phi,\psi+CJ\phi+sC'J\phi)
$$
Then 
$$
N^\sim_0(\sfrac{d\hfill}{ds}\cW'\big|_{s=0};\be;\,X,\vec\rho)
\le \abcst\,Y \, N^\sim_0(\cW;2\be;\,X,\vec\rho)
$$

  }
\prf 
Let 
$$
J^\sim\big((k_0,\k,\si,a),(x_0,\x,\si',a')\big)
= \de_{\si,\si'}e^{i(-1)^a\<k,x\>_-}
    \cases{1& if $a=1,\ a'=0$\cr
          -1& if $a=0,\ a'=1$\cr
           0& otherwise}
$$
be the partial Fourier transform of $J(\et,\xi)$ with respect to its first
argument. Observe that
$$
\int  d\ze\ J^\sim(\check\et,\ze)\,C(\ze,\ze')
=C(k)E_+(\check\et,\ze')\hbox{ where }\check\et=(k_0,\k,\si,a)
$$
Let $f\in\check\cF_m(n)$, $1\le i\le n$ and set,
for $\check\et_{m+1}=\big(k_{m+1},\si_{m+1},a_{m+1}\big)$,
$$\eqalign{
g({\sst\check\et_1,\cdots,\check\et_{m+1}\,;\,\xi_1,\cdots,\xi_{n-1}})
&={\rm Ant_{ext}}\int\!\! d\ze d\ze'\ J^\sim({\sst\check\et_{m+1},\ze})\,C({\sst\ze,\ze'})\,
f({\sst\check\et_1,\cdots,\check\et_m\,;\,
\xi_1\cdots,\xi_{i-1},\ze',\xi_{i},\cdots,\xi_{n-1}})\cr
&={\rm Ant_{ext}}\int\!\!  d\ze\ 
C({\sst k_{m+1}})E_+({\sst \check\et_{m+1},\ze})\,
f({\sst\check\et_1,\cdots,\check\et_m\,;\,
\xi_1\cdots,\xi_{i-1},\ze,\xi_{i},\cdots,\xi_{n-1}})\cr
}$$
for $n=2$ and
$$
g({\sst\check\et_1,\cdots,\check\et_{m+1}})\ (2\pi)^{d+1}
 \de({\sst \check\eta_1+\cdots+\check\eta_{m+1}}) 
={\rm Ant_{ext}}\int\!\! d\ze d\ze'\ J^\sim({\sst\check\et_{m+1},\ze})\,C({\sst\ze,\ze'})\,
f({\sst\check\et_1,\cdots,\check\et_m\,;\,\ze'})
$$
or equivalently
$$\eqalign{
g({\sst\check\et_1,\cdots,\check\et_{m+1}})
&={\rm Ant_{ext}} C({\sst k_{m+1}})\,
f({\sst\check\et_1,\cdots,\check\et_m\,;\,(0,\0,\si_{m+1},a_{m+1})})\cr
}$$
for $n=1$. In both cases, since  $\rD^\de_{m+1}E_+({\sst\check\et_{m+1},\ze})
=\ze^\de E_+({\sst\check\et_{m+1},\ze})$,
$$
\| g \tnorm
\le \|f\tnorm\ \ \| C(k)\tnorm 
$$
so that
$$
\rho_{m+1\,;\,n-1}\| g \tnorm
\le \Ga\,\rho_{m;n}\|f\tnorm\,\fe_0(X)
\EQN\eqnOSTextimpr$$

\Item i)
Write $\cW(\phi,\psi) = \sum_{m,n}\cW_{m,n}(\phi,\psi)$, 
with $\cW_{m,n}$ of degree $m$ in $\phi$ and degree $n$ in $\psi$, and 
$$
\cW(\phi,\psi+\ze) =  \sum_{m,n}\cW_{m,n}(\phi,\psi+\ze)
=\sum_{m,n}\smsum_{\ell=0}^n\cW_{m,n-\ell,\ell}(\phi,\psi,\ze)
$$ 
with $\cW_{m,n-\ell,\ell}$ of degrees $m$ in $\phi$, $n-\ell$ in $\psi$  and $\ell$
in $\ze$. Let $w_{m,n}$ and $w_{m,n-\ell,\ell}$ be the kernels of 
$\cW_{m,n}(\phi,\psi)$ and $\cW_{m,n-\ell,\ell}(\phi,\psi,CJ\phi)$ 
respectively. By the binomial theorem and repeated application 
of (\eqnOSTextimpr),
$$
\fe_0(X)\rho_{m+\ell\,;\,n-\ell}\| w^\sim_{m,n-\ell,\ell}\tnorm
\le (\abcst\,\Ga)^\ell {\tst{n\choose \ell}}\,\fe_0(X)
             \rho_{m;n}\| w^\sim_{m,n}\tnorm
$$
if $\ell\ge 1$.
 Then, 
$$\meqalign{
\cW'(\phi,\psi)-\cW(\phi,\psi)
 = \cW(\phi,\psi+CJ\phi)-\cW(\phi,\psi)
=\sum_{m,n\ge 0}\smsum_{\ell=1}^n\cW_{m,n-\ell,\ell}(\phi,\psi,CJ\phi)
}$$
and
$$\eqalign{
N^\sim_0\big(\cW'-\cW;\be;X,\vec\rho\big)
&\le \fe_0(X)\sum_{m,n\ge 0}\smsum_{\ell=1}^n\be^{m+n}\,
\rho_{m+\ell\,;\,n-\ell}\| w^\sim_{m,n-\ell,\ell}\tnorm
\cr
&\hskip-0.5in\le \fe_0(X)
\sum_{m,n\ge 0}\smsum_{\ell=1}^n
{\tst{n\choose \ell}}(\abcst\,\Ga)^\ell\be^{m+n} \,
                \rho_{m;n}\| w^\sim_{m,n}\tnorm\cr
&\hskip-0.5in= \fe_0(X)
\sum_{m,n\ge 0}\big[(1+\abcst\,\Ga)^n-1\big]\be^{m+n} \,
\rho_{m;n}\| w^\sim_{m,n}\tnorm\cr
}$$
If $\abcst\,\Ga\le\sfrac{1}{3} $,
$$\deqalign{
\big(1+\abcst\,\Ga\big)^{n}-1
&\le \abcst\,\Ga\,n\big(1+\abcst\,\Ga\big)^{n-1}
&\le \abcst\,\Ga\,\big(\sfrac{3}{2}\big)^n\big(1+\abcst\,\Ga\big)^{n-1}\cr
&\le\abcst\,\Ga\,2^{n}
}$$
and
$$
N^\sim_0(\cW'-\cW;\be;\,X,\vec\rho)
\le \abcst\,\Ga\, N^\sim_0(\cW;2\be;\,X,\vec\rho)
$$

\Item ii)
Write 
$$
\cW(\phi,\psi+\ze+\et) 
=\sum_{m,n}
\sum_{n_1,n_2,n_3\ge 0\atop n_1+n_2+n_3=n}\cW_{m,n_1,n_2,n_3}(\phi,\psi,\ze,\et)
$$ 
with $\cW_{m,n_1,n_2,n_3}$ of degrees $m$ in $\phi$, $n_1$ in $\psi$, $n_2$
in $\ze$ and $n_3$ in $\et$. Let $w_{m,n_1,n_2,n_3}$ be the kernel of 
$\cW_{m,n_1,n_2,n_3}(\phi,\psi,CJ\phi,C'J\phi)$. By the binomial theorem,
repeated application of (\eqnOSTextimpr) and the obvious analog of 
(\eqnOSTextimpr) with $C$ replaced by $C'$,
$$
\fe_0(X)\rho_{m+\ell+1\,;\,n-\ell-1}\| w^\sim_{m,n-\ell-1,\ell,1}\tnorm
\le (\abcst\,\Ga)^\ell(\abcst\,Y) {\tst{n\choose n-\ell-1,\ell,1}}\,
\fe_0(X)\rho_{m;n}\| w^\sim_{m,n}\tnorm
$$
Then, 
$$\eqalign{
\sfrac{d\hfill}{ds}\cW'_s(\phi,\psi)\big|_{s=0}
&= \sfrac{d\hfill}{ds}\cW(\phi,\psi+CJ\phi+sC'J\phi)\big|_{s=0}\cr
&=\sum_{m,n\ge 0}\smsum_{\ell=0}^{n-1}
\cW_{m,n-\ell-1,\ell,1}(\phi,\psi,CJ\phi,C'J\phi)
}$$
and
$$\eqalign{
N^\sim_0\big(\sfrac{d\hfill}{ds}\cW'_s\big|_{s=0};\be;X,\vec\rho\big)
&\le \fe_0(X)\sum_{m,n\ge 0}\smsum_{\ell=0}^{n-1}
\be^{m+n}\,\rho_{m+\ell+1\,;\,n-\ell-1}\| w^\sim_{m,n-\ell-1,\ell,1}\tnorm
\cr
&\hskip-0.5in\le \fe_0(X)
\sum_{m,n\ge 0}\smsum_{\ell=0}^{n-1}
{\tst{n\choose n-\ell-1,\ell,1}}(\abcst\,\Ga)^\ell(\abcst\,Y)
\be^{m+n} \,\rho_{m;n}\| w^\sim_{m,n}\tnorm\cr
&\hskip-0.5in= \abcst\,Y\fe_0(X)
\sum_{m,n\ge 0}\smsum_{\ell=0}^{n-1}
n{\tst{n-1\choose \ell}}\big(\abcst\,\Ga)^\ell
\be^{m+n} \,\rho_{m;n}\| w^\sim_{m,n}\tnorm\cr
&\hskip-0.5in=  \abcst\,Y\fe_0(X)
\sum_{m,n\ge 0}n(1+\abcst\,\Ga)^{n-1}
\be^{m+n} \,\rho_{m;n}\| w^\sim_{m,n}\tnorm\cr
}$$
If $\abcst\,\Ga\le\sfrac{1}{3} $,
$$\deqalign{
n\big(1+\abcst\,\Ga\big)^{n-1}
&\le \big(\sfrac{3}{2}\big)^n\big(1+\abcst\,\Ga\big)^{n-1}
&\le 2^{n}
}$$
and
$$
N^\sim_0\big(\sfrac{d\hfill}{ds}\cW'_s\big|_{s=0};\be;X,\vec\rho\big)
\le \abcst\,Y\, N^\sim_0(\cW;2\be;\,X,\vec\rho)
$$
\endproof

In Definition \defNPampGreen\ of [FKTf3], we shall 
amputate a Grassmann function by applying the Fourier transform $\hat A$, 
in the sense of Definition \defOSfourtransII, of $A(k)=ik_0-e(\k)$  to its 
external arguments. Precisely, if $\cW(\phi,\psi)$ is a Grassmann function,
then
$$
\cW^a(\phi,\psi)=\cW(\hat A\phi,\psi)
$$
where
$$
(\hat A\phi)(\xi)=\int d\xi'\ \hat A(\xi,\xi')\phi(\xi')
$$
If $C(\xi,\xi')$ is the covariance associated to  $C(k)$
in the sense of Definition \defOSftcov\ and $J$ is the particle
hole swap operator of (\eqnOSjdef), then,
by parts (i) and (ii) of Lemma \lemOSjhat,
$$
\int d\et d\et'\ C(\xi,\et)J(\et,\et')\hat A(\et',\xi)=\hat E(\xi,\xi')
$$
where
$\hat E$ is the Fourier transform of $\big(ik_0-e(\k)\big)C(k)$ in the sense
of Definition \defOSfourtransII.

\theorem{\STM\thmOSTfirststep}{ 
Fix $j_0\ge 1$ and set, for $\de e\in\cE_\mu$,
$$
C_0(k;\de e) = \sfrac{U(\k) - \nu^{(> j_0)}(k)}{\imath k_0 - e(\k)+\de e(\k)}
$$
Let $C_0(\de e)$ be the Fourier transform of $C_0(k;\de e)$ in the sense of 
Definition  \defOSftcov.
Then there are ($M$ and $j_0$--dependent) constants
$\be_0,\ \veps_0,\ \const\,$ and $\mu>0$ such that, for all 
$\be\ge\be_0$ and $\veps,\veps'\le \veps_0$, the following holds:

{\parindent=.25in\item{}
Choose a system $\big(\rho_{m;n}\big)_{m,n\in\bbbn_0}$, of positive real 
numbers obeying
$\rho_{m;n-1} \le \rho_{m;n}$,  
$\rho_{m+1;n-1} \le \veps'\rho_{m;n}$ and 
$\rho_{m+m';n+n'-2} \le \rho_{m;n}\,\rho_{m';n'}$.
Let  $X \in \fN_{d+1}$ with $X_\0<\sfrac{1}{4}$.
For all $\de e\in\cE_\mu$ with $ \|\de\hat e\|_{1,\infty}\le X$ and all
even Grassmann functions $\cV(\psi)$ with $\,N_0^\sim(\cV \cl 32\be;X,\vec\rho) 
\le \veps\fe_0(X)\,$,
$$
\tilde\cW(\phi,\psi;\de e)
=\tilde\Om_{C_0(\de e)}(\cV)(\phi,\psi) - \cV(\psi)
-\half\phi JC_0(\de e)J\phi
$$
obeys
$$
N_0^\sim\big(\tilde\cW^a(\phi,\psi;\de e)\,\cl  \be;X,\vec\rho\big) 
\le \veps\big(\sfrac{1}{\be}+\sqrt{\veps'}\,\big)\,\fe_0(X) 
$$
and
$$
N_0^\sim\big(\sfrac{d\hfill}{ds}\tilde\cW^a(\phi,\psi;\de e+s\de e')\big|_{s=0}
\cl\be;X,\vec\rho\big) 
\le \veps\big(\sfrac{1}{\be}+\sqrt{\veps'}\big)
\fe_0(X)\,\| \de\hat e'\|_{1,\infty}
$$

}
}

\prf
The proof of this theorem is similar that of Theorem
\thmOSfirststep. Let $V_{\rm ext}$ be the vector space generated by $\phi(\eta),\,\eta\in\cB$, and recall from \S\CHnorms\ that $V$ is 
the vector space generated by $\psi(\xi),\,\xi \in \cB$. Set 
$\tilde V =V_{\rm ext}\oplus V$ and\ $\tilde A =\bbbc$. Then
$$
\tilde A \otimes \tilde V^{\otimes n} \ =\ \tilde V^{\otimes n}\ 
=\ \bigoplus_{m_1+n_1+\cdots+m_r+n_r=n \atop n_1,m_2,\cdots,n_{r-1},m_r \ge 1}
V_{\rm ext}^{\otimes m_1} \otimes V^{\otimes n_1} \otimes \cdots \otimes
V_{\rm ext}^{\otimes m_r} \otimes V^{\otimes n_r} 
$$
Every element $F_{m_1,n_1,\cdots,m_r,n_r}$ of
$\ V_{\rm ext}^{\otimes m_1} \otimes V^{\otimes n_1} \otimes \cdots \otimes
V_{\rm ext}^{\otimes m_r} \otimes V^{\otimes n_r}\ $
can be uniquely written in the form
$$\eqalign{
&F_{m_1,n_1,\cdots,m_r,n_r}   
= 
\int \hskip -2pt {\sst d\eta_1\cdots d\eta_{m_1+\cdots+m_r}\
  d\xi_1\cdots d\xi_{n_1+\cdots+n_r}}\cr
& \hskip 4cm 
 f_{m_1,n_1,\cdots,m_r,n_r}({\sst \eta_1,\cdots,\eta_{m_1+\cdots+m_r};\,
   \xi_1,\cdots ,\xi_{n_1+\cdots+n_r}}) \cr
&\hskip 5cm \phi({\sst \eta_1})\otimes\cdots \otimes\phi({\sst \eta_{m_1}})\otimes
  \psi({\sst \xi_1})\otimes\cdots\otimes \psi({\sst \xi_{n_1}}) \cr
&  \hskip 6cm \otimes \cdots \otimes 
  \psi({\sst\xi_{n_1+\cdots+n_{r-1}+1}})
      \otimes\cdots\otimes 
     \psi({\sst \xi_{n_1+\cdots+n_r}})
}$$
We define
$$
\V F_{m_1,n_1,\cdots,m_r,n_r}  \tV
= \rh_{m_1+\cdots+m_r;n_1+\cdots n_r}\, \big\|  f^\sim_{m_1,n_1,\cdots,m_r,n_r} \Tnorm
$$
and for 
$$
F = \sum_{m_1+n_1+\cdots+m_r+n_r=n \atop n_1,m_2,\cdots,n_{r-1},m_r \ge 1} 
F_{m_1,n_1,\cdots,m_r,n_r}\ \in \tilde V^{\otimes n}
$$
with 
$F_{m_1,n_1,\cdots,m_r,n_r} \in
 V_{\rm ext}^{\otimes m_1} \otimes V^{\otimes n_1} \otimes \cdots \otimes
V_{\rm ext}^{\otimes m_r} \otimes V^{\otimes n_r}\ $
$$
\v F \tv = 
\sum_{m_1,n_1,\cdots,m_r,n_r} 
\V F_{m_1,n_1,\cdots,m_r,n_r}  \tV
$$
Define the covariance $\tilde C$ on $\tilde V$ by
$$\eqalign{
\tilde C\big( \phi({\sst \eta}), \phi({\sst \eta'}) \big) &= 0\cr
\tilde C\big( \phi({\sst \eta}), \psi({\sst \xi}) \big) &= 0\cr
\tilde C\big( \psi({\sst \xi}), \psi({\sst \xi'}) \big) &= 
C_0( \xi, \xi';\de e)\cr
}$$
The restriction of $\tilde C$ to $V$ coincides with the covariance on $V$ 
determined by $C_0(\de e)$ as at the beginning of \S\CHnorms, 
while $V_{\rm ext}$ is isotropic and perpendicular to $V$ with 
respect to $\tilde C$.

We have already observed, in the proof of Theorem \thmOSfirststep,
that there is a constant $\cst{}{1}$ such that
$
\ S\big(C_0(\de e)\big)\le\cst{}{1}\ $ and 
$
\ \| C_0(\de e)\|_{1,\infty}
\ \le \sfrac{1}{4}\cst{}{1}\fe_0({\sst \|\de \hat e\|_{1,\infty}})\ .
$
Let $C^a_0(\de e)$ be the covariance associated with 
$\sfrac{U(\k) - \nu^{(> j_0)}(k)}{\imath k_0 - e(\k)+\de e(\k)}
(\imath k_0 - e(\k))$. Then 
$$
\big\| C_0^a(k;\de e)\Tnorm\le\const \big\|C_0(\de e)\big\|_{1,\infty}
\le \cst{}{1}\fe_0({\sst \|\de \hat e\|_{1,\infty}})
$$
By Lemma \lemOSTscalednorm,  $\ib=4\cst{}{1}$ is an integral bound for 
$C_0(\de e)$, and  hence for $\tilde C$, and $\cb = \cst{}{1}\fe_0(X)$ is a contraction bound for $C_0(\de e)$, and hence for $\tilde C$.
 Furthermore,  Lemma \lemOSTZsourceterm.i, with $C=C^a_0(\de e)$, 
$X=\|\de\hat e\|_{1,\infty}$ and $\Ga=\veps'\ib$ is applicable.

Set $\al=\sfrac{\be}{\ib}$. For any Grassmann function $\cW(\phi,\psi)=\sum_{m,n}\cW_{m,n}(\phi,\psi)$,
with $\cW_{m,n}(\phi,\psi)$ of degree $m$ and $n$ in $\phi$ and $\psi$,
respectively, let
$$
N^\sim(\cW ;\cb,\ib, \al) = \sfrac{1}{\ib^2}\cb\,\smsum_{m,n} 
\al^{m+n}\, \ib^{m+n}\, \v W^\sim_{m,n} \tv
$$
be the norm of Definition \defOSgrnorm\  and of Definition \deffunctnorm\ of [FKTr1],
but with $V$ replaced by $\tilde V$ and $A$ replaced by $\tilde A=\bbbc$. Then
$\ 
N^\sim(\cW ;\cb,\ib, \al)\ 
=\ \sfrac{1}{4\ib}N_0^\sim(\cW \cl \be;X,\vec\rho)
\ $
and, if $\cV = \lw \cV' \rw_{C_0(\de e)}$,
$$\eqalign{
N^\sim(\cV' ;\cb,\ib, \al) &\le \sfrac{1}{4\ib}
                           \,N_0^\sim(\cV \cl 2\be;X,\vec\rho) \cr
N^\sim(\cV'-\cV ;\cb,\ib, \al) &\le \sfrac{\ib}{2\be^2}\,
                           N_0^\sim(\cV \cl 2\be;X,\vec\rho) \cr
}$$
by Corollary \corwicknorm\ of [FKTr1].

To prove the first part of the Theorem, set 
$\,\cV = \lw \cV' \rw_{C_0(\de e)}$. 
Then the hypotheses of Theorem \thmOSroptheorII\ with $C=C_0(\de e)$
 and $\cW=\cV'$ are fulfilled.
Therefore
$$\eqalign{
\sfrac{1}{4\ib}N_0^\sim\big(\Om_{C_0(\de e)}(\cV)-\cV\cl\be;X,\vec\rho\big) 
&\le  N^\sim\big(\Om_{C_0(\de e)}(\lw \cV'\rw_{C_0(\de e)})-\cV';\cb,\ib,\al\big)  
+ N^\sim\big(\cV'-\cV;\cb,\ib,\al\big) \cr
&\le \sfrac{2}{\al^2}\, \sfrac{N^\sim(\cV';\cb,\ib, 8\al)^2}
{1-{4\over\al^2}N^\sim(\cV';\cb,\ib, 8\al ) }
+ N^\sim\big(\cV'-\cV;\cb,\ib,\al\big) \cr
&\le \sfrac{2}{\al^2}\, \sfrac{{1\over 16\ib^2}N_0^\sim(\cV;16\be;X,\vec\rho)^2}
{1-{\ib\over\be^2}N_0^\sim(\cV;16\be;X,\vec\rho ) }
+ \sfrac{\ib}{2\be^2}\,N_0^\sim\big(\cV\cl 2\be;X,\vec\rho\big) \cr
&\le \sfrac{\veps^2}{8\be^2}\, \sfrac{\fe_0(X)^2}
{1-{\veps\ib\over\be^2}\fe_0(X) }
+ \sfrac{\veps\ib}{2\be^2}\fe_0(X)\cr
&\le \sfrac{\veps\ib}{\be^2}\fe_0(X)\cr
}$$
since $\veps_0$ is chosen so that ${\veps\ib\over\be^2}<\sfrac{1}{4}$ and,
by Corollary \corOSappMonoidIV.ii,
$\sfrac{\fe_0^2(X)}{1-{1\over4}\fe_0(X) }\le\abcst\ \fe_0(X)$.
Observe that $\cV$ and hence $\Om_{C_0(\de e)}(\cV)$ are independent of
$\phi$ and consequently are not affected by amputation. So,
by Lemma \lemOStworengrpmaps\ and Lemma \lemOSTZsourceterm,
$$\eqalign{
&N_0^\sim\big(\tilde\cW^a(\phi,\psi;\de e)\cl\be;X,\vec\rho\big)
 = N_0^\sim\big(\Om_{C_0(\de e)}(\cV)(\phi,\psi
               +C^a_0(\de e)J\phi)-\cV(\psi)\cl\be;X,\vec\rho\big)\cr
&\hskip.4in\le N_0^\sim\big(\Om_{C_0(\de e)}(\cV)(\phi,\psi
     +C^a_0(\de e)J\phi)
     -\Om_{C_0(\de e)}(\cV)(\phi,\psi)\cl\be;X,\vec\rho\big)\cr
&\hskip3in
+N_0^\sim\big(\Om_{C_0(\de e)}(\cV)(\phi,\psi)
                    -\cV(\psi)\cl\be;X,\vec\rho\big)\cr
&\hskip.4in
\le\abcst\,\veps'\ib\,N_0^\sim\big(\Om_{C_0(\de e)}(\cV)\cl 2\be;X,\vec\rho\big)
+\sfrac{4\veps\ib^2}{\be^2}\fe_0(X)\cr
&\hskip.4in
\le\abcst\,\veps'\ib\,  
\Big[N_0^\sim\big(\Om_{C_0(\de e)}(\cV)-\cV\cl 2\be;X,\vec\rho\big)
            +N_0^\sim\big(\cV\cl 2\be;X,\vec\rho\big)\Big]
+\sfrac{4\veps\ib^2}{\be^2}\fe_0(X)\cr
&\hskip.4in
\le\abcst\,\veps'\ib\,  
\Big[\sfrac{\veps\ib^2}{\be^2}\fe_0(X)+\veps\fe_0(X)\Big]
+\sfrac{4\veps\ib^2}{\be^2}\fe_0(X)\cr
&\hskip.4in\le \abcst\,(1+\ib)^3\,\veps\big(\sfrac{1}{\be^2}+\veps'\big)\fe_0(X)
\le \veps\big(\sfrac{1}{\be}+\sqrt{\veps'}\,\big)\fe_0(X)\cr
}$$

Finally, we prove the bound on 
$\sfrac{d\hfill}{ds}\tilde\cW^a(\phi,\psi;\de e+s\de e')\big|_{s=0}$.
As
$$\eqalign{
\sfrac{d\hfill}{ds}C_0(k;\de e+s\de e')\big|_{s=0}
&=-\sfrac{U(\k)-\nu^{(> j_0)}(k)}{[\imath k_0-e(\k)+\de e(\k)]^2}\de e'(\k)\cr
\sfrac{d\hfill}{ds}C^a_0(k;\de e+s\de e')\big|_{s=0}
&=-\sfrac{U(\k)-\nu^{(> j_0)}(k)}{[\imath k_0-e(\k)+\de e(\k)]^2}\de e'(\k)
\ \big(\imath k_0-e(\k)\big)\cr
}$$
Proposition \propIntBndsII.i and
Proposition  \propOSrealfirstpropbound\ give that
$$\eqalign{
S\big(\sfrac{d\hfill}{ds}C_0(\de e+s\de e')\big|_{s=0}\big)
                   &\le\cst{}{1}\sqrt{\tn \de\hat e'\tn_{1,\infty}}\cr
\TN \sfrac{d\hfill}{ds}C_0(\de e+s\de e')\big|_{s=0}\TN_\infty
                      &\le \cst{}{1}\tn \de\hat e'\tn_{1,\infty}\cr
\big\| \sfrac{d\hfill}{ds}C_0(\de e+s\de e')\big|_{s=0}\big\|_{1,\infty}
&\le \sfrac{1}{4}\cst{}{1}\fe_0({\sst \|\de \hat e\|_{1,\infty}})\,
              \| \de\hat e'\|_{1,\infty}\cr
\big\| \sfrac{d\hfill}{ds}C^a_0(k;\de e+s\de e')\big|_{s=0}\Tnorm
&\le \cst{}{1}\fe_0({\sst \|\de \hat e\|_{1,\infty}})\,
              \| \de\hat e'\|_{1,\infty}\cr
}$$
Set $\ib'=4\cst{}{1}\sqrt{\tn \de\hat e'\tn_{1,\infty}}$ and 
$\cb' = \cst{}{1}\fe_0(X)\,\| \de\hat e'\|_{1,\infty}$. 
By Lemma \lemOSTscalednorm, $\half \ib'$ is an integral bound for 
$\sfrac{d\hfill}{ds}C_0(\de e+s\de e')\big|_{s=0}$ and $\cb'$
is a contraction bound for $\sfrac{d\hfill}{ds}C_0(\de e+s\de e')\big|_{s=0}$. 
 Furthermore,  Lemma \lemOSTZsourceterm.ii, with
 $C=C^a_0(\de e)$, $C'=\sfrac{d\hfill}{ds}C^a_0(\de e+s\de e')\big|_{s=0}$, 
$X=\|\de\hat e\|_{1,\infty}$, $\Ga=\cst{}{1}\veps'$ 
and $Y=\veps'\ib\| \de\hat e'\|_{1,\infty}$ is applicable.

Define $C_\ka=C_0(\de e+\ka\de e')$ and $\cW_\ka$ by 
$\lw \cW_\ka\rw_{C_\ka}=\cV$. Even when $\al$ is replaced by $2\al$,
the hypotheses of Lemmas  \lemprftwoA.i and \lemprftwoB.i of [FKTr1], 
with $\mu=1$, are satisfied. By these two Lemmas, followed by Corollary 
\corwicknorm.iii of [FKTr1],
$$\eqalign{
&N^\sim\big(\sfrac{d\hfill}{ds}\Om_{C_s}(\cV)\big|_{s=0}\cl\cb,\ib, \al\big) 
=N^\sim\big(\sfrac{d\hfill}{ds}\Om_{C_s}\big(\lw \cW_s\rw_{C_s}\big)\big|_{s=0}
              \cl\cb,\ib, \al\big) \cr
&\hskip1in\le N^\sim\big(\sfrac{d\hfill}{ds}\Om_{C_s}
\big(\lw\cV'\rw_{C_s}\big)\big|_{s=0}\cl\cb,\ib, \al\big)  
+N^\sim\big(\sfrac{d\hfill}{ds}\Om_{C_0}\big(\lw \cW_s\rw_{C_0}\big)\big|_{s=0}
              \cl\cb,\ib, \al\big) \cr
&\hskip1in\le \sfrac{1}{2\al^2}\, \sfrac{N^\sim(\cV';\cb,\ib, 8\al)^2}
{1-{4\over\al^2}N^\sim(\cV';\cb,\ib, 8\al ) }
\cst{}{1}\fe_0(X)\,\| \de\hat e'\|_{1,\infty}\cr
&\hskip2in+\Big\{1+\sfrac{2}{\al^2}\, \sfrac{N^\sim(\cV';\cb,\ib, 8\al)}
                  {1-{4\over\al^2}N^\sim(\cV';\cb,\ib, 8\al ) } \Big\} 
N^\sim\big(\sfrac{d\hfill}{ds}\cW_s\big|_{s=0} \cl\cb,\ib, 2\al\big)\cr
&\hskip1in\le \sfrac{1}{2\al^2}\, \sfrac{N^\sim(\cV';\cb,\ib, 8\al)^2}
{1-{4\over\al^2}N^\sim(\cV';\cb,\ib, 8\al ) }
\cst{}{1}\fe_0(X)\,\| \de\hat e'\|_{1,\infty}\cr
&\hskip2in+\Big\{1+\sfrac{2}{\al^2}\, \sfrac{N^\sim(\cV';\cb,\ib, 8\al)}
                  {1-{4\over\al^2}N^\sim(\cV';\cb,\ib, 8\al ) } \Big\} 
\sfrac{\tn \de\hat e'\tn_{1,\infty}}{(2\al-1)^2}N^\sim\big(\cV \cl\cb,\ib, 4\al\big)\cr
&\hskip1in\le \const\sfrac{\veps}{\al^2}\fe_0(X)\,\| \de\hat e'\|_{1,\infty}\cr
}$$
as above. 
By Lemma \lemOStworengrpmaps\ and Lemma \lemOSTZsourceterm,
$$\eqalign{
&N_0^\sim\big(\sfrac{d\hfill}{ds}\tilde\cW^a(\phi,\psi;\de e+s\de e')\big|_{s=0}\cl\be;X,\vec\rho\big)\cr
&\hskip.25in= 
N_0^\sim\Big(\sfrac{d\hfill}{ds}\Om_{C_0(\de e+s\de e')}(\cV)
                    \big(\phi,\psi+C^a_0(\de e+s\de e')J\phi\big)\big|_{s=0}
                    \cl\be;X,\vec\rho\Big)\cr
&\hskip.25in\le N_0^\sim\Big(\sfrac{d\hfill}{ds}\Om_{C_0(\de e+s\de e')}(\cV)
\big(\phi,\psi+C^a_0(\de e)J\phi\big)\big|_{s=0}\cl\be;X,\vec\rho\Big)\cr
&\hskip1.8in
+N_0^\sim\Big(\sfrac{d\hfill}{ds}\Om_{C_0(\de e)}(\cV)
              \big(\phi,\psi+C^a_0(\de e+s\de e')J\phi\big)\big|_{s=0}
              \cl\be;X,\vec\rho\Big)\cr
&\hskip.25in
\le\big(1+\abcst\,\veps'\ib\big)
 N_0^\sim\big(\sfrac{d\hfill}{ds}\Om_{C_0(\de e+s\de e')}(\cV)\big|_{s=0}
                        \cl 2\be;X,\vec\rho\big)\cr
&\hskip2in+\abcst\,\veps'\ib
N_0^\sim\big(\Om_{C_0(\de e)}(\cV)\cl 2\be;X,\vec\rho\big)
                          \| \de\hat e'\|_{1,\infty}
\cr
&\hskip.25in\le \const\sfrac{\veps}{\be^2}\fe_0(X)\| \de\hat e'\|_{1,\infty}
+\const\veps\veps'\fe_0(X)\| \de\hat e'\|_{1,\infty}\cr
&\hskip.25in\le \veps\big(\sfrac{1}{\be}+\sqrt{\veps'}\big)
\fe_0(X)\,\| \de\hat e'\|_{1,\infty}\cr
}$$
as above.
\endproof

\vfill\eject

\appendix{\APappSymmetries}{Symmetries}\PG\pgOSB

\definition{\STM\defOSsymmetries (Symmetries)}{

\noindent
Let $x=(x_0,\x,\si) \in \bbbr \times \bbbr^d \times \{\uparrow, \downarrow\}$,
$\xi = (x,a) \in 
\cB= \big(\bbbr \times \bbbr^d \times \{\uparrow,\downarrow\}\big)
\times\{0,1\}$ and $t=(t_0,\t) \in  \bbbr \times \bbbr^d$. We set
$$\meqalign{
x+t &= (x_0+t_0,\x+\t,\si) &&\qquad
\xi+t &= (x+t,a) \cr
R_0x &= (-x_0,\x,\si) &&
R_0\xi &= (R_0x,a) \cr
-x&=(-x_0,-\x,\si) &&
-\xi &= (-x,a) \cr
}$$
\Item{(T)} 
A function $f(\xi_1,\cdots,\xi_n)$ on $\cB^n$ is called {\bf translation invariant} if, for all $t \in \bbbr \times \bbbr^d$,
$$
f(\xi_1+t,\cdots,\xi_n+t) = f(\xi_1,\cdots,\xi_n)
$$
In the same way, one defines translation invariance for functions on 
$\big( \bbbr \times \bbbr^d \times \{\uparrow, \downarrow\} \big)^n$.
\Item{(N)}
 A function $f$ on $\cB^n$ {\bf conserves particle number}
if $f\big((x_1,a_1),\cdots,(x_n,a_n)\big)=0$
unless 
$$
\#\set{j}{a_j=0}
\ =\ \#\set{j}{a_j=1}
\ =\ \sfrac{n}{2}
$$
\Item{(S)}
Let $f$ be a function on $\cB^n$. Set, for each $A\in SU(2)$,
$$
f^A\big((\cdot,\si_1,b_1),\cdots,(\cdot,\si_{n},b_{n})\big)
=\sum_{\tau_1,\cdots\tau_{n}}
f\big((\cdot,\tau_1,b_1),\cdots,(\cdot,\tau_{n},b_{n})\big)
\smprod_{j=1}^{n}A^{(b_j)}_{\tau_j,\si_j}
$$
where $A^{(0)}=A$ and $A^{(1)}=\bar A$.
$f$ is called {\bf spin independent} if $f=f^A$ for all $A\in SU(2)$.

\Item{(R)} 
A function $f(\xi_1,\cdots,\xi_n)$ on $\cB^n$ is called 
{\bf $k_0$--reversal real} if 
$$
f(R_0\xi_1,\cdots,R_0\xi_n) = \overline{f(-\xi_1,\cdots,-\xi_n)}
$$
or, equivalently, if its Fourier transform obeys
$$
\check f(R_0\check \xi_1,\cdots,R_0\check \xi_n) 
= \overline{\check f(\check \xi_1,\cdots,\check \xi_n)}
$$
where $R_0(k_0,\k,\si,a)=(-k_0,\k,\si,a)$.
\Item{(B)} 
A function $f(\xi_1,\cdots,\xi_n)$ on $\cB^n$ is called 
{\bf  bar/unbar exchange invariant} if
$$
f\big((x_1,1-b_1),\cdots,(x_n,1-b_n)\big)
=i^n f\big((-x_1,b_1),\cdots,(-x_n,b_{n})\big)
$$
or, equivalently, if its Fourier transform obeys
$$
\check f({\sst(k_1,\si_1,1-b_1),\cdots,(k_n,\si_n,1-b_n)}) 
= i^n\check f({\sst (k_1,\si_1,b_1),\cdots,(k_n,\si_n,b_n)})
$$

\Item{vi)} Let $m\in\bbbn$ and  $\Si_1,\ \cdots,\ \Si_m\in\{{\rm B,N,R,S,T}\}$.
Then $f$ is $\Si_1\cdots \Si_m$--symmetric if $f$ satisfies part ($\Si_i$) 
of this definition for $1\le i\le m$.

\Item{vii)}  Let $\Si\in\{{\rm B,N,R,S,T}\}$. A Grassmann function
$$
\cW(\phi,\psi)=\sum_{m,n}W_{m,n}\phi^m\psi^n
$$
is $\Si$--symmetric if all of the coefficient functions $W_{m,n}$ are.

}

\remark{\STM\remOSgrassmannsymmetries}{
Let $
\cW(\phi,\psi)=\sum_{m,n}W_{m,n}\phi^m\psi^n
$ 
be a Grassmann function. 

\noindent
$\cW(\phi,\psi)$ is translation invariant if and only
if it is invariant under 
$$
\phi(\xi)\rightarrow\phi(\xi+t),\ \psi(\xi)\rightarrow\psi(\xi+t)
\qquad\hbox{for all $t\in\bbbr\times\bbbr^d$}
$$ 
$\cW(\phi,\psi)$ conserves particle number if and only if it is invariant under 
$$
\phi(x,a)\rightarrow e^{i(-1)^a\th}\phi(x,a),\ 
\psi(x,a)\rightarrow e^{i(-1)^a\th}\psi(x,a)
\qquad\hbox{for all $\th\in\bbbr$}
$$
$\cW(\phi,\psi)$ is spin independent if and only if it is invariant under 
$$
\phi(\,\cdot\,,\si,a)\rightarrow \sum_{\tau\in\{\uparrow,\downarrow\}}A^{(a)}_{\si,\tau}\phi(\,\cdot\,,\tau,a),\ 
\psi(\,\cdot\,,\si,a)\rightarrow \sum_{\tau\in\{\uparrow,\downarrow\}}A^{(a)}_{\si,\tau}\psi(\,\cdot\,,\tau,a)
\qquad\hbox{for all $A\in SU(2)$}
$$
$\cW(\phi,\psi)$ is bar/unbar exchange invariant if and only if it is invariant under 
$$
\phi(x,a)\rightarrow i\phi(-x,1-a),\ 
\psi(x,a)\rightarrow i\psi(-x,1-a)
$$
or, equivalently, under
$$
\check\phi(k,\si,a)\rightarrow i\check\phi(k,\si,1-a),\ 
\check\psi(k,\si,a)\rightarrow i\check\psi(k,\si,1-a)
$$
Define $\overline{\cW}(\phi,\psi) = \sum_{m,n}\overline{W_{m,n}}\,\phi^m\psi^n$
and
$$
(R_0\psi)(\xi)=\psi(R_0\xi)\qquad (R\psi)(\xi)=\psi(-\xi)
$$
$\cW(\phi,\psi)$ is $k_0$--reversal real if and only 
$$
\cW(R_0\phi,R_0\psi)=\overline{\cW}(R\phi,R\psi)
$$

}

\remark{\STM\remOSsymmetryConsequences}{
\Item i)
If the function $f(\xi_1,\xi_2)$ on $\cB^2$ is NS--symmetric, then it is of 
the form
$$
f\big((x,\si,a), (y,\tau,b)\big) =\de_{\si,\tau}\tilde f\big((x,a), (y,b)\big)
$$
for some $\tilde f$.

\Item ii)
If the function $f(\xi_1,\xi_2)$ on $\cB^2$ is antisymmetric and
NST--symmetric, then, translating by $-x-y$,
$$
f\big((x,\si,0), (y,\si,1)\big)
= f\big((-y,\si,0), (-x,\si,1)\big)
= - f\big((-x,\si,1), (-y,\si,0)\big)
$$
so that $f$ is also B--symmetric.

\Item iii)
If the function $f(\xi_1,\xi_2)$ on $\cB^2$ is antisymmetric and
NST--symmetric and if $\check f(k)$ is its Fourier transform specified at 
the end of Definition \defOSfourtrans.i, then
$$
f=J\,\widehat{\check f}
$$
The operator $J$ was defined in (\eqnOSjdef) and for a function $\chi(k)$, 
the Fourier transform $\hat\chi(\xi,\xi')$ was defined in Definition \defOSfourtransII. If, in addition, $f$ is $k_0$--reversal real, then
$\check f(-k_0,\k)=\overline{\check f(k_0,\k)}$.

\Item iv) If
$$
\cV(\psi,\bar\psi) = \int_{(\bbbr\times\bbbr^d\times\{\uparrow,\downarrow\})^4} 
\hskip-.7in V_0(x_1,x_2,x_3,x_4)\, \bar\psi(x_1)\psi(x_2)\bar\psi(x_3)\psi(x_4)\
dx_1dx_2dx_3dx_4
$$
with 
$$
V_0\big( (x_{1,0},\x_1,\si_1),\cdots,(x_{4,0},\x_4,\si_4)\big)
= -\half \de(x_1,x_2) \de(x_3,x_4) v(x_{1,0}-x_{3,0},\x_1-\x_3)
$$
where $\de\big((x_0,\x,\si),(x_0',\x',\si'))=\de(x_0-x_0')\de(\x-\x')\de_{\si,\si'}$,
then $\cV$ is BNST-symmetric.
If in addition $v(x_0,-\x)=\overline {v(x_0,\x)}$, then $\cV$ is also
R-symmetric. This is the case if 
$v(x_1-x_3)=\de(x_{1,0}-x_{3,0}) {\bf v}(\x_1-\x_3)$ with {\bf v} having a 
real--valued Fourier transform.

}

\remark{\STM\remOSphiJpsi}{
$\phi J\psi$ is BNRST--symmetric. If  $C(\xi,\xi')$ is the covariance 
associated to a function  $C(k)$ as in Definition \defOSftcov,
then $C(\xi,\xi')$ is antisymmetric and BNST--symmetric. If
$C(-k_0,\k)=\overline{C(k_0,\k)}$ then $C(\xi,\xi')$ is R--symmetric.

}

\remark{\STM\remOSrengrppreserves}{
Assume that $C(\xi,\xi')$ is the covariance associated to the function  
$C(k)$ as in Definition \defOSftcov\  and that $\cW(\phi,\psi)$ is a Grassmann 
function. Let $\Si\in\{{\rm BNST}\}$. If $\cW$ is $\Si$--symmetric, then
$\int\cW(\phi,\psi)\,d\mu_C(\psi)$, $\tilde\Om_C(\cW)$ and $\Om_C(\cW)$
are too. If $C(-k_0,\k)=\overline{C(k_0,\k)}$, then 
$\int\cW(\phi,\psi)\,d\mu_C(\psi)$, $\Om_C(\cW)$ and $\tilde\Om_C(\cW)$
are R--symmetric.
}

\lemma{\STM\lemOSphipsistruct}{
Let $W(\et,\xi)$ be BNST--symmetric. Then
$$
\int d\xi d\et\ \psi(\xi) \big(J\,(\check W)^{\hat{}}\,\big)(\xi,\et)\,\phi(\et)
=\int d\et d\xi\ \phi(\et) W(\et,\xi)\psi(\xi)
$$
}
\prf
The bar/unbar invariance implies
$$\eqalign{
&\check W\big((k,\si,a),(k,\si,1-a)\big)
=-\check W\big((k,\si,1-a),(k,\si,a)\big)\cr
&\Longrightarrow
\check W((k,\si,a),(k,\si,1-a))
=-(-1)^a  \check W\big((k,\si,1),(k,\si,0)\big)
=-(-1)^a  \check W(k)
}$$
for an arbitrary spin $\si$.
Hence, if $x$ and $y$ both have spin component $\si$,
$$\eqalign{
&\big(J\,(\check W)^{\hat{}}\,\big)\big((x,1-a),(y,a)\big)
=(-1)^a\,(\check W)^{\hat{}}\big((x,a),(y,a)\big)\cr
&\hskip.2in=(-1)^a\int\sfrac{d^{d+1}k}{(2\pi)^{d+1}}\ 
e^{(-1)^a\imath<k,x-y>_-}\,\check W(k)\cr
&\hskip.2in=-\int\sfrac{d^{d+1}k}{(2\pi)^{d+1}}\ 
e^{(-1)^a\imath<k,x-y>_-}\,\check W({\sst (k,\si,a),(k,\si,1-a)})\cr
&\hskip.2in=-\int\sfrac{d^{d+1}k}{(2\pi)^{d+1}}\sfrac{d^{d+1}k'}{(2\pi)^{d+1}}\ 
e^{-(-1)^a\imath<k,y>_-}\,e^{-(-1)^{1-a}\imath<k',x>_-}\,\check W({\sst (k,\si,a),(k',\si,1-a)})\ {\sst (2\pi)^{d+1}\de(k-k')}\cr
&\hskip.2in=-W\big((y,a),(x,1-a)\big)
}$$
The lemma follows.
\endproof

\vfill\eject

\appendix{\APappGrass}{ Some Standard Grassmann Integral Formulae}\PG\pgOSC

For a function $C(\xi,\xi')$ on $\cB\times\cB$ we set, as in 
Section \CHamputate 
$$\eqalign{
\psi\phi&=\int \psi(\xi)\phi(\xi)\ d\xi\cr
C\phi&=\int C(\xi,\xi')\phi(\xi')\ d\xi'\cr
\phi C\psi&=\int \phi(\xi)C(\xi,\xi')\psi(\xi')\ d\xi d\xi'\cr
}$$

\lemma{\STM\lemOSappGrassI}{ 
Let $f(\psi)$ be a Grassmann function  and let $C$ be an arbitrary 
antisymmetric covariance.
Then
$$
\int f(\psi)e^{\psi\phi}d\mu_{C}(\psi)
=e^{-{1\over 2} \phi C\phi}\int f(\psi+C\phi)d\mu_{C}(\psi)
$$
}
\prf It suffices to consider $f(\psi)=e^{\psi\ze}$ with $\ze$ another
Grassmann field. Then comparing
$$\deqalign{
\int f(\psi)e^{\psi\phi}d\mu_{C}(\psi)
&=\int e^{\psi(\ze+\phi)}d\mu_{C}(\psi)
&=e^{-{1\over 2}(\ze+\phi)C(\ze+\phi)}\cr
\int f(\psi+C\phi)d\mu_{C}(\psi)
&=\int e^{(\psi+C\phi)\ze}d\mu_{C}(\psi)
&=e^{-\phi C\ze}e^{-{1\over 2}\ze C\ze}
}$$
gives the desired result. 
\endproof

\lemma{\STM\lemOSappGrassII}{
Let $C(\xi,\xi')$ and $U(\xi,\xi')$ be antisymmetric functions such that the
norm of the integral operator with kernel 
$\int d\xi''\ C(\xi,\xi'')U(\xi'',\xi')$ is strictly smaller than one.
Let $C'(\xi,\xi')$ be the kernel of the integral operator $[\bbbone -CU]^{-1}C$
and set
$$\eqalign{
\cU&=\int  U(\xi,\xi)\ \psi(\xi) \psi(\xi)\ d\xi d\xi'\cr
}$$
and $Z= \int e^{{1\over 2}\cU(\psi)}d\mu_C(\psi)$. Then, 
\Item a) For all Grassmann functions  $f(\psi)$
$$
\sfrac{1}{Z}\int f(\psi) e^{{1\over 2}\cU(\psi)}d\mu_C(\psi)
=\int f(\psi)d\mu_{C'}(\psi)
$$
\Item b) For all Grassmann functions $\cW(\psi)$
$$
\sfrac{1}{Z}\int e^{\cW(\psi+\phi)}d\mu_C(\psi)
=e^{{1\over 2} \phi U[\bbbone+  C'U]\phi}
\int e^{(\cW-{1\over 2}\cU)(\psi+[\bbbone+C'U]\phi)}
d\mu_{C'}(\psi)
$$
}
\prf We give the proof for the case that the Grassmann algebra is finite dimensional. The general case then follows by approximation.
\Item a)
It suffices to consider the generating functional
$f(\psi)=e^{\psi \phi}$. By definition
$$\eqalign{
\int e^{\psi \phi}e^{{1\over 2}  \cU(\psi)}d\mu_C(\psi)
&=\int e^{\psi \phi}e^{{1\over 2} \psi  U\psi}d\mu_C(\psi)\cr
&={\rm Pf}(C)\int e^{\psi \phi}e^{{1\over 2} \psi  U\psi}
e^{-{1\over 2} \psi  C^{-1}\psi}d\psi\cr
&={\rm Pf}(C)\int e^{\psi \phi}
e^{-{1\over 2} \psi  {C'}^{-1}\psi}d\psi\cr
&=\sfrac{{\rm Pf}(C)}{{\rm Pf}(C')}\int e^{\psi \phi}d\mu_{ C'}(\psi)\cr
}$$
where ${\rm Pf}(C)$ is the Pfaffian of $C$. In particular, setting $\phi=0$,
$\sfrac{{\rm Pf}(C)}{{\rm Pf}(C')}=\int e^{{1\over 2}  \cU(\psi)}d\mu_C(\psi)
=Z$. 
\Item b) By part a
$$\eqalign{
\int e^{\cW(\psi+\phi)}d\mu_C(\psi)
&=e^{{1\over 2}\cU(\phi)}\int e^{\cW(\psi+\phi)-{1\over 2}\cU(\psi+\phi)
+\psi U\phi+{1\over 2}\cU(\psi)}d\mu_C(\psi)\cr
&=\sfrac{{\rm Pf}(C)}{{\rm Pf}(C')}
e^{{1\over 2}\cU(\phi)}\int e^{\cW(\psi+\phi)-{1\over 2}\cU(\psi+\phi)
+\psi U\phi}d\mu_{C'}(\psi)\cr
&=\sfrac{{\rm Pf}(C)}{{\rm Pf}(C')}
e^{{1\over 2}\cU(\phi)}e^{{1\over 2} \phi U C' U\phi}
\int e^{(\cW-{1\over 2}\cU)(\psi+\phi+C'U\phi)}
d\mu_{C'}(\psi)\cr
}$$
by Lemma \lemOSappGrassI, with the replacements $C\rightarrow C'$ and
$\phi\rightarrow U\phi =-\phi\, U$.
\endproof

\remark{\STM\remOSappGrassII}{
Recall from Lemma \lemOSjhat\ that, if $C(k)$ is a function on $\bbbr\times\bbbr^d$ and $C(\xi,\xi')$ the 
associated covariance in the sense of Definition \defOSftcov, then 
$C =-\widehat{ C(k)} J$. Also recall from Remark \remOSsymmetryConsequences.iii
 that if $U$ is antisymmetric,
particle number conserving, translation invariant and spin independent,
then $U=J\widehat{\check U(k)}$. In this case $C' =-\widehat{ C'(k)} J$
where $C'(k)=\sfrac{C(k)}{1-C(k)\check U(k)}$.

}

\vfill\eject

\titlea{ References}\PG\pgOSIIref

\item{[FKTf1]} J. Feldman, H. Kn\"orrer, E. Trubowitz, 
{\bf A Two Dimensional Fermi Liquid, Part 1: Overview}, preprint.
\smallskip%
\item{[FKTf3]} J. Feldman, H. Kn\"orrer, E. Trubowitz, 
{\bf A Two Dimensional Fermi Liquid, Part 3: The Fermi Surface}, preprint.
\smallskip%
\item{[FKTr1]} J. Feldman, H. Kn\"orrer, E. Trubowitz, 
{\bf Convergence of Perturbation Expansions in Fermionic Models, Part 1: Nonperturbative Bounds}, preprint.
\smallskip%
\item{[FKTo1]} J. Feldman, H. Kn\"orrer, E. Trubowitz, 
{\bf Single Scale Analysis of Many Fermion Systems, Part 1: Insulators}, preprint.
\smallskip%
\item{[FKTo3]} J. Feldman, H. Kn\"orrer, E. Trubowitz, 
{\bf Single Scale Analysis of Many Fermion Systems, Part 3: Sectorized Norms}, preprint.
\smallskip%
\item{[FKTr2]} J. Feldman, H. Kn\"orrer, E. Trubowitz, 
{\bf Convergence of Perturbation Expansions in Fermionic Models, Part 2: Overlapping Loops}, preprint.
\smallskip%
\item{[FMRT]} J.Feldman, J. Magnen, V. Rivasseau, E. Trubowitz, 
{\bf Two Dimensional Many Fermion Systems as Vector Models},
Europhysics Letters, {\bf 24} (1993) 521-526.
\smallskip%

\vfill\eject

\titlea{Notation }\PG\pgOSIInot
\null\vskip-0.8in
\vfil
\titleb{Norms}
\centerline{
\vbox{\offinterlineskip
\hrule
\halign{\vrule#&
         \strut\hskip0.05in\hfil#\hfil&
         \hskip0.05in\vrule#\hskip0.05in&
          #\hfil\hfil&
         \hskip0.05in\vrule#\hskip0.05in&
          #\hfil\hfil&
           \hskip0.05in\vrule#\cr
height2pt&\omit&&\omit&&\omit&\cr
&Norm&&Characteristics&&Reference&\cr
height2pt&\omit&&\omit&&\omit&\cr
\noalign{\hrule}
height2pt&\omit&&\omit&&\omit&\cr
&$\tn\ \cdot\ \tn_{1,\infty}$&&no derivatives, external positions, acts on functions&&Example \exOSSymmNorm&\cr
height4pt&\omit&&\omit&&\omit&\cr
&$\|\ \cdot\ \|_{1,\infty}$&&derivatives, external positions, acts on functions&&Example \exOSSymmNorm&\cr
height4pt&\omit&&\omit&&\omit&\cr
&$\|\ \cdot\ \cnorm_\infty$&&derivatives, external momenta, acts on functions
&&Definition \defOSderivmom&\cr
height4pt&\omit&&\omit&&\omit&\cr
&$\tn\ \cdot\ \tn_{\infty}$&&no derivatives, external positions, acts on functions&&Example \exOSelloneinftycontr&\cr
height4pt&\omit&&\omit&&\omit&\cr
&$\|\ \cdot\ \cnorm_1$&&derivatives, external momenta, acts on functions
&&Definition \defOSderivmom&\cr
height4pt&\omit&&\omit&&\omit&\cr
&$\|\ \cdot\ \cnorm_{\infty,B}$&&derivatives, external momenta, $B\subset\bbbr\times\bbbr^d$
&&Definition \defOSderivmom&\cr
height4pt&\omit&&\omit&&\omit&\cr
&$\|\ \cdot\ \cnorm_{1,B}$&&derivatives, external momenta, $B\subset\bbbr\times\bbbr^d$
&&Definition \defOSderivmom&\cr
height4pt&\omit&&\omit&&\omit&\cr
&$\|\ \cdot\ \|$&&$\rho_{m;n}\|\ \cdot\ \|_{1,\infty}$&&Lemma \lemOSscalednorm&\cr
height4pt&\omit&&\omit&&\omit&\cr
&$N(\cW;\cb,\ib,\al)$&&$\sfrac{1}{\ib^2}\,\cb\!\sum_{m,n\ge 0}\,
\al^{n}\,\ib^{n} \,\|\cW_{m,n}\|$&&Definition \defOSgrnorm&\cr
height4pt&\omit&&\omit&&\omit&\cr
& && &&Theorem  \thmOSinsulators&\cr
height4pt&\omit&&\omit&&\omit&\cr
&$N_0(\cW;\be;X,\vec\rho)$&&$\fe_0(X)\ \sum_{m+n\in 2\bbbn}\,
\be^{n}\rho_{m;n} \,\|\cW_{m,n}\|_{1,\infty}$
&&Theorem  \thmOSfirststep&\cr
height4pt&\omit&&\omit&&\omit&\cr
&$\|\ \cdot\ \|_{L^1}$&&derivatives, acts on functions on $\bbbr\times\bbbr^d$
&&before Lemma \lemOSprepintup&\cr
height4pt&\omit&&\omit&&\omit&\cr
&$\|\ \cdot\ \tnorm$&&derivatives, external momenta, acts on functions
&&Definition \defOSdiffdecaynorm&\cr
height4pt&\omit&&\omit&&\omit&\cr
&$N^\sim_0(\cW ;\be;X,\vec\rho)$&&$\fe_0(X)
\!\sum_{m+n\in 2\bbbn}\,\be^{m+n}\rho_{m;n} \,\| W^\sim_{m,n}\tnorm$
&&before Lemma \lemOSTZsourceterm&\cr
height4pt&\omit&&\omit&&\omit&\cr
}\hrule}}

\vfil\vfil
\goodbreak
\titleb{Other Notation}
\vfill
\centerline{
\vbox{\offinterlineskip
\hrule
\halign{\vrule#&
         \strut\hskip0.05in\hfil#\hfil&
         \hskip0.05in\vrule#\hskip0.05in&
          #\hfil\hfil&
         \hskip0.05in\vrule#\hskip0.05in&
          #\hfil\hfil&
           \hskip0.05in\vrule#\cr
height2pt&\omit&&\omit&&\omit&\cr
&Not'n&&Description&&Reference&\cr
height2pt&\omit&&\omit&&\omit&\cr
\noalign{\hrule}
height2pt&\omit&&\omit&&\omit&\cr
&$\Om_S(\cW)(\phi,\psi)$
&&$\log\sfrac{1}{Z} \int  e^{\cW(\phi,\psi+\ze)}\,d\mu_{S}(\ze)$
&&before (\eqnOSintroI)&\cr
height2pt&\omit&&\omit&&\omit&\cr
&$J$&&particle/hole swap operator&&(\eqnOSjdef)&\cr
height2pt&\omit&&\omit&&\omit&\cr
&$\tilde \Om_C(\cW)(\phi,\psi)$
&&$\log \sfrac{1}{Z}\int e^{\phi J\ze}\,e^{\cW(\phi,\psi +\ze)} d\mu_C(\ze)$
&&Definition \defOSrengrpmap&\cr
height2pt&\omit&&\omit&&\omit&\cr
&$r_0$&&number of $k_0$ derivatives tracked&&\S\CHintroII&\cr
height2pt&\omit&&\omit&&\omit&\cr
&$r$&&number of $\k$ derivatives tracked&&\S\CHintroII&\cr
height2pt&\omit&&\omit&&\omit&\cr
&$M$&&scale parameter, $M>1$&&before Definition \defOSscales&\cr
height2pt&\omit&&\omit&&\omit&\cr
&$\const$&&generic constant, independent of scale&& &\cr
height2pt&\omit&&\omit&&\omit&\cr
&$\abcst$&&generic constant, independent of scale and $M$&& &\cr
height2pt&\omit&&\omit&&\omit&\cr
&$\nu^{(j)}(k)$&&$j^{\rm th}$ scale function&&Definition \defOSscales&\cr
height2pt&\omit&&\omit&&\omit&\cr
&$\tilde\nu^{(j)}(k)$&&$j^{\rm th}$ extended scale function
&&Definition \defOSextendedshell.i&\cr
height2pt&\omit&&\omit&&\omit&\cr
&$\nu^{(\ge j)}(k)$&&$\varphi\big(M^{2j-1}(k_0^2+e(\k)^2)\big)$&&Definition \defOSscales&\cr
height2pt&\omit&&\omit&&\omit&\cr
&$\tilde\nu^{(\ge j)}(k)$
&&$\varphi\big(M^{2j-2}(k_0^2+e(\k)^2)\big)$
&&Definition \defOSextendedshell.ii&\cr
height2pt&\omit&&\omit&&\omit&\cr
&$\bar\nu^{(\ge j)}(k)$
&&$\varphi\big(M^{2j-3}(k_0^2+e(\k)^2)\big)$
&&Definition \defOSextendedshell.iii&\cr
height2pt&\omit&&\omit&&\omit&\cr
&$S(C)$&&$\sup_m\sup_{\xi_1,\cdots,\xi_m \in \cB}\
\Big(\ \Big| \int \psi(\xi_1)\cdots\psi(\xi_m)\,d\mu_C(\psi) \Big|\ \Big)^{1/m}$&&Definition \defIntBndsS&\cr
height2pt&\omit&&\omit&&\omit&\cr
&$\check f$&&Fourier transform&&Definition \defOSfourtrans.i&\cr
height2pt&\omit&&\omit&&\omit&\cr
&$f^\sim$&&partial Fourier transform&&Definition \defOSfourtrans.ii&\cr
height2pt&\omit&&\omit&&\omit&\cr
&$\hat\chi$&&Fourier transform&&Definition \defOSfourtransII&\cr
height2pt&\omit&&\omit&&\omit&\cr
&$\cB$&&$\bbbr \times \bbbr^d \times \{\uparrow, \downarrow\}\times\{0,1\}$ 
viewed as position space&&beginning of \S\CHnorms&\cr
height2pt&\omit&&\omit&&\omit&\cr
&$\check \cB$&&$\bbbr\times\bbbr^d\times\{\uparrow, \downarrow\}\times\{0,1\}$ 
viewed as momentum space&&beginning of \S \CHfourier&\cr
height2pt&\omit&&\omit&&\omit&\cr
&$\check \cB_m$&&$\set{(\check \eta_1,\cdots,\check \eta_m)\in \check \cB^m}
{\check \eta_1+\cdots+\check \eta_m=0}$&&before Definition \defOSamptransinv&\cr
height2pt&\omit&&\omit&&\omit&\cr
&$\cF_m(n)$&&functions on $\cB^m \times \cB^n$, antisymmetric in $\cB^m$
arguments&&Definition \defOSFmn&\cr
height2pt&\omit&&\omit&&\omit&\cr
&$\check\cF_m(n)$&&functions on $\check\cB^m \times \cB^n$, antisymmetric in $\check\cB^m$
arguments&&Definition \defOScheckcF&\cr
height2pt&\omit&&\omit&&\omit&\cr
}\hrule}}

\vfill

\end